\documentclass[12pt]{article}

\usepackage[utf8]{inputenc}
\usepackage[english]{babel}
\usepackage{amsmath, amssymb}
\usepackage[range-units=brackets, range-phrase=-]{siunitx}[=v2]
\usepackage[scale=0.97]{XCharter}
\usepackage[xcharter, bigdelims, vvarbb, scaled=1.05]{newtxmath}
\usepackage{fullpage}
\usepackage{graphicx}
\usepackage[sort&compress,numbers]{natbib}
\usepackage[total={6.5in,9in}, top=1in, headsep=0.1in, headheight=1in]{geometry}
\usepackage{setspace}
\usepackage[hidelinks, linktocpage=true]{hyperref}
\usepackage[babel=true]{microtype}

\usepackage{authblk}

\newcommand\snowmass{
\begin{center}
  \rule[-0.2in]{\hsize}{0.01in}\\
  \rule{\hsize}{0.01in}\\
  \vskip 0.1in
  Submitted to the Proceedings of the US Community Study\\ 
  on the Future of Particle Physics (Snowmass 2021)\\
  \rule{\hsize}{0.01in}\\
  \rule[+0.2in]{\hsize}{0.01in}\\[-2em]
\end{center}
}

\usepackage[firstpage=true]{background}
\backgroundsetup{contents={\parbox{6.5in}{\snowmass}}, scale=1, placement=top, opacity=1, color=black, position={3.25in,0.95in}}

\usepackage{fancyhdr}
\fancypagestyle{plain}{%
  \fancyhf{}%
  \fancyhead[C]{}
}

\fancypagestyle{empty}{%
  \fancyhf{}%
  \fancyhead[C]{\textit{Inflation: Theory and Observations}}
  \fancyfoot[C]{\thepage}
}
\numberwithin{equation}{section}
\allowdisplaybreaks[1]
\DeclareMathAlphabet{\pazocal}{OMS}{zplm}{m}{n} 
\renewcommand{\mathcal}[1]{\pazocal{#1}}

\renewcommand{\d}{\mathrm{d}}
\newcommand{\ee}{\mathrm{e}}
\newcommand{\ii}{\mathrm{i}}
\newcommand{\As}{A_\mathrm{s}}
\newcommand{\ns}{n_\mathrm{s}}
\newcommand{\At}{A_\mathrm{t}}
\newcommand{\nt}{n_\mathrm{t}}
\newcommand{\Mp}{M_\mathrm{p}}
\newcommand{\fnl}{f_\mathrm{NL}}
\DeclareSIUnit{\parsec}{pc}
\DeclareSIUnit{\Mpc}{\mega\parsec}
\DeclareSIUnit{\h}{\mathit{h}}

\title{Inflation: Theory and Observations\\[12pt]{\fontsize{13.5}{15}\selectfont Editors: Guilherme~L.\ Pimentel, Benjamin Wallisch and W.~L.~Kimmy Wu}\vspace{-10pt}}
\date{}

\author[1,2]{Ana Ach\'ucarro,}
\author[3,4,5,6]{Matteo~Biagetti,}
\author[7,8]{Matteo~Braglia,}
\author[9]{Giovanni~Cabass,}
\author[10]{Robert~Caldwell,}
\author[11]{Emanuele~Castorina,}
\author[12]{Xingang~Chen,}
\author[13]{William~Coulton,}
\author[14]{Raphael~Flauger,}
\author[7,15]{Jacopo~Fumagalli,}
\author[9]{Mikhail~M.~Ivanov,}
\author[16]{Hayden~Lee,}
\author[17]{Azadeh~Maleknejad,}
\author[18]{P.~Daniel~Meerburg,}
\author[19]{Azadeh~Moradinezhad~Dizgah,}
\author[20]{Gonzalo~A.~Palma,}
\author[1,21]{Guilherme~L.~Pimentel,}
\author[22]{S\'ebastien~Renaux-Petel,}
\author[9,14]{Benjamin Wallisch,}
\author[22,13]{Benjamin~D.~Wandelt,}
\author[22]{\\Lukas~T.~Witkowski\hskip1pt}
\author[23,24]{W.~L.~Kimmy Wu\hskip1pt}

\affil[1]{Lorentz Institute for Theoretical Physics, Leiden University, 2333~CA Leiden, The Netherlands}
\affil[2]{Department of Physics, University of the Basque Country~UPV-EHU, 48940~Bilbao, Spain}
\affil[3]{Institute for Fundamental Physics of the Universe (IFPU), 34014~Trieste, Italy}
\affil[4]{International School for Advanced Studies~(SISSA), 34136~Trieste, Italy}
\affil[5]{INAF -- Osservatorio Astronomico di Trieste, 34143~Trieste, Italy}
\affil[6]{National Institute for Nuclear Physics (INFN), 34127~Trieste, Italy}
\affil[7]{Instituto de F\'isica Teorica UAM-CSIC, Universidad Aut\'onoma de Madrid, 28049~Madrid, Spain}
\affil[8]{INAF -- Osservatorio di Astrofisica e Scienza dello Spazio di Bologna, 40129~Bologna, Italy}
\affil[9]{School of Natural Sciences, Institute for Advanced Study, Princeton, NJ~08540, USA}
\affil[10]{Department of Physics and Astronomy, Dartmouth College, Hanover, NH~03755, USA}
\affil[11]{Dipartimento di Fisica ``Aldo Pontremoli'', Universit\`a{} degli Studi di Milano, 20133~Milano, Italy}
\affil[12]{Harvard-Smithsonian Center for Astrophysics, Cambridge, MA~02138, USA}
\affil[13]{Center for Computational Astrophysics, Flatiron Institute, New York, NY~10010, USA}
\affil[14]{Department of Physics, University of California San Diego, La Jolla, CA~92093, USA}
\affil[15]{Departamento de F\'isica Teorica, Universidad Aut\'onoma de Madrid, 28049~Madrid, Spain}
\affil[16]{Kavli Institute for Cosmological Physics, University of Chicago, Chicago, IL~60637, USA}
\affil[17]{Theoretical Physics Department, CERN, 1211~Gen\`eve, Switzerland}
\affil[18]{Van Swinderen Institute for Particle Physics and Gravity, University of Groningen, 9747~AG~Groningen, The~Netherlands}
\affil[19]{D\'epartement de Physique Th\'eorique, Universit\'e de Gen\`eve, 1211~Gen\`eve, Switzerland}
\affil[20]{Departamento de F\'isica, Facultad de Ciencias F\'isicas y Matem\'aticas, Universidad de Chile, Santiago, Chile}
\affil[21]{Institute of Physics, University of Amsterdam, 1098~XH~Amsterdam, The Netherlands}
\affil[22]{Institut d'Astrophysique de Paris (IAP), CNRS \& Sorbonne University, 75014~Paris, France}
\affil[23]{SLAC National Accelerator Laboratory, Menlo Park, CA~94025, USA}
\affil[24]{Kavli Institute for Particle Astrophysics and Cosmology, Stanford, CA~94305, USA}

\begin{document}

\pagenumbering{roman}

\maketitle

\begin{abstract}
\noindent
Cosmic inflation provides a window to the highest energy densities accessible in nature, far beyond those achievable in any realistic terrestrial experiment. Theoretical insights into the inflationary era and its observational probes may therefore shed unique light on the physical laws underlying our universe. This white paper describes our current theoretical understanding of the inflationary era, with a focus on the statistical properties of primordial fluctuations. In particular, we survey observational targets for three important signatures of inflation: primordial gravitational waves, primordial non-Gaussianity and primordial features. With the requisite advancements in analysis techniques, the tremendous increase in the raw sensitivities of upcoming and planned surveys will translate to leaps in our understanding of the inflationary paradigm and could open new frontiers for cosmology and particle physics. The combination of future theoretical and observational developments therefore offer the potential for a dramatic discovery about the nature of cosmic acceleration in the very early universe and physics on the smallest scales.
\end{abstract}

\clearpage
\begin{spacing}{0.80}\begin{center}{\small\textbf{Endorsers}\vspace{-2.0ex}}\end{center}{\fontsize{9.5}{12}\selectfont Lars Aalsma, 
Kevork Abazajian, 
Tom Abel, 
Aliakbar Abolhasani, 
Peter Adshead, 
Fruzsina Julia Agocs, 
Zeeshan Ahmed, 
Kazuyuki Akitsu, 
Yashar Akrami, 
Soner Albayrak, 
Mustafa Amin, 
Luis Anchordoqui, 
Adam Anderson, 
Behzad Ansarinejad, 
Amjad Ashoorioon, 
Tasos Avgoustidis, 
Anton Baleato Lizancos, 
Denis Barkats, 
Alexandre Barreira, 
Darcy Barron, 
Nicola Bartolo, 
Ritoban Basu Thakur, 
Daniel Baumann, 
Dominic Beck, 
Amy Bender, 
Charles Bennett, 
Bradford Benson, 
Jose Luis Bernal, 
Florian Beutler, 
Shubham Bhardwaj, 
Federico Bianchini, 
Colin Bischoff, 
Lindsey Bleem, 
James Bock, 
Christian Boehmer, 
Boris Bolliet, 
J.~Richard Bond, 
Julian Borrill, 
François~R.\ Bouchet, 
Jonathan Braden, 
Rafael Bravo, 
Philippe Brax, 
Samuel Brieden, 
Thejs Brinckmann, 
Marco Bruni, 
Sean Bryan, 
Cliff Burgess, 
Christian Byrnes, 
Guadalupe Cañas-Herrera, 
Graeme Candlish, 
John Carlstrom, 
John Joseph Carrasco, 
Sean Carroll, 
Julien Carron, 
Jorge~L.\ Cervantes-Cota, 
Sebastian Cespedes, 
Anthony Challinor, 
Clarence Chang, 
Tzu-Ching Chang, 
Jie-Wen Chen, 
Thomas~Y.\ Chen, 
James Cheshire, 
Kolen Cheung, 
Jens Chluba, 
Susan Clark, 
Katy Clough, 
Timothy Cohen, 
Thomas Colas, 
Alex Cole, 
Edmund Copeland, 
James Cornelison, 
Nathaniel Craig, 
Paolo Creminelli, 
Ari Cukierman, 
Francis-Yan Cyr-Racine, 
Guido D'Amico, 
Neal Dalal, 
Ruth Daly, 
Anne-Christine Davis, 
Roger de Belsunce, 
Jacques Delabrouille, 
Nicholas DePorzio, 
Vincent Desjacques, 
Eleonora Di Valentino, 
Marion Dierickx, 
Konstantinos Dimopoulos, 
Xi Dong, 
Olivier Doré, 
Carlos Duaso Pueyo, 
Jo Dunkley, 
Ruth Durrer, 
Reza Ebadi, 
Lorenz Eberhardt, 
Josquin Errard, 
Misael Arnaldo Espinal Valladares, 
Angelo Esposito, 
Thomas Essinger-Hileman, 
Giulio Fabbian, 
JiJi Fan, 
Richard Feder, 
Chang Feng, 
Andrea Ferrara, 
Simone Ferraro, 
Elisa Ferreira, 
Pedro~G.\ Ferreira, 
Jeffrey Filippini, 
Hassan Firouzjahi, 
Thomas Flöss, 
Emanuele Fondi, 
Simon Foreman, 
Sébastien Fromenteau, 
Nicholas Galitzki, 
Silvia Galli, 
Jose Tomas Galvez Ghersi, 
Mauricio Gamonal, 
Juan Garcia-Bellido, 
Martina Gerbino, 
Hector Gil-Marin, 
Neil Goeckner-Wald, 
Jinn-Ouk Gong, 
Victor Gorbenko, 
Peter Graham, 
Tanguy Grall, 
Daniel Green, 
Daniel Grin, 
Alessandro Gruppuso, 
Riccardo Gualtieri, 
Jon~E.\ Gudmundsson, 
Federica Guidi, 
Eraj Gulraiz, 
Yi Guo, 
ChangHoon Hahn, 
Jiashu Han, 
Shaul Hanany, 
Jonte Hance, 
Will Handley, 
Daniel Harlow, 
Katrin Heitmann, 
Sophie Henrot-Versillé, 
Brandon Hensley, 
Thomas Hertog, 
Mark Hertzberg, 
J.~Colin Hill, 
Kurt Hinterbichler, 
Renee Hlozek, 
William Holzapfel, 
Anson Hook, 
Sina Hooshangi, 
Shaun Hotchkiss, 
Selim Hotinli, 
Bin Hu, 
Kevin Huffenberger, 
Ayodeji Ibitoye, 
Sadra Jazayeri, 
Matthew Johnson, 
William Jones, 
Austin Joyce, 
Shamit Kachru, 
David Kaiser, 
Alba Kalaja, 
Renata Kallosh, 
Marc Kamionkowski, 
Jae Hwan Kang, 
Arun Kannawadi, 
Kirit Karkare, 
Brian Keating, 
Ryan Keeley, 
Alex Kehagias, 
Stephen Kent, 
Ali Rida Khalifeh, 
Justin Khoury, 
William Kinney, 
Lloyd Knox, 
Kazunori Kohri, 
Nickolas Kokron, 
John Kovac, 
Ely Kovetz, 
Soubhik Kumar, 
Chao-Lin Kuo, 
Julius Kuti, 
King Lau, 
Louis Legrand, 
Qiuyue Liang, 
Andrew Liddle, 
Robert Lilow, 
Eugene Lim, 
Andrei Linde, 
Marilena Loverde, 
Kaloian Lozanov, 
Jiyu Lu, 
Xiao-Han Ma, 
Yin-Zhe Ma, 
Mathew Madhavacheril, 
Sagar Kumar Maity, 
Juan Maldacena, 
Ameek Malhotra, 
Karim Malik, 
Abhishek~S.\ Maniyar, 
Gustavo Marques-Tavares, 
Jerome Martin, 
Liam McAllister, 
Evan McDonough, 
Scott Melville, 
Joel Meyers, 
Marius Millea, 
Vivian Miranda, 
Swagat Saurav Mishra, 
Zahra Molaee, 
Pierluigi Monaco, 
Gabriele Montefalcone, 
Paulo Montero-Camacho, 
Maria Elena Monzani, 
Jakob Moritz, 
Edward Morvan, 
Tony Mroczkowski, 
Suvodip Mukherjee, 
Moritz Münchmeyer, 
Elena Murchikova, 
Johanna Nagy, 
Mohammad Hossein Namjoo, 
Laura Newburgh, 
Michael Niemack, 
Christian Nitschelm, 
Gustavo Niz, 
Yasunori Nomura, 
Mahdiyar Noorbala, 
Kourosh Nozari, 
Ken Olum, 
Alain Omont, 
Yuuki Omori, 
Giorgio Orlando, 
Ogan Özsoy, 
Sonia Paban, 
Antonio Padilla, 
Lyman Page, 
Enrico Pajer, 
Zhaodi Pan, 
Kevin Pardede, 
Julio Parra-Martinez, 
Vivek Pathak, 
Subodh Patil, 
Hiranya Peiris, 
Leandros Perivolaropoulos, 
Patrick Peter, 
Matthew Petroff, 
Oliver~H.~E.\ Philcox, 
Michel Piat, 
Massimo Pietroni, 
Lucas Pinol, 
Anna Porredon, 
Marieke Postma, 
Karthik Prabhu, 
Akhil Premkumar, 
Clement Pryke, 
Davide Racco, 
Benjamin Racine, 
Antonio Racioppi, 
Alexandra Rahlin, 
Fazlu Rahman, 
Marco Raveri, 
Christian Reichardt, 
Mathieu Remazeilles, 
Xin Ren, 
Gerasimos Rigopoulos, 
Antonio Riotto, 
Simón Riquelme, 
Natalie Roe, 
Diederik Roest, 
Ashley Ross, 
Sandip Roy, 
John Ruhl, 
Augusto Sagnotti, 
Pankaj Saha, 
Noah Sailer, 
Mairi Sakellariadou, 
Benjamin Saliwanchik, 
Domenico Sapone, 
Murali Saravanan, 
Nikolina Sarcevic, 
Misao Sasaki, 
V.~H.\ Satheeshkumar, 
Marco Scalisi, 
Emmanuel Schaan, 
Koenraad Schalm, 
Fabian Schmidt, 
Benjamin Schmitt, 
David Seery, 
Emiliano Sefusatti, 
Neelima Sehgal, 
Leonardo Senatore, 
Edgar Shaghoulian, 
Sarah Shandera, 
Mikhail Shaposhnikov, 
M.~M.\ Sheikh-Jabbari, 
Chia-Hsien Shen, 
Stephen Shenker, 
Blake Sherwin, 
Gary Shiu, 
Joseph Silk, 
Maximiliano Silva-Feaver, 
Eva Silverstein, 
Sara Simon, 
Rajeev Singh, 
Charlotte Sleight, 
Anže Slosar, 
Wuhyun Sohn, 
Adam Solomon, 
David Spergel, 
Antony Stark, 
David Stefanyszyn, 
Andre Steklain, 
Radek Stompor, 
Michael Strauss, 
Ganesh Subramaniam, 
Raman Sundrum, 
Aritoki Suzuki, 
Spyros Sypsas, 
David Tacero, 
Alireza Talebian, 
Ting Tan, 
Massimo Taronna, 
Grant Teply, 
Ayngaran Thavanesan, 
Peter Timbie, 
Andrew Tolley, 
Jesus Torrado, 
Cynthia Trendafilova, 
Enrico Trincherini, 
Matthieu Tristram, 
Oem Trivedi, 
Mark Trodden, 
Yu-Dai Tsai, 
Gregory Tucker, 
Gansukh Tumurtushaa, 
Gustavo Joaquin Turiaci, 
Cora Uhlemann, 
Caterina Umilta, 
Jean-Philippe Uzan, 
Jorinde van de Vis, 
Jan Pieter van der Schaar, 
Alexander van Engelen, 
Thomas Van Riet, 
Mariana Vargas-Magaña, 
Vincent Vennin, 
Clara Vergès, 
Filippo Vernizzi, 
Nelson Videla, 
Digvijay Wadekar, 
Bob Wagoner, 
Yidun Wan, 
David Wands, 
Dong-Gang Wang, 
Frank Wang, 
Gensheng Wang, 
Yi Wang, 
Zun Wang, 
Scott Watson, 
Duncan Watts, 
Zachary Weiner, 
Denis Werth, 
Gilles Weymann-Despres, 
Michael Wilson, 
Mark Wise, 
Edward Wollack, 
Zhong-Zhi Xianyu, 
Zhilei Xu, 
Sheng-Feng Yan, 
Zhao Yaqi, 
Cyndia Yu, 
Matias Zaldarriaga, 
Ivonne Zavala, 
David Zegeye, 
Cristóbal Zenteno, 
Yunlong Zheng, 
Siyi Zhou, 
Zihan Zhou, 
Houri Ziaeepour and
Andrea Zonca}\end{spacing}

\clearpage
\tableofcontents

\clearpage
\pagenumbering{arabic}
\setcounter{page}{1}

\clearpage
% !TEX root = whitepaper.tex

\section{Introduction}

The origin of the universe is one of the most fascinating questions in science. In its first moments, the universe appears to have been very flat and filled with a hot plasma. This plasma has small inhomogeneities which grow under the influence of gravity to seed the formation of structure. The origin of this primordial era, and of the initial conditions for the cosmological history of the universe, is still a matter of very active research. Excitingly, we live in times when this question can be addressed quantitatively. Through powerful theoretical ideas and exquisitely precise observations, which are tied together by major advances in modeling and data analysis, we can probe the earliest moments of the universe. The leading paradigm to explain its beginning posits that, prior to the hot big bang, the universe underwent a phase of exponential expansion that sets up its very special initial conditions. This era is known as cosmic inflation.\medskip

Although the detailed mechanism driving inflation is still unknown, the underlying framework makes predictions that are supported by cosmological observations to a striking degree. The apparently acausal long-range correlations and the mentioned plasma inhomogeneities at the onset of the hot phase of the universe are elegantly explained as originating from quantum fluctuations that are sourced throughout inflation and get stretched to enormous distances as the universe expands exponentially fast. Moreover, the self-similar behavior of an expanding cosmology manifests itself in the almost scale invariance of these fluctuations. Importantly, the predictions of inflation can often be calculated reliably since specific models of inflation are amenable to weakly coupled descriptions.\medskip

Ushering us into the era of precision cosmology are technological advances which have enabled a huge influx of observational data that can be used to probe these predictions. In particular, the almost scale-invariant initial conditions have been inferred with high statistical significance from multiple cosmological probes, notably the observed distribution of matter and radiation in the universe. These observational tests are possible because matter and radiation trace the earliest phase of the universe: the initial conditions provide the seeds for structure formation, around which dark matter forms a big scaffolding to which galaxies, stars and planets ultimately attach.\medskip

\begin{figure}[t]
		\centering
		\includegraphics{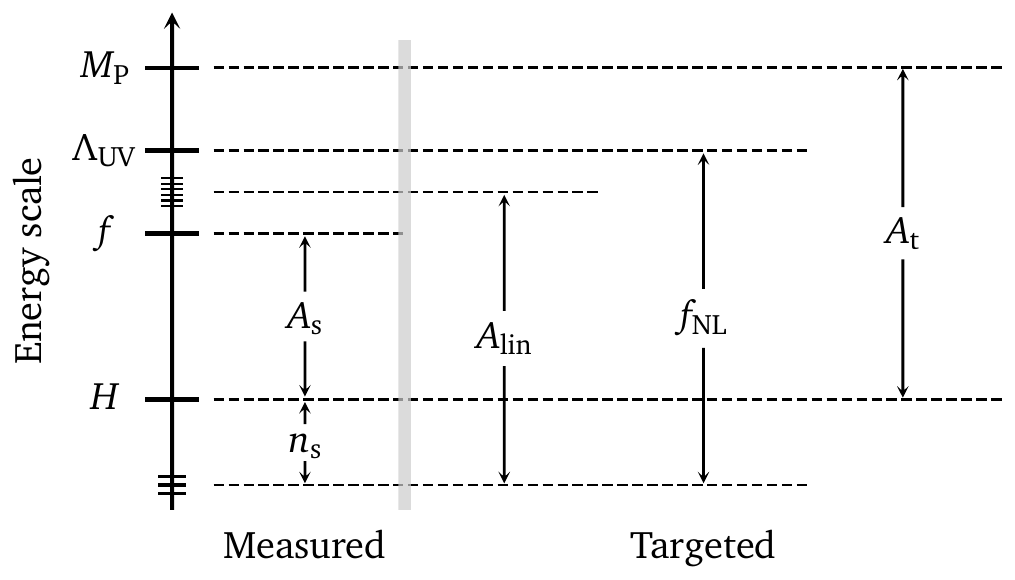}
		\caption{Sketch of the known energy scales relevant to inflationary cosmology, and how they are related to observables that have been measured and will be targeted in the next decade. The energy scales in descending order are the Planck scale~$\Mp$, the scale beyond which the scalar fluctuations become strongly coupled,~$\Lambda_\mathrm{UV}$, the scale which controls the size of the scalar primordial fluctuations,~$f$, and the Hubble scale~$H$ during inflation. Additional scales that are determined by observations, but are more model dependent, are also included. The amplitude~$\As$ of the scalar power spectrum of initial fluctuations and its spectral tilt~$\ns$ have already been measured. In addition, we indicate primordial features, primordial non-Gaussianity and primordial tensor modes by the amplitude~$A_\mathrm{lin}$ of linear oscillatory features as a proxy for more general features, the relative bispectrum amplitude~$\fnl$ and the tensor amplitude~$A_t$, respectively. A detection of these three prime observables, which are targeted in the next decade and the main subject of this white paper, will be sensitive to higher energy scales, giving us access to the ``energy frontier'' in cosmology.}
		\label{fig:scales}
\end{figure}

To study the beginning of the universe and learn about the physics of inflation, much effort has been focused on the detailed study and observational characterization of the statistical two-point function of primordial density~(scalar) perturbations. Since observations are consistent with these fluctuations being Gaussian and almost scale invariant, we can encapsulate its statistics well using a power-law power spectrum characterized by two parameters. These are two of the seven\hskip1pt\footnote{The remaining five parameters are associated with the geometry and composition of the universe: The matter content of the universe is described by the baryon and cold dark matter densities, the radiation content is parameterized by the photon temperature, and dark energy is included via its energy density, completed by the optical depth due to reionization.} cosmological parameters describing the standard model of cosmology,~$\Lambda$CDM: the scalar amplitude $\As$,~which parameterizes the amplitude, and the spectral tilt~$\ns$, which parameterizes the scale dependence of the density perturbation power spectrum. In fact, with these two numbers, we can already gain insights today into some aspects of inflation and probe some of the energy scales which play an important role in this era of the universe. This is illustrated in Fig.~\ref{fig:scales}, which shows a model-agnostic~(but necessarily fuzzy) picture of these scales. %
We see that the spectral tilt~$\ns$ is generated from a ratio of different low-energy scales compared to the Hubble rate~$H$ during inflation. (For instance, in single-field inflation, these ratios are the slow-roll parameters and the new energy scales come from the time dependence of~$H$.) On the other hand, the scalar amplitude~$\As$ is directly related to an unknown scale~$f$ relative to the Hubble scale~$H$.\medskip
 
To understand even more aspects of the primordial universe, it is useful to consider other observables beyond the nearly scale-invariant scalar power spectrum. In our quest to decode the physics of inflation, the statistics of primordial fluctuations are the main tool. While generating the inflationary background~(a quasi-de Sitter era of exponential expansion) is also an important outstanding problem, the dynamics of the background and the fluctuations can be treated separately. This is emphasized, for example, in the effective field theory approach to inflation,\footnote{In the language of effective field theory, the separation between the dynamics of the background and the fluctuations is evident since background quantities only affect induced interactions. This relatively recent development provides a framework for consistently studying more general models of inflation. While it is possible to write down effective theories for the background and fluctuations together, the fluctuations tend to interact more weakly than theoretically allowed.} and the reason why we focus on the observational and theoretical aspects of the fluctuations in this white paper. In the context of a specific inflationary model, the background dynamics can also be constrained with an observation of the fluctuations.

More specifically, we center our attention on the simplest imprints of new physics in the statistics of the initial conditions:
\begin{itemize}
	\item Inflation predicts that \textbf{primordial gravitational waves} are produced by quantum fluctuations just like the primordial density fluctuations. They are tensor fluctuations of the metric, characterized in particular by their two-point correlation function~(or power spectrum);
	\item Inflation also predicts small deviations from Gaussianity. The simplest statistical measure of \textbf{primordial non-Gaussianity} is the three-point function~(the bispectrum) of density fluctuations;
	\item Many models of inflation introduce new scales in the inflationary era. These new scales manifest themselves as \textbf{primordial features} in the power spectrum and higher-point spectra of primordial fluctuations.
\end{itemize}
The detection of any of these observables would give us access to new energy scales and the detailed dynamics of inflation, while also carving the space of viable models. At the same time, even in the absence of any detection, more stringent limits on these intrinsic predictions of the inflationary paradigm~(and its alternatives) will severely restrict the vast landscape of models and teach us powerful lessons about the primordial universe.

A schematic view on the relationship of primordial gravitational waves, non-Gaussianity and features to the various inflationary energy scales is presented in Fig.~\ref{fig:scales}. These scales can be sharply determined in a specific model of inflation which therefore predicts correlations between different observables. In single-field slow-roll inflation, for example, the ratio of the tensor and the scalar power spectra, referred to as the tensor-to-scalar ratio~$r$, and the spectral index of tensor fluctuations are proportional to each other, and the level of non-Gaussianity is controlled by the size of the scalar spectral tilt. Similarly, the introduction of features in the scalar power spectrum often comes from a new energy scale during inflation leaving similar imprints in other observables, e.g.\ a feature in the shape of non-Gaussianities. This shows that studying inflationary models is important to sharpen the energy scales relevant during inflation, and to predict the size of and potential correlations between observables.\medskip

Since inflation sets the initial conditions for the entire causal evolution of the universe, signatures of this primordial era are imprinted in all cosmological observations. So far, observations of the cosmic microwave background~(CMB) have provided the strongest constraints on all three observables, and it is expected to continue to provide significant improvements to the measurements and serve as an important anchor to the other observational probes in the coming decade. Similar to the~CMB, tracers of the large-scale structure~(LSS) of the universe are sensitive to the largest scales of the primordial spectrum. With upcoming observations, constraints from~LSS will be competitive with CMB~constraints for primordial non-Gaussianity and features. On smaller scales, spectral distortions of the CMB~black body spectrum and direct observations of the stochastic gravitational wave background~(SGWB) will be valuable in advancing our understanding of inflation on a broad range of scales and are expected to have significant improvements in sensitivity in the coming decade.\medskip

While theoretically studying and observationally probing inflation through these three observables might be the most promising path in the next decade, the landscape of phenomena and research questions related to inflation is much broader. This means that we are not able to cover many relevant topics, such as non-adiabatic fluctuations, primordial black holes, primordial magnetic fields, (p)reheating and general inflationary model building, and open questions, such as how inflation started and how inflation ended. At the same time, we focus on the observational signatures of inflation, but we will only be able to mention some of the far-reaching advances and remaining open problems in cosmological observations that are being actively tackled, such as astrophysical foregrounds and modeling, theoretical descriptions, instrumental systematics and new experimental probes.

For in-depth discussions on many of these topics, we refer to dedicated Snowmass~2021 white papers on theoretical cosmology~\cite{Snowmass2021:CosmoTheory}, the cosmological bootstrap~\cite{Snowmass2021:Bootstrap}, the application of effective field theory to cosmology~\cite{Snowmass2021:CosmoEFT}, primordial black holes~\cite{Snowmass2021:PBH}, early-universe model building~\cite{Snowmass2021:EarlyModels}, data-driven cosmology~\cite{Snowmass2021:CosmoData} and the stochastic gravitational wave background~\cite{Snowmass2021:SGWB}. The white papers on light relics~\cite{Snowmass2021:LightRelics} and cosmological tensions~\cite{Snowmass2021:CosmologyIntertwined} also review topics related to inflation. On the observational front, the experimental design of~\mbox{CMB-S4}, as described in the Snowmass~2021 white paper~\cite{Snowmass2021:CMBs4}, will be of particular importance to probe primordial gravitational waves, and primordial non-Gaussianity and primordial features, respectively. Additional CMB-related white papers provide an overview of CMB~measurements~\cite{Snowmass2021:CMBobs} and a description of~\mbox{CMB-HD}~\cite{Snowmass2021:CMBhd}. Future large-scale structure surveys, as detailed in the white papers on observational probes from the three-dimensional large-scale structure~\cite{Snowmass2021:3dLSS}, \SI{21}{cm}~and millimeter-wavelength line intensity mapping~\cite{Snowmass2021:21cm, Snowmass2021:LIM}, and cross-correlations of observables~\cite{Snowmass2021:StaticProbes}, will be instrumental to push the sensitivity to non-Gaussian and feature signatures of inflation. Gravitational wave observatories will complement these observational efforts, cf.~the Snowmass~2021 white paper~\cite{Snowmass2021:GWO}. The cross-disciplinary approach and importance of inflationary research is also evident from the list of related white papers submitted to the Decadal Survey on Astronomy and Astrophysics~2020~(Astro2020)~\cite{Shandera:2019ufi, Meerburg:2019qqi, Slosar:2019gvt, Chluba:2019kpb, Bandura:2019uvb, Schlegel:2019eqc, Baker:2019pnp, Reitze:2019iox, Ferraro:2019uce}, which included dedicated papers on primordial gravitational waves~\cite{Shandera:2019ufi}, primordial non-Gaussianity~\cite{Meerburg:2019qqi} and primordial features~\cite{Slosar:2019gvt}. Finally, we refer the reader to review articles on various aspects of inflationary research~\cite{Bartolo:2004if, Komatsu:2009kd, Baumann:2009ds, Liguori:2010hx, Chen:2010xka, Maleknejad:2012fw, Wang:2013zva, Baumann:2014nda, Alvarez:2014vva, Chluba:2015bqa, Renaux-Petel:2015bja, Kamionkowski:2015yta, Desjacques:2016bnm, Caprini:2018mtu, Staggs:2018gvf, Martin:2018ycu, Martin:2019zia, Biagetti:2019bnp, Domenech:2021ztg, Komatsu:2022nvu}.\bigskip

The outline of this white paper is as follows: We conclude this section with a summary of the observational status and prospects. In Sections~\ref{sec:pgw},~\ref{sec:png} and~\ref{sec:features}, we discuss primordial gravitational waves, primordial non-Gaussianity and primordial features, respectively. In each section, we provide a review of the theoretical state of the art of these observables and describe their imprints in cosmological observations. In addition, we summarize current constraints, mention analysis and modeling challenges, and examine future directions. We conclude in Sec.~\ref{sec:conclusion}.

\setlength{\fboxsep}{10pt}
\noindent\fbox{%
	\parbox{0.95\textwidth}{%
		\begin{center} \textbf{Status and Prospects of Observational Constraints}\end{center}

		With many current and upcoming surveys, it is an exciting time for precision cosmology---general predictions of the inflationary paradigm can be put to test given the upcoming data. We outline below a few highlights of what an observational detection of each of the three main observables would tell us about the early universe:
		\begin{itemize}
			\item For primordial gravitational waves, the current upper limit on the tensor-to-scalar ratio, $r < 0.035$ at~95\%~C.L., along with the observed spectral tilt~$\ns$ strongly disfavors single-field monomial models. We can look forward to data from CMB~experiments in the next decade to cross important theory thresholds of $r \simeq 0.01$ and $r \simeq 0.001$, with the former being associated with a super-Planckian excursion of the inflaton field and the latter being associated with classes of models that naturally predict the observed spectral tilt~$\ns$.
			
			\item For primordial non-Gaussianity, the current limits on the amplitude of the three major bispectrum shapes are $\fnl^\mathrm{local} = -0.9\pm 5.1$, $\fnl^\mathrm{equil} = -26 \pm 47$ and $\fnl^\mathrm{ortho} = -38 \pm 24$, showing no evidence for primordial non-Gaussianity. With data from upcoming and proposed CMB~and LSS~experiments, we anticipate $\sigma(\fnl^\mathrm{local}) < 1$, enabling the differentiation of models that include the existence of extra light species during or after inflation. We also anticipate an improvement on bounds for $\fnl^\mathrm{equil,ortho}$, which will constrain the symmetry breaking patterns of inflation.
			
			\item For primordial features, cosmological data have constrained the departures from the almost scale-invariant primordial spectrum to be less than one percent of the scalar amplitude~$\As$. The future influx of CMB~and especially LSS~data is projected to improve these limits by one to two orders of magnitude~(or make a detection at these levels). These insights will be complemented by constraints on small scales from measurements targeting CMB~spectral distortions and the stochastic gravitational wave background. The combination of theory and observations may offer an exciting opportunity to not only reveal a portion of rather detailed evolutionary history of inflation, but also provide direct model-independent evidence for the inflationary paradigm.
		\end{itemize}
		With the tremendous increase in data volume and complexity, dedicated analysis, modeling and simulation efforts are essential to ensure robust measurements and interpretation of these datasets. Given the strong theoretical foundation, with many new exciting ongoing developments, influx of high-quality data, and dedicated analysis and interpretation, allowable model spaces will continue to shrink, qualitative and quantitative features of the inflationary era will be better understood, and groundbreaking discoveries could be just around the corner. 
	}
}
\clearpage
% !TEX root = whitepaper.tex

%%%%%%%%%%%%%%%%
\section{Primordial Gravitational Waves}
\label{sec:pgw}
%%%%%%%%%%%%%%%%

Many of the simplest models of inflation predict the existence of a stochastic background of primordial gravitational waves~(PGWs), which are tensor perturbations of the metric from the very early universe. Since these perturbations are quantum fluctuations, a detection of these~PGWs would constitute a glimpse of quantum gravity at work. In addition, the amplitude of the tensor perturbation spectrum is related to the expansion rate during inflation which can be related to the energy scale of inflation in the simplest models. For a range of amplitudes of the tensor perturbation spectrum that are measurable in the near future, a detection would point to an energy scale near the scale of grand unification. This would open new frontiers of accessible energies for cosmology and particle physics~(see e.g.~\cite{Kamionkowski:2015yta, Caprini:2018mtu, Staggs:2018gvf} for previous reviews).\medskip

Inflation was originally proposed to solve the horizon and flatness problems.\footnote{Since then, alternative scenarios that solve the horizon and flatness problems have been proposed~(see e.g.~\cite{Brandenberger:2018wbg} for a recent review).} The simplest realization relies on a single scalar field whose energy density is dominated by its potential energy density. As inflation proceeds, the field slowly rolls down its potential. The slowly-varying potential energy density leads to a period of accelerated expansion. Inflation ends once the kinetic and potential energy densities become comparable. Subsequently, the inflaton decays, filling the universe with a hot plasma of relativistic Standard Model particles (and possibly particles beyond the Standard Model). This general picture is referred to as single-field slow-roll inflation. As the scalar field rolls down, its quantum-mechanical fluctuations are then responsible for sourcing the inhomogeneities in the early universe, producing the seeds for structure formation. Moreover, fluctuations of the energy density of the vacuum lead to the rippling of spacetime. In other words, the perturbations of the inflaton can be identified with the scalar perturbations of the metric. Similarly, tensor perturbations of the metric are gravitational waves, which is the topic of this section.\medskip

Given the profound consequences of a detection of~PGWs from inflation, many current and planned experiments are designed to search for their signatures which can be categorized in two classes: (i)~measuring the effects of~PGWs on other cosmological observations and (ii)~directly measuring the strain of spacetime induced by~PGWs. The first class of measurements make use of observations of the cosmic microwave background, large-scale structure and astronomical objects on smaller scales. The second class of measurements are pursued by interferometric gravitational wave~(GW) observatories. While we will touch on many of these observations, we will mainly focus on the reach of CMB~polarization because many of the simplest inflationary models predict tensor modes at levels detectable by upcoming CMB~experiments.

Upper limits on the tensor spectrum amplitude, parameterized by the tensor-to-scalar ratio~$r$, have already been set by multiple CMB~experiments. The BICEP/Keck collaboration published the current tightest upper limit, $r < 0.035$~(95\%~C.L.), which has strongly disfavored a class of single-field monomial models when combined with the measurement of the spectral tilt of the scalar spectrum,~$\ns$. With the advances in experimental sensitivities and analysis techniques in the next decade, these limits will improve by more than an order of magnitude or potentially make a detection, crossing critical thresholds~($r \simeq 0.01$ and $r \simeq 0.001$). Moreover, future data from experiments with increasing sensitivities could shed light on non-vacuum mechanisms of PGW~production which can arise when embedding inflation in particle physics and string theory setups, revealing valuable information about the particle physics of inflation. Future CMB~experiments could therefore make great leaps in constraining the available inflationary model space.
% !TEX root = whitepaper.tex

% Primordial Gravitational Waves

\subsection{Theoretical Background}

As was already mentioned in the introduction, according to inflation, the very early universe underwent a period of nearly exponential expansion. According to the simplest models of inflation, this expansion is driven by the energy density in a scalar field that is described by the action
\begin{equation}
	S[\phi] = \int \d^4x \sqrt{-g} \left[-\frac{1}{2} g^{\mu\nu} \partial_\mu\phi \partial_\nu\phi - V(\phi)\right],
\end{equation}
where the potential~$V(\phi)$ is an a-priori arbitrary function that characterizes the model. In the $3+1$~(or ADM)~decomposition~\cite{Arnowitt:1962hi}, the line element is parameterized as
\begin{equation}
	\d s^2 = -N^2 \d t^2 + a^2 h_{ij} (\d x^i + N^i \d t)(\d x^j + N^j \d t)\, ,
\end{equation}
and we can choose a gauge so that all dynamical degrees of freedom are described by components of the spatial metric~$h_{ij}$. The fluctuations describing the density~(scalar;~$\mathcal{R}$) and gravitational wave~(tensor;~$\gamma$) fluctuations can be taken as
\begin{equation}
	h_{ij} = \ee^{2\mathcal{R}} (\ee^\gamma)_{ij} = \left(\delta_{ij} + 2\mathcal{R} \delta_{ij} + \gamma_{ij} + 2\mathcal{R}^2 \delta_{ij} + \frac{\gamma_{ik}\gamma_{kj}}{2} + \cdots\right),\quad \partial_i\gamma_{ij} = 0\,,\quad \gamma_{ii}=0\,,	\label{eq:hij}
\end{equation}
with the lapse~$N$ and the shift~$N^i$ being determined in terms of the dynamical degrees of freedom by the Hamiltonian and momentum constraints. In general, vector perturbations may also be present, but these rapidly decay and can be neglected unless they are actively sourced, e.g.~by cosmic strings or other defects.

The time evolution of the scale factor~$a(t)$ is described by the Friedmann equation,
\begin{equation}
	H^2 = \frac{8\pi G}{3}\rho\, ,
\end{equation}
where the Hubble rate~$H$ is defined by $H = \dot{a}/a$, with the overdot indicating a time derivative, and the energy density is $\rho = \frac{1}{2}\dot{\phi}^2+V(\phi)$. The nearly exponential growth of the scale factor characteristic for inflation corresponds to a Hubble rate that is approximately constant or, more precisely, a Hubble rate whose fractional rate of change is small compared to the Hubble rate itself,
\begin{equation}
	\epsilon \equiv -\frac{\dot{H}}{H^2} \ll1 \, .
\end{equation}
For the simplest models of inflation this holds, provided that the kinetic energy density is small compared to the potential energy density. In order to solve the horizon and flatness problems~\cite{Guth:1980zm, Linde:1981mu, Albrecht:1982wi, Kazanas:1980tx, Sato:1980yn}, this condition must hold sufficiently long so that the potential must be sufficiently flat to keep the fractional rate of change of the inflaton velocity small compared to the Hubble rate, at least on average.\medskip

To discuss the time evolution of the perturbations, it is convenient to introduce their Fourier expansions
\begin{equation}
	\mathcal{R}(t,\mathbf{x}) = \int \frac{\d^3 k}{(2\pi)^3}\mathcal{R}(t,\mathbf{k})\ee^{\ii \mathbf{k}\cdot\mathbf{x}} + \mathrm{h.c.}\, ,
\end{equation}
and
\begin{equation}
\gamma_{ij}(t,\mathbf{x}) = \sum_\lambda \int\frac{\d^3k}{(2\pi)^3} \gamma_\lambda(t,\mathbf{k}) e_{ij}(\mathbf{k},\lambda) \ee^{\ii \mathbf{k}\cdot\mathbf{x}} + \mathrm{h.c.}\,,
\end{equation}
where~$e_{ij}(\mathbf{k},\lambda)$ is the polarization tensor for a graviton with comoving wavenumber~$\mathbf{k}$ and polarization~$\lambda$, and ``h.c.'' denotes the Hermitian conjugate. In an inflationary universe, the physical wavenumber~$k/a$ of any mode at early times far exceeds the expansion rate~$H = \dot{a}/a$. In this limit, the modes are ``inside the horizon'' and oscillate rapidly. It is assumed that the modes are only excited to the extent required by quantum mechanics at these early times. At later times, the physical wavenumber drops below the expansion rate, and the modes ``exit the horizon'', $k\ll aH$. It can be shown that~$\mathcal{R}(t,\mathbf{k})$ and~$\gamma_\lambda(t,\mathbf{k})$ become time-independent in this limit~\cite{Weinberg:2003sw}. In particular, they are not affected by unknown physics associated with the end of inflation or subsequent epochs about which very little is known, such as dark matter decoupling.

At some point, inflation ends and the universe becomes filled with ordinary matter. The expansion rate decreases more rapidly than the wavenumber and eventually the modes ``enter the horizon'' again,~$k\gg aH$, and begin to oscillate. The conservation of~$\mathcal{R}$ and~$\gamma_{ij}$ outside the horizon ensures that the statistical properties of the scalar and tensor fluctuations are preserved and allow us to infer the conditions of the inflationary epoch from late-time observations. If the probability distribution governing the primordial perturbations is statistically homogeneous, isotropic and parity-invariant, the two-point correlation functions can be parameterized as
\begin{equation}
	\begin{split}
		\langle \mathcal{R}(\mathbf{k}) \mathcal{R}(\mathbf{k}^{\prime}) \rangle &= (2\pi)^3 \delta^3(\mathbf{k} + \mathbf{k}^{\prime})\, \frac{2\pi^2}{k^3}\, \Delta^2_{\mathcal{R}}(k)\, , \\
		\langle\gamma_\lambda(\mathbf{k})\gamma_{\lambda^{\prime}}(\mathbf{k}^{\prime})\rangle &= (2\pi)^3 \delta_{\lambda\lambda^{\prime}}\, \delta^3(\mathbf{k} + \mathbf{k}^{\prime})\, \frac{2\pi^2}{k^3}\, \frac{1}{2} \Delta^2_\gamma(k)\, ,
	\end{split}
\end{equation}
where the factor of~$1/2$ in the last line reflects the fact that the measured power includes contributions from each of the two graviton polarizations. 

Observationally, we know that the power spectrum of primordial density perturbations is well-described by
\begin{equation}
	\Delta^2_{\mathcal{R}}(k) = \As \left(\frac{k}{k_\ast}\right)^{\ns-1} ,	\label{eq:primordial_power_spectrum}
\end{equation}
with a spectral index~$\ns$ that is nearly, but not exactly unity. The amplitude of primordial density perturbations is constrained at the percent level and all observations are currently consistent with $\Delta^2_h(k) = 0$. The fact that the power spectrum of primordial density perturbations is however well-described by a power law suggests an ansatz of the form
\begin{equation}
	\Delta^2_h(k)=\At\left(\frac{k}{k_\ast}\right)^{\nt} .
\end{equation}
Since the power spectrum of primordial density perturbations has been measured, it is common to quantify the amplitude of the primordial gravitational wave signal by the tensor-to-scalar ratio,
\begin{equation}
	r = \frac{\At}{\As}\, .
\end{equation}

For the simplest models of inflation, the primordial power spectrum of density perturbations is given by
\begin{equation}
	\Delta^2_\mathcal{R}(k) = \frac{1}{2\epsilon \Mp^2} \left(\frac{H}{2\pi}\right)^2 ,	\label{eq:DR2}
\end{equation}
where~$\Mp$ is the (reduced)~Planck mass, and the Hubble rate~$H$ and slow-roll parameter~$\epsilon$ should be evaluated at a time when $k = aH$. Since both $\epsilon$ and~$H$ are slowly varying functions of time by construction, we expect a nearly (but not exactly) scale-invariant spectrum. More specifically, since the rate of change of the Hubble rate is negative definite and the slow-roll parameter increases monotonically during inflation, we expect the power to decrease with increasing wavenumber so that $\ns \lesssim 1$. This is sometimes referred to as a red spectrum and is consistent with observations. Over the next decade, constraints on the scalar spectral index~$\ns$ will improve by a factor of two. Together with improved constraints on the running of the scalar spectral index and on primordial gravitational waves, which we will focus on here, this significantly reduces the space of available models and may also distinguish between different reheating histories after inflation. Additional discussion of the implications of improved measurements of the primordial density perturbations can be found in the Snowmass~2021 White Paper~\cite{Snowmass2021:CosmoData}.\medskip

The primordial power spectrum of gravitational waves is 
\begin{equation}
	\Delta^2_h(k) = \frac{8}{\Mp^2} \left(\frac{H}{2\pi}\right)^2 ,	\label{eq:Dh2}
\end{equation}
with~$H$ again being evaluated at a time when $k = aH$. Since~$H$ varies slowly as inflation proceeds, we expect a nearly scale-invariant spectrum of primordial gravitational waves. More specifically, since~$H$ decreases with time, we also expect the primordial gravitational wave spectrum to be red so that $\nt \lesssim 0$. The tensor-to-scalar ratio for these models then is $r = 16\epsilon$.

Rather remarkably, this is the power spectrum associated with quantum fluctuations in the metric. A detection of this signal would therefore provide evidence for quantum gravity. The form of the primordial gravitational wave power spectrum highlights that a detection would yield a measurement of the expansion rate of the universe during inflation. In addition, since the energy density of the scalar field is dominated by the potential energy density, the Friedmann equation can be used to infer the energy scale of inflation. This relation is often used to express the energy scale in terms of the tensor-to-scalar ratio as
\begin{equation}
	V^{1/4} = \SI{1.04e16}{GeV} \left(\frac{r}{0.01}\right)^{1/4} .
\end{equation}
This highlights that inflation would have occurred near the energy scale associated with grand unified theories for a tensor-to-scalar ratio within reach of CMB~observations, which will be discussed in~\textsection\ref{sec:pgw_observations}. A detection would therefore provide evidence for new physics at energy scales far beyond the reach of any terrestrial experiment.\medskip

For the simplest models of inflation, the tensor-to-scalar ratio furthermore constrains the distance traversed by the inflaton~\cite{Lyth:1996im},
\begin{equation}
	\frac{\Delta\phi}{\Mp} \gtrsim \left(\frac{r}{8}\right)^{1/2} \mathcal{N}_\ast\, ,
\end{equation}
where~$\mathcal{N}_\ast$ measures the number of ``e-folds'', or the natural logarithm of the ratio of the scale factor at the end of inflation and when the pivot scale~$k_\ast$ exits the horizon, $k_\ast = a H$. The precise value of~$\mathcal{N}_\ast$ depends on the details of reheating, the epoch when inflation ends and the universe becomes filled with ordinary matter. A typical number for reheating that occurs nearly instantaneously is of order sixty~(with details depending on the choice of model as well as pivot scale). Less-efficient reheating scenarios can typically lead to a reduction in this number of order ten, but reheating can also be further delayed. Taking a very conservative value of $\mathcal{N}_\ast = 30$, we see that any detection of a tensor-to-scalar ratio above $r \gtrsim 0.01$ would imply that the inflaton must have traveled over a distance larger than~$\Mp$. 

Theories of quantum gravity are expected to contain new degrees of freedom at or below the Planck scale. In the absence of symmetries, the inflaton is expected to interact with these degrees of freedom. These interactions would imply features in the inflaton potential on sub-Planckian scales that prevent the inflaton from rolling slowly over a super-Planckian distance. This implies that a detection of gravitational waves with $r \gtrsim 0.01$ would provide evidence for a symmetry of nature that forbids these interactions. Whether such symmetries and, more generally, super-Planckian excursions can occur in a fundamental theory of quantum gravity continues to be the subject of an active debate~\cite{Obied:2018sgi}.\medskip

It is natural to ask whether theoretical considerations single out additional scales. To highlight such an additional scale, we will follow an argument laid out in~\cite{CMB-S4:2016ple}. Provided $\epsilon \ll 1$, the tensor-to-scalar ratio obeys an ordinary first-order differential equation~\cite{Mukhanov:2013tua, Roest:2013fha, Creminelli:2014nqa},
\begin{equation}
	\frac{\d\log r}{\d \mathcal{N}} = \left[\ns(\mathcal{N})-1\right] + \frac{r}{8}\, .	\label{eq:drdn}
\end{equation}
The observed departure from exact scale invariance is numerically close to~$1/\mathcal{N}_\ast$. Of course, this might just be a numerical coincidence, but it implies that models of inflation for which the functional dependence of the spectral index on the number of e-folds is
\begin{equation}
	\ns(\mathcal{N})-1 = -\frac{p+1}{\mathcal{N}}\, ,
\end{equation}
with~$p$ being a positive number of order unity, naturally predict the observed departure from scale invariance. The general solution~\eqref{eq:drdn} can then be written as
\begin{equation}
	r(\mathcal{N}) = \frac{8p}{\mathcal{N}} \frac{1}{1+(\mathcal{N}/\mathcal{N}_\mathrm{eq})^p}\, ,
\end{equation}
where~$\mathcal{N}_\mathrm{eq}$ is an integration constant. In the absence of additional hierarchies in the theory, we expect one of the terms in the denominator to dominate and expect the solution to be well-approximated by one of the two limiting cases,
\begin{equation}
	r(\mathcal{N}) = \frac{8p}{\mathcal{N}} \qquad \text{and} \qquad r(\mathcal{N}) = \frac{8p}{\mathcal{N}} \left(\frac{\mathcal{N}_\mathrm{eq}}{\mathcal{N}}\right)^p,	\label{eq:r_vs_N}
\end{equation}
with~$\mathcal{N}_\mathrm{eq}$ now expected to be of order unity.

We saw that the simplest single-field models are fully characterized by the potential~$V(\phi)$ and that we can derive the functional form of the potential that naturally explains the observed value of the spectral index~$\ns$. The first limiting form in~\eqref{eq:r_vs_N} corresponds to so-called monomial potentials, $V(\phi) = \mu^{4-2p} \phi^{2p}$. For reasonable choices of~$p$, these models predict $r>0.01$, which is well within the reach of upcoming experiments, and in their simplest form are already strongly disfavored.

For the second limiting form in~\eqref{eq:r_vs_N}, the qualitative behavior changes depending on whether~$p$ is smaller or larger than unity. Larger values, $p > 1$, describe the so-called ``hilltop'' models~\cite{Boubekeur:2005zm}. In this case, the potential near the origin in field space approaches a constant from below like a power of the field set by~$p$ and inflation takes place as the field rolls down the hill toward a minimum. Smaller values, $p < 1$, correspond to the so-called ``plateau'' models. In this case, the potential tends to a constant at large field values, again with a power set by~$p$, and the inflaton rolls down from the plateau toward the origin. Interestingly, data appear to favor $p = 1$, a special choice of plateau models in which the plateau is approached exponentially. 

All potentials associated with the second limiting form describing hilltop and plateau models are characterized by the distance in field space over which the potential appreciably departs from the hilltop or plateau. A priori, this scale, which is referred to as the ``characteristic scale'' of the potential, is unconstrained. However, we may expect this characteristic scale to be of order~$\Mp$ in the most economic scenario. This expectation is realized in many well-known models, such as Starobinsky's $R^2$~inflation~\cite{Starobinsky:1980te}, models in which the Higgs boson takes on the additional role of the inflaton~\cite{Bezrukov:2007ep, Ballesteros:2016xej} or, more generally, models with non-minimally coupled scalar fields~\cite{Fakir:1990iu, Fakir:1990eg}. In these models, the characteristic scale is of order the Planck scale because both are determined by the coefficient of the Einstein-Hilbert term. This class also includes $\alpha$-attractors~\cite{Kallosh:2013yoa, Kallosh:2014rga, Carrasco:2015pla}, fibre inflation~\cite{Cicoli:2008gp}, and Poincar\'e disk models~\cite{Ferrara:2016fwe, Kallosh:2017ced}.

The class of hilltop and plateau models with Planckian characteristic scale provides the target of $r = 0.001$ for the next generation of CMB~experiments~\cite{CMB-S4:2016ple}, to be discussed in the following. We see that this target is interesting both because the class contains many well-known models and because the absence of a detection would exclude the simplest models of inflation that naturally explain the observed value of the spectral index~$\ns$ with a super-Planckian characteristic scale.
% !TEX root = whitepaper.tex

% Primordial Gravitational Waves

\subsection{Observational Imprints}
\label{sec:pgw_observations}

Observations of the~CMB have provided model-independent constraints on the primordial perturbations that any theory of the early universe must obey. The primordial perturbations are dominated by density perturbations. Within observational uncertainties, these perturbations are adiabatic, Gaussian, and their power spectrum is well-approximated by a power law and nearly, but not exactly scale-invariant~($\ns \lesssim 1$). All these properties are predicted by the simplest models of inflation. As we reviewed, inflation also predicts a nearly scale-invariant background of~PGWs. Because the energy density in gravitational waves rapidly redshifts after modes enter the horizon, observations of CMB~polarization provide the most promising avenue to detect this characteristic signature. Furthermore, many of the simplest models of inflation, in particular those based on symmetries in which inflation occurs at high energies and with a large field displacement, predict amplitudes of~PGWs within reach of upcoming CMB~experiments. As a consequence, we will focus on the imprint of gravitational waves on the primary CMB~polarization anisotropies, but will also touch on other ideas to search for this signal, such as intrinsic alignments, shear and clustering of galaxies, circular polarization of \SI{21}{cm}~radiation, polarized Sunyaev-Zel'dovich~(SZ) tomography, and direct detection of primordial gravitational waves with gravitational wave observatories.

\subsubsection{Cosmic Microwave Background}
In the early universe, the baryon-photon plasma underwent acoustic oscillations seeded by the primordial perturbations. When the universe was around \num{400000}~years old, it became cool enough for hydrogen to form. During this period of ``recombination'', the photons decoupled from the baryon-photon plasma leading to a transparent universe filled with cosmic black body radiation that now peaks at microwave frequencies and is referred to as the~CMB. Some of these CMB~photons have interacted with free electrons after the first stars and galaxies reionized the universe, but most of them experience their first non-gravitational interaction since recombination as they are measured by CMB~experiments.\medskip

At the time the~CMB was released, the universe was homogeneous and isotropic to one part in $\numrange{e4}{e5}$. The small anisotropies in the observed CMB~temperature arise predominantly from photon density fluctuations at the time of recombination. For practical reasons, it is convenient to expand the observed temperature perturbations in terms of spherical harmonics~$Y_{\ell m}$,
\begin{equation}
	\Delta T(\hat{n}) = \sum\limits_{\ell,m} a^T_{\ell m} Y_{\ell m}^{}(\hat{n})\, ,
\end{equation}
and work with the so-called multipole coefficients~$a^T_{\ell m}$.

Individual photons are linearly polarized by their last scattering event: the CMB~radiation observed in a particular direction acquires a net polarization through scattering off electrons that experience a local temperature quadrupole, which is generated by dissipative processes in the presence of velocity gradients in the baryon-photon fluid. This means that polarization anisotropies provide an observational handle on the velocity of the medium, while the temperature anisotropies predominantly probe the intrinsic temperature perturbations.

The polarization of the~CMB is conventionally encoded by the Stokes~$Q$ and $U$~parameters, which characterize the linear polarization of the radiation field. These can be expanded in terms of spin-weighted spherical harmonics~${}_{2}Y_{\ell m}$,
\begin{equation}
	Q(\hat{n}) + \ii\hskip1pt U(\hat{n}) = -\sum_{\ell,m} \left(a^E_{\ell m} + \ii\hskip0.5pt a^B_{\ell m}\right) {}_{2}^{}Y_{\ell m}^{}(\hat{n})\, .
\end{equation}
The information about the temperature and polarization anisotropies are therefore encoded in the multipole coefficients~$a^T_{\ell m}$, $a^E_{\ell m}$ and~$a^B_{\ell m}$. Since observations have demonstrated that the~CMB is remarkably Gaussian, the ensemble averages are well-characterized by the angular power spectra
\begin{equation}
	C^{XY}_{\ell} \delta_{\ell\ell'} \delta_{mm'} = \left\langle a^X_{\ell m} a^{Y*}_{\ell' m'} \right\rangle\,,	\label{eq:Cl}
\end{equation}
where $X, Y \in \{T, E, B\}$ and the star denotes the complex conjugate. It can be shown that density perturbations only generate temperature and curl-free (parity-even) E-mode anisotropies at linear order in perturbation theory. In the absence of primordial gravitational waves, the angular power spectra are then given by
\begin{equation}
	C_{XX,\ell} = \int \frac{\d k}{k}\hskip1pt \Delta^2_\mathcal{R}(k) \left|\int\limits_0^{\tau_0} \d\tau\, S_X(k,\tau)\, u_{X,\ell}[k(\tau_0-\tau)]\right|^{\hskip1pt 2}\, ,	\label{eq:clscalar}
\end{equation}
where $X=T,E$, $S_X(k,\tau)$ are the so-called source functions that encode the evolution of the modes, and~$u_{X,\ell}$ are special functions that encode the geometry of the universe. If the universe is spatially flat, as we have implicitly assumed in writing~\eqref{eq:hij}, then $u_{T,\ell}=j_\ell$, where~$j_\ell$ are spherical Bessel functions. 

\begin{figure}[t]
	\centering
	\includegraphics{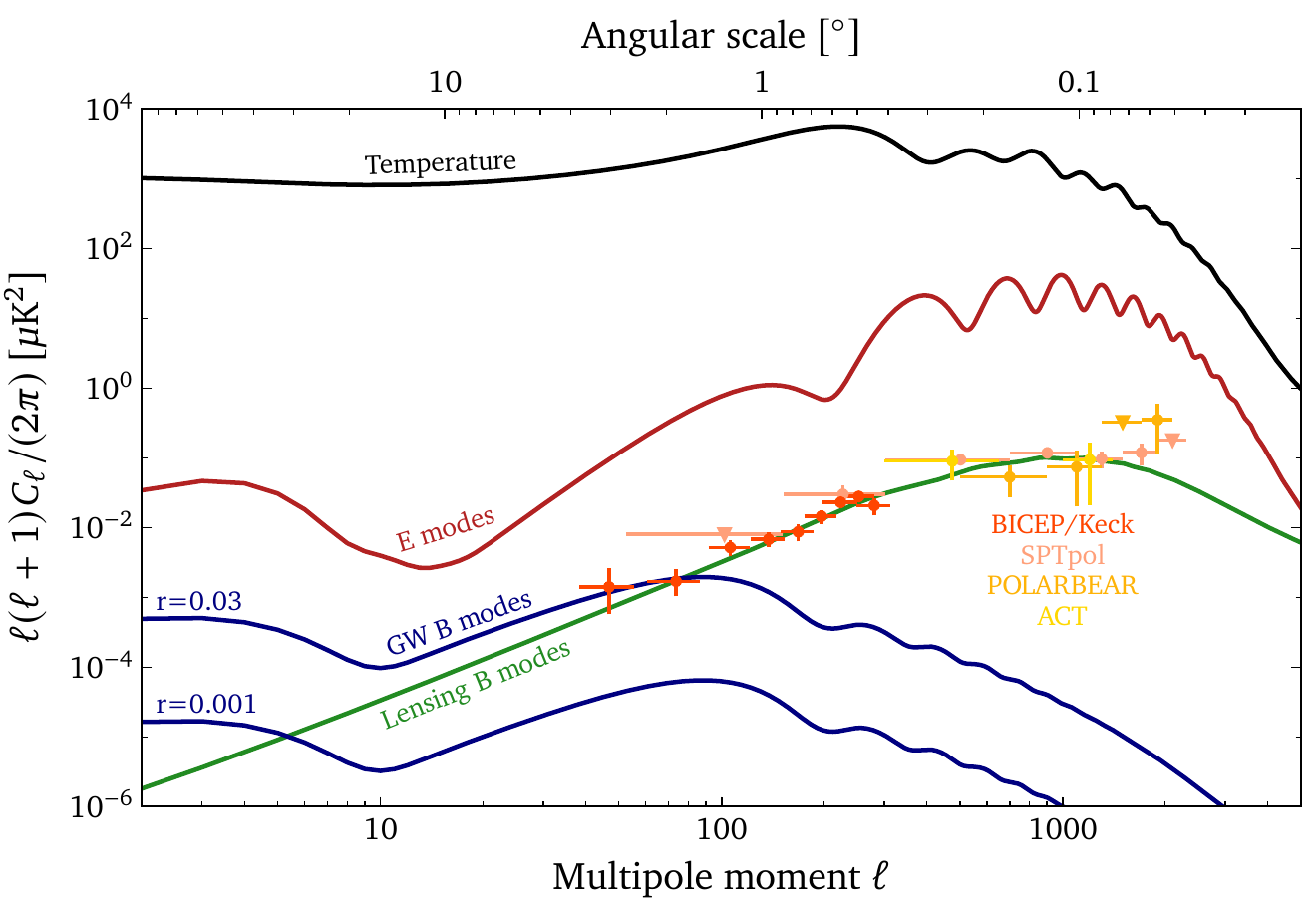}
	\caption{Theoretical predictions for the CMB~temperature~(black), E-mode~(red), and tensor B-mode~(blue) power spectra. Primordial B-mode spectra are shown for two representative values of the tensor-to-scalar ratio: $r=0.001$ and $r=0.03$. The contribution to tensor B~modes from scattering during the recombination epoch peaks around $\ell \approx 80$ and from reionization at $\ell < 10$. The expected values for the contribution to B~modes from gravitationally lensed E~modes are shown in green. Current measurements of the B-mode spectrum from ground-based experiments are displayed for BICEP/Keck~(dark orange)~\cite{BICEP:2021xfz}, SPTpol~(light orange)~\cite{SPT:2019nip}, POLARBEAR~(yellow)~\cite{POLARBEAR:2017beh}, and ACT~(light yellow)~\cite{ACT:2020frw}. The BICEP/Keck experiment has produced the most sensitive measurements of degree-scale B~modes, which are relevant for constraining the recombination peak from PGW~B~modes. The lensing contribution to the B-mode spectrum can be partially removed~(``delensed'') by measuring the E-mode polarization and exploiting the non-Gaussian statistics of the lensing signal.}
	\label{fig:clall}
\end{figure}

Primordial gravitational waves also create a local temperature quadrupole and leave an imprint in the temperature and polarization anisotropies. In particular, PGWs~generate divergence-free~(parity-odd) B-mode polarization and are the only source of B~modes at recombination at linear order. As such, B~modes of the~CMB provide a unique window to~PGWs~\cite{Kamionkowski:1996zd, Seljak:1996gy}. Figure~\ref{fig:clall}%
shows theoretical and observed angular power spectra for CMB~temperature and polarization anisotropies. The B-mode power spectrum created by gravitational waves generated during inflation are shown for two representative values of the tensor-to-scalar ratio.\footnote{We note that there could be matter fields that modify the PGW B-mode signal beyond the simplest models of inflation. These new observable signatures include chirality, non-Gaussianity and a blue-tilted spectrum~(see~\cite{Maleknejad:2011sq, Maleknejad:2011jw, Adshead:2012kp, Maleknejad:2012fw, Komatsu:2022nvu} for reviews and early works), and can generate cosmological~$C_\ell^{TB}$ and~$C_\ell^{EB}$~spectra~\cite{Lue:1998mq}.} We see that the angular power spectrum of B~modes generated by~PGWs peaks at angular multipoles around $\ell \approx 80$, which corresponds to the light horizon scale. This peak is often referred to as the recombination peak. In addition, rescattering of CMB~photons during reionization leads to another peak in the B-mode spectrum at $\ell < 10$. This is often referred to as the reionization bump. Ground-based experiments typically target the recombination peak, whereas satellite experiments target both the recombination peak and reionization bump.\medskip

\begin{figure}[t]
	\centering
	\includegraphics[width=5.5in, trim=15 3 7 23, clip]{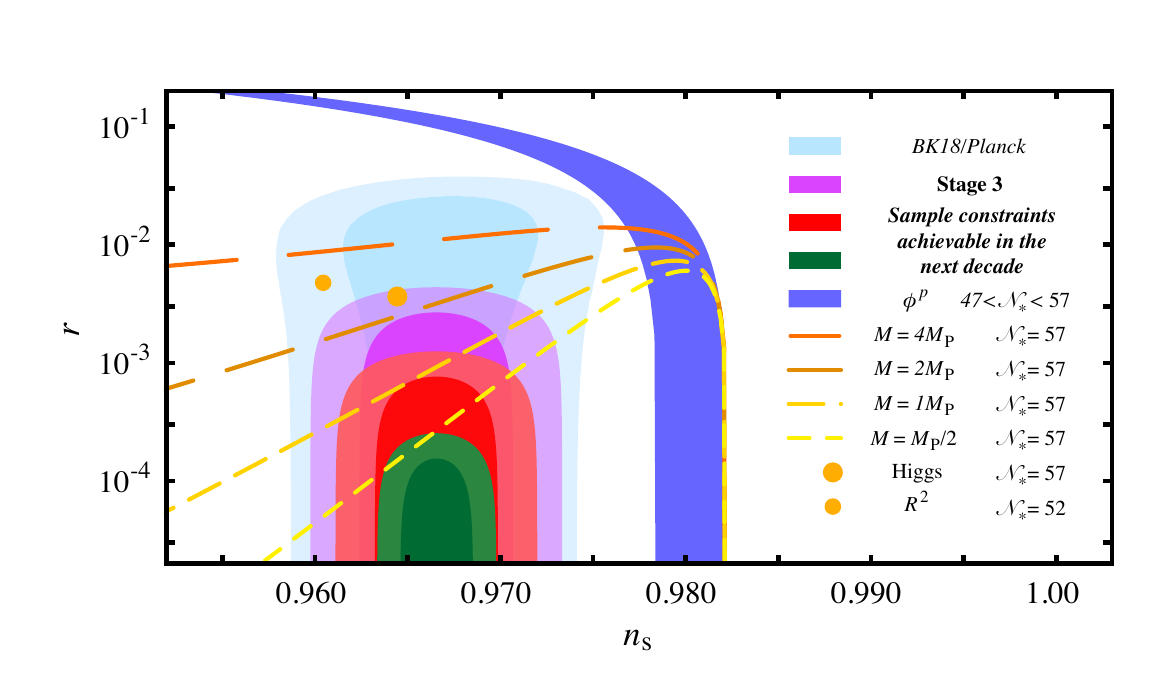}\vspace{-3pt}
	\caption{Predictions for the tensor-to-scalar ratio~$r$ and spectral index~$\ns$ for the class of single-field inflationary models in which \mbox{$\ns(\mathcal{N})-1\propto-1/\mathcal{N}$}. This class includes monomial models, $V(\phi)\propto \phi^p$~(dark blue), the Starobinsky~($R^2$) model and models in which inflation is driven by the Higgs field~(orange filled circles). The dashed lines show the predictions of models for different values of the characteristic scale in the potential. All models with a characteristic scale that exceeds the Planck scale can be detected or excluded in the next decade. In addition, the current constraints are shown in light blue~\cite{BICEP:2021xfz}, strongly disfavoring the single-field monomial models. The purple contours labeled~`Stage~3' show representative constraints of a combination of the Simons~\cite{SimonsObservatory:2018koc} and South Pole Observatories~\cite{Hui:2018cvg, SPT-3G:2021vps}. The red contours give an indication of the constraints achievable by~CMB-S4~\cite{CMB-S4:2020lpa} and LiteBIRD~\cite{LiteBIRD:2022cnt}, and the green contour shows the potential reach of a more futuristic high-resolution satellite mission, such as PICO~\cite{NASAPICO:2019thw}.}
	\label{fig:r_ns}
\end{figure}

\paragraph{Observational Progress}~\\
Given the significance of the implications of a detection of primordial gravitational waves, many experiments are designed to go after its so-called B-mode signature in the CMB~\cite{WMAP:2010qai, QUIET:2012szu, Harrington:2016jrz, Kusaka:2018yzq, Planck:2018vyg, POLARBEAR:2019kzz, SPT:2019nip, SPIDER:2021ncy, BICEP:2021xfz, POLARBEAR:2022dxa}. In the last about 10~years, the uncertainty on the amplitude of~PGWs as parameterized by the tensor-to-scalar ratio~$r$ has tightened by about two orders of magnitude, driven mainly by results from the ground-based experiment~BICEP/Keck located at the South Pole~\cite{BICEP:2021xfz}. Looking forward, the telescopes targeting~PGWs within the planned experiment~\mbox{CMB-S4}~\cite{CMB-S4:2020lpa, Snowmass2021:CMBs4}, as described in its dedicated Snowmass~2021 White Paper~\cite{Snowmass2021:CMBs4}, has adopted many of its design features from the BICEP/Keck~experiment to further drive the constraining power on~$r$. In fact, as illustrated in Fig.~\ref{fig:r_ns},%
current experiments, such as the Simons~\cite{SimonsObservatory:2018koc} and South Pole Observatories~\cite{Hui:2018cvg, SPT-3G:2021vps}, and planned experiments, such as LiteBIRD~\cite{LiteBIRD:2022cnt} and CMB-S4~\cite{CMB-S4:2020lpa, Snowmass2021:CMBs4}, are projected to cross two important thresholds over the next decade, respectively: (i)~the threshold around $r \simeq 0.01$, which is associated with monomial models and a super-Planckian excursion in field space that would provide strong evidence for the existence of an approximate shift symmetry in quantum gravity, and (ii)~the threshold at $r\simeq 0.001$, which is associated with the simplest models of inflation that naturally predict the observed value of the scalar spectral index~$\ns$ and have a characteristic scale that exceeds the Planck scale.

Before we discuss the challenges of this measurement, let us note again that the detection of the imprint of a PGW~background on the~CMB would have profound implications. It would constitute an indirect observation of quantum fluctuations in the spacetime metric and, therefore, of the quantum nature of gravity. In addition, it would provide evidence for new physics at the energy scale associated with grand unified theories. Finally, it would also have important implications for high-energy physics more generally, for example, by constraining axion physics and moduli, which are the fields that control the shapes and sizes of the internal manifold in string theory. For a detailed discussion of inflation in string theory and implications of a PGW~detection, we refer to the Snowmass~2021 White Paper~\cite{Snowmass2021:CosmoTheory}.

\paragraph{Measurement Challenges}~\\
The PGW-sourced B-mode power is orders of magnitude below the now well-measured temperature anisotropy spectrum, as shown in Fig.~\ref{fig:clall}. This poses the first challenge of the search of primordial B~modes in the~CMB: it is extremely faint. Given the latest published 95\%~C.L.~upper limit of $r < 0.035$~\cite{BICEP:2021xfz}, these B-mode fluctuations are at most 10s of~\si{nK}. Apart from this signal being extremely faint, there are two known astrophysical sources of B~modes of comparable or larger amplitude than the primordial B~modes which hinder the search.

The first contaminating astrophysical source of B~modes is from within our galaxy: thermal dust and synchrotron emission produce polarized-foreground B-mode patterns. Both mechanisms of polarized emissions depend on the galactic magnetic field. Synchrotron radiation is generated by cosmic rays moving through this magnetic field, with the emission being polarized perpendicular to the projection of the magnetic field onto the line of sight~\cite{Orlando:2013ysa}. Likewise, thermal emission from interstellar grains is polarized perpendicular to the local magnetic field orientation since grains tend to align with their short axes parallel to field lines~\cite{Purcell:1975}.

The second source, called ``lensing B~modes'', is produced by weak gravitational lensing of the~CMB. To first order, density perturbations from inflation produce only E-mode polarization. This means that the polarization pattern of the~CMB would be purely E~mode in~$\Lambda$CDM without lensing and in the absence of~PGWs. However, during their propagation to us, the paths of the polarized CMB~photons are deflected by the intervening gravitational potentials along the line of sight~\cite{Lewis:2006fu}. This produces the lensing B-mode component which must be accounted for or removed~(``delensed'') in order to tighten the constraint on or potentially detect~$r$. Delensing typically involves estimating the lensing potential and subsequently the lensing B~modes in the map. This estimate could then be subtracted from the observed CMB~maps or otherwise incorporated in a likelihood analysis for~$r$ to reduce \mbox{the sample-variance contribution from lensing.}\medskip

Controlling both sources of PWG~B-mode contaminants is key to obtaining robust $r$~results. Different survey configurations will have different fractional contributions of foreground and lensing uncertainties to their total uncertainty budget for~$r$. For instance, full-sky satellite surveys typically cover higher-foreground amplitude and complexity regions, while also having access to higher frequency bands for foreground characterization compared to ground-based surveys. This is particularly advantageous given that dust emission is brighter at higher frequencies. In contrast, ground-based surveys are designed to cover a low-foreground, as-small-as-practical patch of the sky to maximize the per-mode signal-to-noise ratio. In the limit that instrument noise is subdominant, the respective strategies tend to result in a larger fraction of the total uncertainties on~$r$ coming from foregrounds for satellite surveys and from lensing for ground-based surveys.

To reduce the impact of foregrounds on~$r$ in the upcoming high-sensitivity regimes, better modeling and simulations of the foreground components, and improved foreground cleaning methods are needed. While both dust and synchrotron emission could be sufficiently modeled as a single component with their spectral dependence following simple spectral energy density~(SED) descriptions in current data~\cite{Planck:2018gnk, Sheehy:2017gfx, BICEP:2021xfz}, it would be unsurprising if each of them demonstrates more complex spectral and spatial dependence in future data. Indeed, frequency decorrelation of dust and spatial variation in the dust~SED are expected and are already modeled in current analyses and forecasts~\cite{Errard:2018ctl, BICEP:2021xfz, Azzoni:2020hpw, CMB-S4:2020lpa, Pelgrims:2021gqi}. To better probe the physical origin of the observed spectral and spatial complexities of foregrounds, new approaches using ancillary observations of neutral-hydrogen~(HI) and starlight polarization are providing new ways to trace the three-dimensional dust distribution in the interstellar medium and the galactic magnetic field~\cite{2018arXiv181005652T, Clark:2019gap, Clark:2021kze, Puglisi:2021hqe}. Additionally, more accurate models and simulations of the polarized galactic foregrounds are being developed, leveraging improved observations and computational resources~\cite{Kritsuk:2017aab, Martinez-Solaeche:2017mgz, Kim:2019xov, Krachmalnicoff:2020rln, CMB-S4:2020lpa, Hervias-Caimapo:2021zue, Thorne:2021nux, Zonca:2021row}. In terms of foreground-cleaning approaches, various methods have been developed mostly for previous satellite experiments~\cite{Eriksen:2007mx, WMAP:2003cmr, Tegmark:2003ve, Delabrouille:2008qd, Planck:2015mvg, Svalheim:2020zud, SPIDER:2021ncy, BICEP:2021xfz}. For upcoming and future experiments, which have much lower noise regimes and unique survey configurations, active research is underway to optimize the signal-to-noise ratio and reduce residuals in each of the survey scenarios~\cite{Stompor:2008sf, Katayama:2011eh, Remazeilles:2018vaf, SimonsObservatory:2018koc, Errard:2018ctl, Thorne:2019mrd, Azzoni:2020hpw, Puglisi:2021hqe, CMBS4:PBDR}, with particular care being taken to account for impacts of instrumental systematics~\cite{Verges:2020xug, Abitbol:2020fvn, Giardiello:2021uxq, Snowmass2021:CMBs4, CMB-S4:2020lpa, CMBS4:PBDR}.

Delensing, unlike foreground cleaning, has only recently been implemented on data~\cite{Larsen:2016wpa, Carron:2017vfg, SPT:2017ddy, Planck:2018lbu, POLARBEAR:2019snn, ACT:2020goa, SPTpol:2020rqg}. Specifically, a reduction in the uncertainty on~$r$ using delensing was demonstrated for the first time in~\cite{SPTpol:2020rqg}. As mentioned, ground-based surveys typically focus their observing time on a small fraction of the sky when targeting primordial B~modes~\cite{Harrington:2016jrz, Kusaka:2018yzq, SimonsObservatory:2018koc, BICEP:2021xfz, CMB-S4:2020lpa, CMBS4:PBDR}, which is an advantageous strategy until a detection is made. This can however lead to the lensing uncertainty contributing a larger fraction of total uncertainty compared to foregrounds. For current experiments such as~BICEP/Keck, the lensing sample variance already dominates the uncertainty on~$r$~\cite{BICEP:2021xfz}; for planned experiments such as~CMB-S4, delensing is required to reach its~$r$~science goal. On this front, some open questions include whether small-scale galactic foregrounds might bias the lensing estimate and, therefore, bias the delensing procedure, and whether realistic instrument effects would degrade the delensing efficiency at relevant levels~\cite{Beck:2020dhe, Mirmelstein:2020pfk, Nagata:2021pvc}. Active development of delensing methods are underway and will be necessary to ensure that the required level of delensing efficiency \mbox{can be achieved for the planned experiments~\cite{Carron:2017mqf, Millea:2020cpw, BaleatoLizancos:2020mic, Namikawa:2021gyh, Millea:2021had}.}

Given the increased data volume and complexity, continued development in analysis, modeling and simulations are necessary to keep pace with experimental advancements in order to extract maximal information about our universe from these upcoming data sets. This includes the modeling and simulations of galactic and extragalactic foregrounds, and lensing~\cite{Takahashi:2017hjr, Stein:2020its, Han:2021unz}, modeling and approximations of the likelihood and covariance matrices~\cite{Balkenhol:2021gih, Beck:2022efr}, and theoretical work advancing our understanding of inflation.

\subsubsection{Large-Scale Structure}

Primordial gravitational waves can also alter cosmic shear and galaxy clustering spectrum measurements~\cite{Dodelson:2003bv, Dodelson:2010qu, Masui:2010cz, Jeong:2012df, Schmidt:2012ne, Jeong:2012nu, Schmidt:2012nw, Dai:2013kra, Schmidt:2013gwa, Dimastrogiovanni:2014ina, Emami:2015uva, Vlah:2019byq, Vlah:2020ovg}. This is because the effect of very long-wavelength perturbations on the scale of galaxy formation is an effective tidal field which results in non-zero E~and B~modes on large scales. The prospects of detecting primordial gravitational waves indirectly using this observable are challenged primarily by two aspects. First, gravitational nonlinearities can source both E~and B~modes at relatively large scales. This corresponds to a noise component that is much larger than the PWG~signal unless inflation breaks parity and sources an \mbox{E-B}~correlation in the galaxy shape power spectrum which cannot be sourced by late-time nonlinearities~\cite{Biagetti:2020lpx}. The second issue is related to shape noise, i.e.~the stochasticity produced by small-scale perturbations that affects the intrinsic ellipticity of galaxies. Forecasts show that competitive constraints with respect to the~CMB will be hard to reach in the near future~\cite{Masui:2010cz, Schmidt:2012nw, Dai:2013kra, Schmidt:2013gwa, Chisari:2014xia, Dimastrogiovanni:2014ina}. 

There are a few other (futuristic)~ideas of using LSS~and CMB~observations to probe signatures of~PGWs,\footnote{We can even employ astronomical observations on smaller scales to hunt for~PGWs, e.g.~via stellar astrometry~\cite{Book:2010pf} and pulsar timing arrays~\cite{RoperPol:2022iel}.} some of which are generated from the PGW-sourced quadrupole anisotropies. In a futuristic, tomographic \SI{21}{cm}~survey focused on the dark ages~($z \sim 20$), the remote quadrupole of the~CMB can source a small circular-polarization component in the emitted \SI{21}{cm}~radiation~\cite{Hirata:2017dku, Mishra:2017lpz}. In addition, PGWs~can lens the \SI{21}{cm}~fluctuations, from which one can reconstruct a curl component of the deflection field and infer the PGW~amplitude~\cite{Book:2011dz}. Moreover, cross-correlations of galaxy surveys with polarized SZ~signals in small-scale CMB~measurements will probe the remote quadrupole field and, consequently,~PGWs~\cite{Alizadeh:2012vy, Deutsch:2018umo}.

\subsubsection{Gravitational Wave Background}
\label{sec:pgw_gwb}

So far, we discussed probes that are designed to detect the effects of primordial gravitational waves on a set of observables. In the following, we summarize the direct sensitivity of current and planned gravitational wave observatories to a stochastic gravitational wave background of primordial origin~(see also the dedicated Snowmass~2021 White Papers~\cite{Snowmass2021:SGWB, Snowmass2021:GWO} for more detailed discussions).\medskip

The direct detection of gravitational waves proceeds through interferometry. The fractional change in the phase of light along two different paths, between pairs of freely falling test masses, is directly related to the amplitude and phase of a passing gravitational wave. Astrophysical GW~sources produce a finite-duration signal that is characterized by a gravitational waveform. In contrast, a stochastic background is characterized by a signal power or spectral density that is manifest as a frequency-dependent, irreducible noise in interferometer phase measurements. The spectral density is $\Omega_\mathrm{GW} \equiv {\d\hskip-1pt\log \rho_\mathrm{GW}}/{\d\hskip-1pt\log f}$, i.e.\ the energy density in gravitational waves as a fraction of the critical density per logarithmic frequency interval, with the energy density in~GWs being $\rho_\mathrm{GW} = \langle \dot h_{ij}(\tau,\vec x) \dot h^{ij}(\tau,\vec x)\rangle /(32 \pi G)$, where the angle brackets indicate averaging over a time interval much larger than the period of oscillation.\footnote{In the case of the inflationary stochastic background, the spatial/time average needed to formally define~$\rho_\mathrm{GW}$ corresponds, in practice, to the ensemble average of the stochastic variable~$h_{ij}$, which ultimately corresponds to the graviton correlators at late time.} In a universe filled with matter and radiation, the inflationary prediction for modes that entered the horizon before matter-radiation equality is $\Omega_\mathrm{GW}(f) = ({r \As}/{24}) \left(f/f_\mathrm{CMB}\right)^{\nt} \Omega_r$, with the radiation density parameter~$\Omega_r$ and the CMB~pivot frequency~$f_\mathrm{CMB} = \SI{1.94e-17}{Hz}$~\cite{Boyle:2005se}. It is a scientific goal of all GW~observatories to either observe a~PGW directly or characterize any possible foregrounds and place an upper limit on a primordial~$\Omega_\mathrm{GW}$.

The LIGO/Virgo/KAGRA~network of detectors is currently sensitive to gravitational waves in the frequency range $f \sim \SIrange{e1}{e2}{Hz}$. The most recent search for a~SGWB yields a 95\%~C.L.~upper limit of $\Omega_\mathrm{GW} \le \num{5.8e-9}$ on a flat, frequency-independent~SGWB~\cite{KAGRA:2021kbb}. The Laser Interferometer Space Antenna~(LISA)~\cite{LISA:2017pwj}, a proposed space-borne gravitational wave observatory on track for launch in the mid-2030s, is expected to significantly improve this bound. LISA~will operate primarily in the \si{mHz}~band, spanning $\SIrange{e-4}{e-1}{Hz}$. Current forecasts suggest that~LISA will be able to reach a sensitivity level better than $\Omega_\mathrm{GW} = \num{e-12}$ over the course of a nominal four-year mission~\cite{Smith:2019wny, Boileau:2021sni}. Comparing this limit to the CMB~limit, we see that the current CMB~limit on~$r$ under the assumption of a scale-invariant tensor spectrum would translate to gravitational wave spectral energy densities which are several orders of magnitude more stringent than set by current and planned interferometric GW~observatories~\cite{Shandera:2019ufi, Lasky:2015lej}. However, the frequencies probed by GW~observatories are distinct from those accessible to the~CMB and could probe physics of post-inflationary epochs. The process of inflationary preheating, characterized by rapid field variations and particle production, typically produces an additional spectrum of primordial gravitational waves~(see~\cite{Khlebnikov:1997di, Easther:2006vd, Dufaux:2007pt, Garcia-Bellido:2007nns}, and more recently~\cite{Antusch:2016con, Lozanov:2019ylm, Adshead:2019lbr, Snowmass2021:SGWB} and references therein). These are high-frequency~GWs that are beyond the reach of current detectors unless the inflationary energy scale is unusually low, although their presence may be indirectly felt as dark radiation~(see the Snowmass~2021 White Papers~\cite{Snowmass2021:LightRelics, Snowmass2021:SGWB}). In addition, features in the primordial spectra are ubiquitously predicted in scenarios beyond the simplest inflationary models, as addressed in Sec.~\ref{sec:features}, which could be probed at these other frequencies and, therefore, open up new windows to inflation.

Other currently planned or proposed GW~observatories include: $\mu$Ares~\cite{Sesana:2019vho}~(\si{\mu\Hz}); Taiji~\cite{Ruan:2020smc} and successors~\cite{Baker:2019pnp}~($\sim\!\si{mHz}$), DECIGO~\cite{Kawamura:2011zz}, TianQin~\cite{Mei:2020lrl} and TianGO~\cite{Kuns:2019upi} \mbox{($\sim\!\si{dHz}$)}, Cosmic Explorer~\cite{Reitze:2019iox} and the Einstein Telescope~\cite{Punturo:2010zz}. New detector technologies, such as atom interferometry, have been proposed across a range of frequencies, e.g.~MAGIS~\cite{Graham:2017pmn} and AEDGE~\cite{AEDGE:2019nxb}. These GW~observatories will contribute significantly to the search for primordial gravitational waves.
\clearpage
% !TEX root = whitepaper.tex

%%%%%%%%%%%%%%%%
\section{Primordial Non-Gaussianity}
\label{sec:png}
%%%%%%%%%%%%%%%%

Current observations of primordial fluctuations are consistent with Gaussian statistics. At the same time, deviations from Gaussianity are necessarily present even in the simplest models of inflation. More generally, primordial non-Gaussianity~(PNG) is a robust probe of interaction dynamics during inflation beyond the free propagation of curvature fluctuations. Detecting and characterizing~PNG would be a fantastic triumph of theoretical and observational cosmology, probing the dynamics of the early universe and providing clues about physics at very high energy densities, much higher than those achievable in particle colliders, and potentially uncovering new degrees of freedom beyond curvature perturbations~(see e.g.~\cite{Bartolo:2004if, Komatsu:2009kd, Baumann:2009ds, Liguori:2010hx, Chen:2010xka, Wang:2013zva, Baumann:2014nda, Alvarez:2014vva, Renaux-Petel:2015bja, Desjacques:2016bnm, Meerburg:2019qqi, Biagetti:2019bnp} for previous reviews).\medskip

A detection of~PNG could probe new degrees of freedom around the inflationary Hubble scale~$H$, all the way to heavier physics around the strong coupling scale~$\Lambda_\mathrm{UV}$, through self-interactions of the curvature perturbations~(see Fig.~\ref{fig:scales}). Moreover, we could infer the dynamical properties of the various degrees of freedom that are active during inflation, such as their dispersion relations, spectroscopic properties and interactions. For example, depending on the field content and symmetry breaking pattern of inflation, the predicted size and shape of~PNG changes. Moreover, the theoretical characterization of the allowed shapes and sizes of~PNG is an interesting endeavor with many recent developments.\medskip

Any level of non-Gaussianity in the statistics of the primordial fluctuations will be transfered to the maps of the cosmic microwave background and large-scale structure. To extract this information from observations, efficient estimators for the CMB~bispectrum have been developed. Recently, the first analyses involving the bispectrum have also been performed in~LSS. Apart from the bispectrum~(and higher-point functions), PNG~of the `local' type also results in an enhancement of power of biased LSS~tracers on large scales, referred to as the scale-dependent bias. While this signal is comparably easy to extract from LSS~data, bispectrum analyses are more complicated to perform than in the~CMB. Nevertheless, recent theoretical advances have enabled the first such analyses. Having said that, there is no evidence for~PNG for any shape or probe so far, with the tightest constraints being derived from Planck~data. At the same time, there is a continued effort to mitigate and model astrophysical and nonlinear effects, spanning analytic, numerical and simulation-based approaches, to more efficiently extract primordial information from late-time observables. Together with the dramatic increases in observational sensitivity, future CMB~and LSS~analyses are projected to significantly improve the constraints on all PNG~types, with potentially decisive implications for our understanding of the inflation.\medskip

In this section, we will review the theoretical status of~PNG, the most important shapes of the bispectrum, what physics they encode, and current bounds on their sizes. We also review the state of the art in techniques to analyze data from CMB~and LSS~surveys, together with promising future directions, and an outlook in both theory and observations.
% !TEX root = whitepaper.tex

% Primordial Non-Gaussianity

\subsection{Theoretical Background}
\label{sec:png_theory}

Current observational data support the $\Lambda$CDM~assumption that primordial fluctuations have Gaussian statistics. This is consistent with the simplest single-field inflationary models, in which fluctuations only self-interact gravitationally, predicting a very small level of~PNG, which is currently beyond reach~\cite{Acquaviva:2002ud, Maldacena:2002vr}. At the same time, well-motivated inflationary models beyond the simplest ones have been shown to be able to generate~PNG with larger amplitudes and different profiles~\cite{Bartolo:2004if, Liguori:2010hx, Chen:2010xka, Wang:2013zva, Renaux-Petel:2015bja}. These levels of~PNG are potentially detectable in upcoming cosmological surveys. Measuring~PNG will allow us to answer fundamental questions about the primordial universe, such as:
\begin{itemize}
	\item How many scalar degrees of freedom were light during inflation?
	\item Were there degrees of freedom with masses comparable to the Hubble scale of inflation? What were their mass and spin spectra?
	\item What were the initial states of these quantum fluctuations? What were their interactions and how fast were they propagating?
	\item Was the background spacetime of the primordial universe quasi-de Sitter?
\end{itemize}

There are various measures of non-Gaussianity. We focus on the scalar three-point correlation function or bispectrum. It has been the most-studied and analyzed observable in the literature since it is often the dominant non-Gaussian signature in weakly coupled models of inflation. For translation-, rotation- and scale-invariant perturbations, the bispectrum is
\begin{equation}
	\langle \mathcal{R}_{\mathbf{k}_1} \mathcal{R}_{\mathbf{k}_2} \mathcal{R}_{\mathbf{k}_3} \rangle = (2\pi)^3 \delta^3(\mathbf{k}_1 + \mathbf{k}_2 + \mathbf{k}_3)\, \frac{18}{5} \fnl^\mathrm{type} \As^2\, \frac{S^\mathrm{type}(k_1, k_2, k_3)}{(k_1\hskip1pt k_2\hskip1pt k_3)^2}\, ,
\end{equation}
where $\fnl$~parameterizes the size of~PNG and the dimensionless shape function~$S^\mathrm{type}$ controls the overall size of~PNG as a function of the triangle formed by the momenta. The shape dependence encodes information about the specific dynamical mechanism that generated the non-Gaussian signal and, therefore, serves as a discriminator between various inflationary models.

Studies of various inflationary models have demonstrated several broad classes of scale-invariant~PNG\hskip1pt\footnote{Scale-dependent~PNG is closely associated with primordial features, which we will review in the next section. Due to the nature of the mechanisms generating scale-dependent signals, many models typically predict correlated oscillatory signatures in correlation functions at different orders, which motivates the search for such features in the primordial spectra.} with large, potentially detectable~$\fnl$.\footnote{As a rough rule of thumb, models that predict $\fnl > 1$ with a shape function that is larger for squeezed triangles than for equilateral triangles has better observational prospects~(see~\textsection\ref{sec:png_observations}). For single-field slow-roll inflation, the prediction is that the shape function peaks around the equilateral configuration with an amplitude proportional to the scalar spectral tilt~(see~\cite{Acquaviva:2002ud, Maldacena:2002vr, Tanaka:2011aj, Pajer:2013ana, Dai:2015jaa, dePutter:2015vga, Bartolo:2015qva, Cabass:2016cgp} for discussions), which is beyond reach of near-future surveys.} In the following, we list them with an emphasis on the main characteristics of the function~$S$ and the physics that they probe.
\begin{itemize}
	\item \textbf{Equilateral/orthogonal~PNG and single-field inflation.} The bispectra peak in the equilateral configuration, $k_1 \sim k_2 \sim k_3$, and the shape function vanishes as $S \sim q$ for a soft momentum denoted by~$q$. This type of~PNG arises if the inflaton has derivative, local self-interactions~\cite{Creminelli:2003iq, Alishahiha:2004eh, Gruzinov:2004jx, Chen:2006nt, Cheung:2007st, Langlois:2008wt, Langlois:2008qf, Senatore:2009gt, Renaux-Petel:2011lur, Pajer:2020wxk}. Since derivatives are suppressed on superhorizon scales, these interactions contribute the most to the PNG~signal when all modes have similar wavenumbers around the horizon exit. Oftentimes, derivative self-interactions produce several shapes of equilateral~PNG that are slightly different from each other which means that finding an orthogonal basis of these shapes leads to new PNG~profiles in this category.

	\item \textbf{Local~PNG and multi-field inflation.} The bispectra peak in the squeezed limit, scaling as $S \sim q^{-1}$ as~$q$ becomes soft. A large local PNG~signal indicates the presence of more than one light~($m \ll H$) field during inflation~\cite{Salopek:1990jq, Gangui:1993tt, Verde:1999ij, Komatsu:2001rj}. Physically, the fluctuations of massless scalars freeze after horizon exit and open up a multi-field space for superhorizon evolution. Using the $\delta N$~formalism~\cite{Starobinsky:1986fxa, Sasaki:1995aw, Lyth:2005fi}, on superhorizon scales, patches of the universe of Hubble size evolve independently of each other, leading to nonlinearities that are local in these Hubble patches. This gives rise to~PNG that is local in real space and peaks in the squeezed limit in momentum space.

	\item \textbf{Non-analytic~PNG and cosmological collider physics.} The squeezed-limit behavior of these bispectra follows a non-analytic power-law scaling of the soft momentum~$q$, $S\sim q^{\alpha} P_s(\cos\theta)$, where~$\theta$ is the angle between the soft~$\mathbf{q}$ and the hard mode, $\alpha$~is either a real number $\alpha \in [-1, \frac{1}{2}]$ or a complex number $\alpha = \frac{1}{2} + \ii\mu$ with real~$\mu$, depending on the mass of the mediating state, and $P_s$ is the Legendre polynomial of degree~$s$ that captures the angular dependence, with~$s$ being the spin of the mediating state~\cite{Chen:2009zp, Baumann:2011nk, Noumi:2012vr, Arkani-Hamed:2015bza, Lee:2016vti}. Note that local~PNG is a special limit of this general case. Due to the analogy between the squeezed limit behavior of~PNG and how mass/spin spectra are measured at energy thresholds in particle colliders, the general classification of these shapes is dubbed ``cosmological collider physics''. Since the resonances are efficiently produced up to states of mass comparable to the Hubble scale, the detection of such PNG~shapes may be used to probe the nature of particle physics at energies up to or exceeding the Hubble energy during inflation~\cite{Chen:2009we, Chen:2009zp, Baumann:2011nk, Assassi:2012zq, Sefusatti:2012ye, Norena:2012yi, Chen:2012ge, Pi:2012gf, Noumi:2012vr, Gong:2013sma, Emami:2013lma, Kehagias:2015jha, Arkani-Hamed:2015bza, Dimastrogiovanni:2015pla, Chen:2015lza, Chen:2016nrs, Lee:2016vti, Meerburg:2016zdz, Chen:2016uwp, Chen:2016hrz, Chen:2017ryl, Kehagias:2017cym, An:2017hlx, Iyer:2017qzw, An:2017rwo, Kumar:2017ecc, Tong:2018tqf, MoradinezhadDizgah:2018ssw, Chen:2018sce, Saito:2018omt, Chen:2018xck, Achucarro:2018ngj, Chua:2018dqh, Arkani-Hamed:2018kmz, Kumar:2018jxz, Wu:2018lmx, Alexander:2019vtb, Lu:2019tjj, Hook:2019zxa, Hook:2019vcn, Kumar:2019ebj, Liu:2019fag, Wang:2019gbi, Baumann:2019oyu, Wang:2019gok, Wang:2020uic, Li:2020xwr, Wang:2020ioa, Baumann:2020dch, Fan:2020xgh, Kogai:2020vzz, Bodas:2020yho, Aoki:2020zbj, Lu:2021gso, Lu:2021wxu, Wang:2021qez, Tong:2021wai, Pinol:2021aun, Cui:2021iie, Pimentel:2022fsc, Jazayeri:2022kjy}.

	\item \textbf{Folded~PNG and non-Bunch-Davies vacuum.} The bispectra peak in the folded configuration, $k_1 + k_2 \sim k_3$. This shape can arise from non-Bunch-Davies initial states for quantum scalar fluctuations~\cite{Chen:2006nt, Holman:2007na, Meerburg:2009ys, Meerburg:2009fi}, from classically excited states~\cite{Green:2020whw} or a strongly non-geodesic motion in multi-field inflation~\cite{Garcia-Saenz:2018ifx, Garcia-Saenz:2018vqf, Fumagalli:2019noh}. Physically, the peak in the folded configuration is due to the decay of the modified initial state~\cite{Jiang:2015hfa}.
\end{itemize}

The conventional approach of deriving the templates of these~PNG profiles is through Lagrangians and explicit time evolution with the ``in-in'' formalism. In the following, we instead present a new perspective of the ``cosmological bootstrap''~\cite{Arkani-Hamed:2015bza, Arkani-Hamed:2018kmz, Sleight:2019hfp, Baumann:2019oyu, Baumann:2020dch, Pajer:2020wnj, Snowmass2021:Bootstrap}. This method allows us to make theoretically controlled predictions based on assumptions of weak coupling and symmetries. As long as the underlying model satisfies these assumptions, the PNG~form follows directly from consistency conditions such as unitarity and analyticity. This not only dramatically simplifies many computational steps, but also provides a natural language to describe the properties of shape functions without the need to invoke a Lagrangian. In what follows, we will describe three distinct classes of~PNG in this language.

\paragraph{Single-Field Inflation}~\\
In single-field inflation, consistency conditions imply that the most general bispectrum from derivative interactions is captured by the ansatz
\begin{equation}
	S^\mathrm{EFT}(k_1, k_2, k_3) = \frac{1}{k_1 k_2 k_3} \sum_{n=3}^\infty \frac{\mathrm{Poly}_{3+n}(k_1, k_2, k_3)}{(k_1 + k_2 + k_3)^n}\, ,	\label{eq:SEFT}
\end{equation}
where~$\mathrm{Poly}_{3+n}$ is a symmetric polynomial of degree~$3+n$. For the Bunch-Davies initial state, the pole at $k_1 + k_2 + k_3 = 0$ is the only allowed singularity for the bispectrum in single-field inflation, where the degree of the pole is related to the number of derivatives of the interaction. The polynomial in the numerator is not arbitrary, but is largely fixed by demanding locality and the correct soft-limit behavior imposed by the single-field consistency relation, see~\eqref{eq:SFCR}~\cite{Pajer:2020wxk, Jazayeri:2021fvk}.

To gain physical intuition into this shape function, it is useful to interpret it in the context of the effective field theory of single-clock inflation~\cite{Cheung:2007st}. In this framework, the dynamics of scalar fluctuations is captured by the Goldstone boson~$\pi$ which is related to the curvature perturbation as $\mathcal{R} = -H \pi$ at linear order and nonlinearly realizes the spontaneously broken time translations due to a single clock driving the inflationary expansion. At lowest order in derivatives, corresponding to $n = 3$ in~\eqref{eq:SEFT}, there are two cubic interactions~$\dot\pi^3$ and~$\dot\pi(\partial_i\pi)^2$ which come with two independent parameters: $\tilde{c}_3$~which controls the size of~$\dot\pi^3$ and the sound speed~$c_\mathrm{s}$. These precisely translate to the two free coefficients that fix~$S^\mathrm{EFT}$ at $n = 3$~\cite{Pajer:2020wxk}.

For data analyses, it is convenient to introduce simpler templates which approximate the exact shapes. These are conventionally called the ``equilateral'' and ``orthogonal'' shapes,\footnote{For a review on the derivation of these shapes, and other developments related to~EFTs in inflation, we refer to the dedicated Snowmass~2021 White Paper~\cite{Snowmass2021:CosmoEFT}.} and defined as~\cite{Creminelli:2005hu, Senatore:2009gt}
\begin{align}
	S^\mathrm{equil}(k_1, k_2, k_3) &= \frac{(k_1 + k_2 - k_3)\,(k_2 + k_3 - k_1)\,(k_3 + k_1 - k_2)}{k_1\hskip1pt k_2\hskip1pt k_3}\, ,	\\[2pt]
	S^\mathrm{ortho}(k_1, k_2, k_3) &= (1+p)\hskip1pt S^\mathrm{equil}(k_1, k_2, k_3) - p\frac{\Gamma(k_1, k_2, k_3)^3}{k_1\hskip1pt k_2\hskip1pt k_3}\, ,
\end{align}
where $p \approx 8.52$ and $\Gamma(k_1, k_2, k_3) \equiv \frac{2}{3} \sum_{a<b}^3 k_a k_b - \frac{1}{3} \sum_a^3 k_a^2$. The two EFT~parameters therefore get linearly transformed into the basis of~$\fnl^\mathrm{equil}$ and~$\fnl^\mathrm{ortho}$, with $\fnl^\mathrm{equil} \sim 1/c_\mathrm{s}^2$ for small~$c_\mathrm{s}$. Due to locality of the derivative self-interactions, the shape function peaks when all wavenumbers are comparable, which is the origin of the name ``equilateral''. We however note that the equilateral shape is not exactly degenerate with the EFT~shapes and the orthogonal shape therefore captures the direction that is orthogonal to the equilateral shape. Since the inflationary background dynamics admits a weakly coupled description~(in the sense of the derivative expansion for the inflaton) for $\fnl^\mathrm{equil} \lesssim 1$, reaching $\fnl^\mathrm{equil} \sim 1$ is an important observational target~\cite{Baumann:2011su, Baumann:2014cja, Baumann:2015nta}.

\paragraph{Multi-Field Inflation}~\\
From a model-building perspective, it is natural to consider models of inflation involving additional particles beyond the inflaton. In multi-field models, one often considers massless spectator degrees of freedom. They can generate significant isocurvature perturbations~(fluctuations orthogonal to the multi-field trajectory in field space), which in turn are directly converted to curvature perturbations, while still giving a subdominant contribution to the background energy density. Popular models of this type include the ``curvaton''~\cite{Enqvist:2001zp, Lyth:2001nq, Moroi:2001ct, Sasaki:2006kq} or ``modulated reheating'' scenarios~\cite{Dvali:2003em, Kofman:2003nx, Dvali:2003ar}. This transfer of non-Gaussianity occurs on superhorizon scales, which can be approximated by the Taylor expansion $\mathcal{R} = \mathcal{R}_g + \frac{3}{5} \fnl^\mathrm{local} \hskip1pt\mathcal{R}_g^2 + \ldots$ around the Gaussian perturbation~$\mathcal{R}_g$. The ``local'' shape generated by this nonlinearity then is
\begin{equation}
	S^\mathrm{local}(k_1, k_2, k_3) = \frac{1}{3} \frac{k_1^2}{k_2 k_3} + \text{2 perms}\, .
\end{equation}
As a consequence of being generated by local interactions on superhorizon scales, this shape peaks locally at coincidental points in real space. From a bootstrap point of view, this type of locally peaked signal corresponds to correlations formed by exchanging extra massless particles during inflation~(see below). Spectator fields typically generate $\fnl^\mathrm{local} \sim 1$, which provides a natural observational target for upcoming surveys.

A distinctive feature of the local shape is that it maximally violates the single-field consistency relation~\cite{Maldacena:2002vr, Creminelli:2004yq, Cheung:2007sv, Creminelli:2011rh, Creminelli:2012ed, Hinterbichler:2013dpa},
\begin{equation}
	\langle \mathcal{R}_{\mathbf{q}}\, \mathcal{R}_{\mathbf{k}-\mathbf{q}/2}\, \mathcal{R}_{-\mathbf{k}-\mathbf{q}/2} \rangle = (2\pi)^3 \delta^3 (\mathbf{k}_1 + \mathbf{k}_2 + \mathbf{k}_3)\, P_\mathcal{R}(q)\, P_\mathcal{R}(k) \left[(1-\ns) + O\!\left(\frac{q^2}{k^2}\right)\right].	\label{eq:SFCR}
\end{equation}
In single-field inflation, this puts a fully kinematic constraint on the leading and subleading part of the squeezed shape function. On the other hand, a model with more than one light field produces extra contributions to the local shape, violating the consistency relation at leading order. With the exception of models with certain non-Bunch-Davies vacua or non-attractor solutions~\cite{Namjoo:2012aa, Martin:2012pe, Chen:2013aj, Huang:2013oya, Chen:2013eea, Bravo:2017wyw, Cai:2018dkf, Bravo:2020hde, Suyama:2021adn}, a detection of the bispectrum in the squeezed limit would rule out single-field models of inflation and is therefore a smoking gun for additional light particles during inflation. A large local~PNG on large scales may also induce fluctuations that change statistical assumptions about the density perturbations on smaller scales~\cite{Erickcek:2008sm, Erickcek:2008jp, Nelson:2012sb, LoVerde:2013xka, Dai:2013kfa, Lyth:2013vha, Wang:2013lda, Namjoo:2013fka, Abolhasani:2013vaa, Lyth:2014mga, Namjoo:2014nra, Baytas:2015nja, Bonga:2015urq}.

\paragraph{Cosmological Collider Physics}~\\
If the additional particles have masses comparable to the Hubble scale of inflation, then they can lead to a distinct non-analytic behavior in the squeezed limit~\cite{Chen:2009zp, Arkani-Hamed:2015bza}. Assuming a weakly coupled background dynamics, the bispectrum from the exchange of a heavy particle can be efficiently computed by exploiting the isometries of the background de Sitter space. This allows us to construct a differential representation of the bispectrum, which has the advantage that it can systematically incorporate particles of any spin. The basic building block in this framework is the four-point function of a conformally coupled scalar~$\varphi$ exchanging a massive scalar, which satisfies the differential equations~\cite{Arkani-Hamed:2015bza, Arkani-Hamed:2018kmz}
\begin{equation}
	(\Delta_u - \Delta_v)\, F^\mathrm{ex}_\varphi(u,v) = 0\, ,	\qquad	\big(\Delta_u + \mu^2 + \tfrac{1}{4}\big)\hskip1pt F^\mathrm{ex}_\varphi(u,v) = \frac{u\,v}{u + v}\, ,
\end{equation}
where $u \equiv k_I/(k_1 + k_2)$ and $v \equiv k_I/(k_3 + k_4)$, with $k_I = |\mathbf{k}_1 + \mathbf{k}_2|$, $\Delta_u$~is a second-order differential operator, and $\mu = \sqrt{m^2/H^2 - 9/4}$ is a dimensionless mass parameter. The solution, given by a two-variable generalization of the hypergeometric series, can be uniquely fixed by imposing the absence of the folded singularity $u, v = 1$ and the normalization in the partial energy singularity $u, v= -1$ as boundary conditions. The massive-exchange shape has a rich analytic structure, mixing an infinite sum of EFT~contributions that give an equilateral-like shape, with non-analytical contributions from on-shell massive fields not captured by the~EFT. In particular, for $m/H > 3/2$, the shape develops oscillations---a fingerprint of particle production---in the squeezed limit, which become a sharp resonance in position space~\cite{DiPietro:2021sjt}, in close analogy with collider physics for particle accelerators.

The exchange bispectrum can then be expressed in terms of the above building block as~\cite{Arkani-Hamed:2018kmz}
\begin{equation}
	S^\mathrm{ex}(k_1,k_2,k_3) = \frac{k_3^2}{k_1k_2} \sum_s \left. \lambda_s\hskip1pt U_{s,m}(u,\partial_u,\alpha)\hskip1pt F^\mathrm{ex}_\varphi(u,1)\hskip1pt P_s(\alpha)\right|_{k_4\to 0} + \text{2 perms}\, ,
\end{equation}
where~$\lambda_s$ parameterizes the coupling strengths, $\alpha \equiv (k_1 - k_2)/k_3$, and $U_{s,m}$~is the differential operator that transforms the exchange shape of~$\varphi$ to that of a particle of mass~$m$ and spin~$s$. This formula provides an elegant way to classify exchange bispectra with arbitrary mass and spin from the soft limit of a simple scalar-exchange four-point function. Taking the squeezed limit of this general solution, it follows that $S \sim q^{\frac{1}{2} + \ii\mu} P_s(\cos\theta)$, with~$\theta$ being the angle between the soft and hard modes. The bulk particle production in a weakly coupled background however necessarily implies that the amplitude is both slow-roll- and Boltzmann-suppressed, $\fnl^\mathrm{ex} \sim \varepsilon \mu^n \ee^{-\pi\mu}$. Various scenarios have been considered to compensate for these suppression factors, including an EFT~construction~\cite{Lee:2016vti, Bordin:2018pca} and a chemical-potential enhancement~\cite{Chen:2018xck, Hook:2019vcn, Wang:2019gbi, Wang:2020ioa, Bodas:2020yho}. These lead to a slight modification in the overall shape function, but both the oscillatory features and the angular dependence remain robust spectroscopic information of these shapes in the squeezed limit.

Apart from the particle spectra, the functional form of the non-analytical dependence of the soft momentum, $S \propto \ee^{\ii\mu \log q}$, directly encodes the time dependence of the scale factor of the background spacetime. This property is particularly clear in the context of the quantum primordial standard clock~\cite{Chen:2015lza, Chen:2016qce, Chen:2018sce, Chen:2018cgg, Wang:2020aqc}: Quantum fluctuations of heavy fields with constant masses can be regarded as standard clocks and the $q$~dependence of the oscillation phase in the PNG~shape is determined by the inverse function of~$a(t)$.

\begin{figure}[t]
	\centering
	\includegraphics{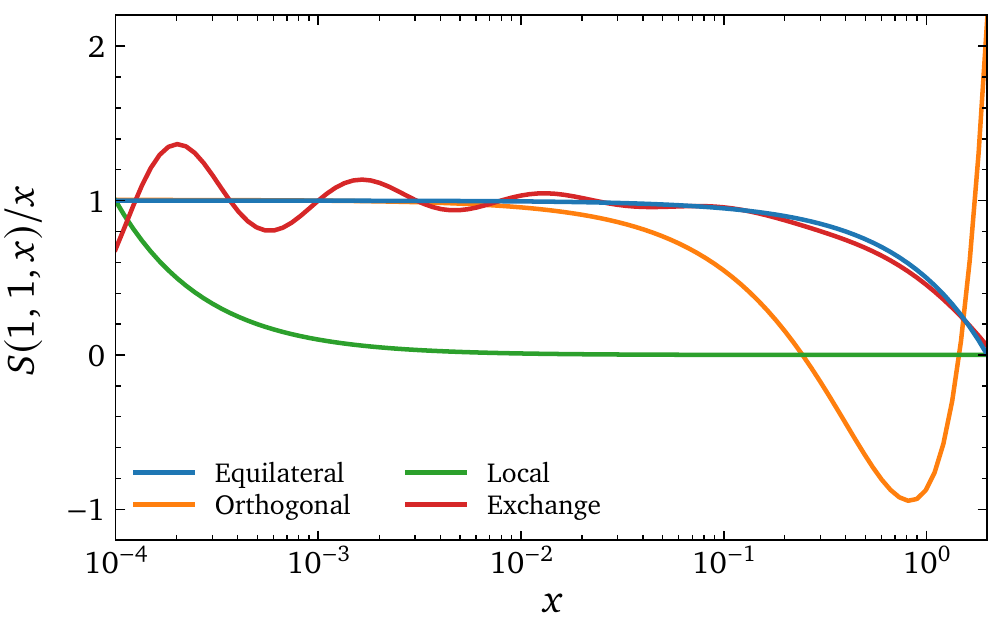}
	\caption{The shape functions~$S^\mathrm{equil}$, $S^\mathrm{ortho}$, $S^\mathrm{local}$ and $S^\mathrm{ex}$ discussed in this section. The $x$~values scan isosceles triangle configurations between the squeezed~($x = \num{e-4}$) and folded~($x = 2$) limits. The normalization is chosen such that the~(asymptotic) amplitudes are set to unity at $x = \num{e-4}$. For the exchange shape, we have chosen $\mu = 3$ and $s = 0$.}
	\label{fig:png_theory}
\end{figure}

Figure~\ref{fig:png_theory}%
shows a one-dimensional projection of the shape functions discussed in this section, highlighting their squeezed-limit behaviors. The local shape has a dominant scaling in the squeezed limit, making it an ideal~PNG type to be constrained from large-scale structure observables~(see~\textsection\ref{sec:png_observations}). While the equilateral and orthogonal shapes have the same soft scaling, the latter peaks in the folded configuration, making it also a useful template for the folded-type~PNG~\cite{Meerburg:2009ys}. For the exchange shape, its oscillatory period in~$\log x$ is precisely fixed by~$\mu$, the mass of the particle in Hubble units. In the large-$\mu$ limit, the non-analytic contribution from the on-shell massive field is Boltzmann suppressed and the exchange shape becomes degenerate with the equilateral shape. This is the familiar low-energy-EFT limit in which the mediating massive particle can effectively be integrated out. Accessing the full two-dimensional shape space can further help to distinguish various bispectra, in particular higher-derivative-EFT or spin-exchange shapes with a unique angular dependence.

\paragraph{Beyond the Bispectrum}~\\
Non-Gaussianity might also be stored in the primordial statistics beyond that captured by three-point functions. Large bispectra generically imply enhanced trispectra (four-point functions), which contain complementary information and can be large even if the bispectra are small~\cite{Seery:2006vu, Huang:2006eha, Seery:2006js, Arroja:2008ga, Seery:2008ax, Gao:2009gd, Adshead:2009cb, Chen:2009bc, Arroja:2009pd, Lehners:2009ja, Gao:2009at, Mizuno:2009mv, Renaux-Petel:2009jdf, Bartolo:2009kg, ValenzuelaToledo:2009nq, Renaux-Petel:2013wya, Renaux-Petel:2013ppa, Baumann:2017jvh}. Another example, in which leading information cannot be captured by any single polyspectra, is found in multi-field inflation, where the statistical properties of isocurvature fields can be transferred to the curvature perturbations which can potentially lead to~PNG being spread over a large number of $n$-point correlation functions. The resummation of these $n$-point functions yields a density probability distribution for~$\mathcal{R}$ with a non-Gaussian shape that is determined by the multi-field potential orthogonal to the inflationary trajectory~\cite{Chen:2018uul, Chen:2018brw, Palma:2019lpt}.

Yet another context, in which an analysis based on the bispectrum fails, is the description of large~(but rare) primordial fluctuations which are parameterized by the tails of their probability distribution. A compelling motivation to study the statistics of rare fluctuations is offered by primordial black holes, as their formation from the collapse of fluctuations is an unlikely event whose occurrence is dictated by the shape of the tail of their distribution. It is now well understood that standard perturbative techniques fail to correctly parameterize non-Gaussian tails~\cite{Celoria:2021vjw}. The use of non-perturbative techniques has allowed the computation of non-Gaussian tails in the context of both single-field~\cite{Celoria:2021vjw, Cohen:2021jbo, Hooshangi:2021ubn, Cai:2021zsp} and multi-field inflation~\cite{Panagopoulos:2019ail, Panagopoulos:2020sxp, Achucarro:2021pdh, Hooshangi:2022lao}. These works have focused on the computation of one-point distributions, although two-point~(or higher-point) distributions are needed in order to accurately predict the abundance and clustering properties of primordial black holes, as noted in~\cite{DeLuca:2022rfz}.
% !TEX root = whitepaper.tex

% Primordial Non-Gaussianity

\subsection{Observational Imprints}
\label{sec:png_observations}

Based on our theoretical assumptions and as suggested by observations, a primordial non-Gaussian signal will be small. As a consequence, a detection of the signal will heavily rely on how well we can remove sources of noise and confusion. In the following, we broadly define everything that is not intrinsic to the sky as noise and refer to everything that is on the sky, but is not our target of interest as confusion. For example, in the~CMB, noise could refer to instrumental noise, while sources of confusion could be galactic foregrounds or CMB~secondary signals, such as weak gravitational lensing and the Sunyaev-Zel'dovich effects. However, even after dealing with all sources of noise and confusion, cosmic variance remains as a limitation which means that a PNG~measurement ultimately relies on the number of available modes that we can reliably extract from the sky. If we were able to directly constrain the matter field, there would in principle be more than \num{e12}~modes between us and the last-scattering surface. Improving current bounds on primordial non-Gaussianity therefore critically depends on how many of these modes can be observed and how accurately we are able to model the relation between the matter and the tracer fields.\medskip

We will focus on the prospects of constraining~PNG with measurements of the CMB~anisotropies and~LSS, but note that there exist additional possibilities, in particular in cross-correlations of different probes and length scales.\footnote{For instance, cross-correlating the primary CMB~anisotropies with CMB~spectral distortions could potentially lead to very tight constraints on local~PNG~\cite{Pajer:2012vz, Ganc:2012ae, Emami:2015xqa, Khatri:2015tla, Bartolo:2015fqz, Ota:2016mqd, Ravenni:2017lgw, Remazeilles:2018kqd, Cabass:2018jgj, Orlando:2021nkv} and their cross-correlations with GW~data~(e.g.\ from~LISA) may allow for interesting constraints on tensor-scalar and tensor-tensor interactions in the early universe~\cite{Dimastrogiovanni:2021mfs}.} While CMB~measurements currently put the tightest and most-robust bounds on~PNG, further improvements with the~CMB will be limited~(with some exceptions) because of its effective two-dimensional nature and the fact that the number of measured primordial modes cannot be substantially increased. On the other hand, LSS~surveys allow us to access a large three-dimensional comoving volume and, in principle, reach smaller comoving scales, but constraining~PNG using~LSS is hard and many challenges need to be overcome for LSS~measurements to reach their full potential. As we will now discuss for CMB~and LSS~observables, a future detection of primordial non-Gaussianity will therefore rely on several key factors: the size of the primordial signal, realistic modeling of the signal, modeling and mitigating sources of confusion, observing a larger comoving volume down to smaller scales, and identifying new observational channels~(multi-tracer cosmology).

\subsubsection{Cosmic Microwave Background}

The projection of the primordial fluctuations~$\{\mathcal{R}, \gamma\}$ on the temperature and polarization fluctuations in the~CMB through transfer functions is almost linear which means that the statistics of the~CMB are directly related to the statistics of the primordial field. This is the main reason why the most stringent constraints on~PNG are currently derived from the~CMB.

\paragraph{CMB Bispectrum}~\\
The bispectrum is generically the most sensitive statistic in the limit of weak~PNG. This mentioned projection from the initial conditions to the last-scattering surface is however computationally prohibitive for $n$-point correlation functions beyond $n = 2$. PNG~analyses of CMB~data therefore rely on optimal estimators that are applied to heavily processed CMB~maps. In particular, unlike for inferences from the power spectra, the cosmology is fixed in these bispectrum analyses and the parameter~$\fnl$ is constrained once for each bispectrum shape, with the uncertainty on~$\fnl$ being derived from simulations of a noisy Gaussian sky. To establish confidence in these analyses, several independent bispectrum estimators have been developed and the Planck Collaboration for example applied the KSW~estimator, the modal estimator and the binned estimator to their data~(for details see~\cite{Planck:2015zfm, Planck:2019kim}).

\begin{figure}[t]
	\centering
	\includegraphics{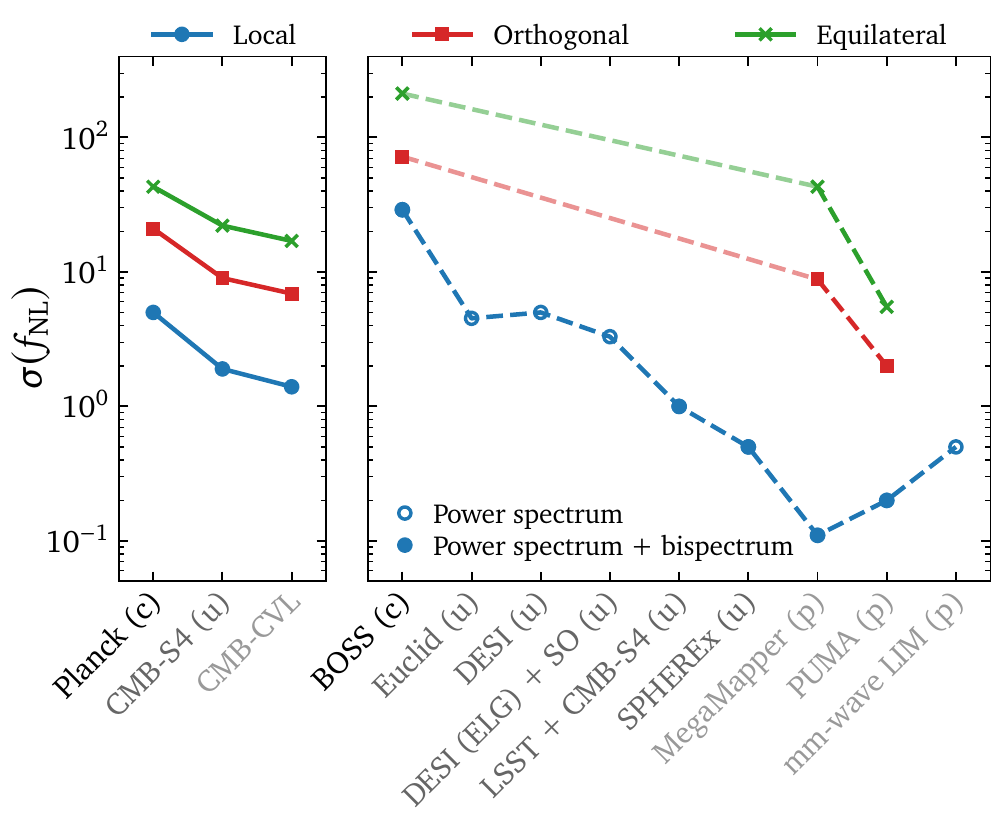}
	\caption{Comparison of constraints on three types of primordial non-Gaussianity from a small subset of completed~(`c'), upcoming~(`u') and proposed~(`p') experiments~(see~\cite{Schmittfull:2017ffw, MoradinezhadDizgah:2018lac, Munchmeyer:2018eey, Ferraro:2019uce, Castorina:2019wmr, Planck:2019kim, Abazajian:2019eic, Bandura:2019uvb, Chen:2021vba, Sailer:2021yzm, Mueller:2021tqa, Cabass:2022wjy, DAmico:2022gki, Snowmass2021:LIM, SPHEREx:web} for the underlying data analyses and forecasts). We also forecast a conservative cosmic-variance-limited~(CVL) CMB~experiment up to $\ell_\mathrm{max}^T=3000$ and $\ell_\mathrm{max}^P=5000$, but note that these limits could be further improved by the use of delensing or the inclusion of Rayleigh-scattering anisotropies. The constraints on local~PNG from~(e)BOSS, Euclid, DESI and~\mbox{mm-LIM} assume only power-spectrum information. The DESI+SO forecast includes the cross-correlation between the emission-line-galaxy sample of~DESI and SZ~maps of the Simons Observatory. The forecast for LSST+CMB-S4 includes power spectra and bispectra, including the cross-correlations between galaxy and lensing maps to remove sample variance. All other LSS~probes include bispectrum information. We note that there are two important caveats to these displayed results: (i)~scale-dependent-bias measurements hinge on the ability to measure the largest scales at high precision and most of the forecasts contain only a limited assessment of the impact of observational systematics; (ii)~there remains a large degree of uncertainty over several aspects of these forecasts despite a lot of theoretical progress in recent years which means that the achievable constraints may become better or worse as these issues are resolved.}
	\label{fig:png_forecasts}
\end{figure}

While it would be ideal to constrain the full bispectrum, i.e.\ including its shape, instead of just the amplitude of a fixed shape, this unfortunately remains out of reach~(see e.g.~\cite{Verde:2013gv}). Having said that, the modal estimator has sufficient flexibility to measure a very large number~(\num{30000}) of different shapes and, in principle, allows a~(noisy) reconstruction of the true bispectrum~\cite{Fergusson:2006pr, Fergusson:2008ra, Fergusson:2010gn}. In addition, the binned bispectrum estimator provides an efficient and model-independent reconstruction within the class of smooth bispectra~\cite{Bucher:2009nm, Bucher:2015ura}, and efficient estimators for linear and logarithmic oscillatory bispectra~(cf.~Section~\ref{sec:features}) are available and have been applied to Planck data~\cite{Munchmeyer:2014cca, Munchmeyer:2014nqa, Meerburg:2015yka}. So far, no analyses with any estimator has found significant evidence for a non-zero primordial bispectrum, with the best constraint from Planck being $\fnl^\mathrm{local} = -0.9\pm 5.1$, $\fnl^\mathrm{equil} = -26 \pm 47$ and $\fnl^\mathrm{ortho} = -38 \pm 24$ using the KSW~estimator~\cite{Planck:2019kim}. Having said that, constraints from just polarization appear tentative for orthogonal non-Gaussianity and future CMB~polarization data could shed further light on these hints in the near future. We present a comparison of these current constraints from Planck with forecasts for~CMB-S4 and a conservative cosmic-variance-limited CMB~experiment in Figure~\ref{fig:png_forecasts}.%

Analyses of bispectra from~PNG of scalar interactions, i.e.\ from tensor-scalar and tensor-tensor couplings, are more challenging since their shapes are intrinsically non-factorizable which makes them hard to constrain using the default estimator~\cite{Komatsu:2003iq}. Nevertheless, the first such constraints were derived from WMAP~temperature data in~\cite{Shiraishi:2017yrq}. Moreover, this challenge has since been resolved with only limited additional computational cost~\cite{Duivenvoorden:2019ses}, which opens up the possibility to make significant advances in the near future to constrain these types of bispectra using precision measurements of B-mode polarization.\medskip

The main source of nuisance in CMB~bispectrum analyses at current levels of sensitivity is the ISW-lensing bispectrum. This observable originates from correlations between small-scale temperature modes that are lensed by the gravitational potential and the integrated Sachs-Wolfe~(ISW) effect which is sourced by the same potential. This bispectrum primarily affects the local shape and was projected out in the Planck analyses, which is possible since we know the shape and amplitude of this bispectrum exactly. Other sources of confusion are point sources and galactic foregrounds which can however be dealt with effectively using foreground cleaning, masking and in-painting.

\paragraph{Beyond the Bispectrum}~\\
Higher-point correlation functions can in principle be constrained as well, but are computationally more challenging than the bispectrum. As shown in~\cite{Kalaja:2020mkq}, constraints on the CMB~trispectrum can in principle be competitive to the bispectrum due to favorable scaling of the signal-to-noise with the number of modes. To date, bounds on the primordial trispectrum are however only provided for two shapes~\cite{Planck:2018lbu}. More generally, higher-order spectra could be enhanced~(despite the natural expectation that the bispectrum provides the leading constraints in the limit of small~PNG) due to physics that affects the shape of the distribution of primordial fluctuations, as discussed in~\textsection\ref{sec:png_theory}. A dedicated attempt to look for these effects in data was undertaken in~\cite{Munchmeyer:2019wlh}~(see also~\cite{WMAP:2003xez, Buchert:2017uup}).

\paragraph{Prospects and Challenges}~\\
The expected improvements on PNG~measurements from the~CMB depend on the bispectrum shape. Due to projection effects~\cite{Babich:2004yc, Bordin:2019tyb, Kalaja:2020mkq}, the signal-to-noise scales as $(S/N)_{\fnl} \propto \ell \log(\ell_\mathrm{max}/\ell_\mathrm{min})$ for local temperature bispectra and as $(S/N)_{\fnl} \propto \sqrt{\ell}$ for equilateral shapes. This scaling behavior can be improved by combining temperature with E-mode polarization data. Based on current constraints, this implies that future CMB~data can in principle reach $\fnl^\mathrm{local} \sim O(1)$, while this level is out of reach for equilateral-type~PNG. For more detailed forecasts, including ISW-lensing and reionization-lensing deprojection, we refer to~\cite{Abazajian:2019eic}. 

Improvements by another factor of two may be obtained by including Rayleigh anisotropies, which are produced shortly after recombination when CMB~photons scatter off of neutral hydrogen and helium~\cite{Coulton:2020oxw}. To include this information in an analysis however requires a careful removal of foregrounds. Since constraints on large-scale B~modes will dramatically improve this decade~(cf.~\textsection\ref{sec:pgw_observations}), we note that current bounds on bispectra sourced by tensor-scalar and tensor-tensor interactions could be even more significantly tightened than those from purely scalar couplings~(see e.g.~\cite{Duivenvoorden:2019ses} for a forecast).\medskip

As CMB~observations push to smaller scales, the primary CMB~anisotropies produced at recombination become increasingly dominated by CMB~secondary anisotropies. These are signals that arise from weak lensing of the CMB~photons by, their scattering off of and additional emission from the intervening large-scale structure between the last-scattering surface and us. These CMB~secondaries are highly non-Gaussian, and can impact PNG~measurements as a bias and as a source of additional confusion.

The largest biases arise from correlations between lensing and the ISW~effect, the thermal Sunyaev-Zel'dovich~(tSZ) effect or the cosmic infrared background~(CIB)~\cite{Curto:2014bna, Hill:2018ypf}. As already mentioned, the ISW-lensing bispectrum must be modeled and removed from PNG~measurements since the ISW~effect cannot be addressed through multi-frequency CMB~observations. On small-enough scales, the bispectra from the tSZ~and CIB~effects become important. Since these effects produce anisotropies with a frequency dependence that differs from the~CMB, their induced PNG~biases can be effectively suppressed by combining CMB~observations at multiple frequencies. This multi-frequency cleaning must however be performed at very high precision because the size of these late-time non-Gaussianities is much larger than any primordial signal.

The primary sources of additional confusion arise from CMB~lensing and the kinetic Sunyaev-Zel'dovich~(kSZ) effect. Lensing of the~CMB introduces a large trispectrum which affects the covariance of the bispectrum, with the main impact on local-type~PNG given the nature of the lensed~CMB. While this nonlinear covariance can be mitigated by first delensing the~CMB~\cite{Coulton:2019odk}, not adequately addressing this effect can entirely reverse the projected improvements over current constraints. Although the CMB~temperature anisotropies imprinted by the kSZ~effect are non-Gaussian, the projection of these non-Gaussianities onto the primordial bispectra is very small. The main issue induced by the kSZ~effect is that it limits the gains from improved measurements. On the other hand, CMB~polarization is significantly less obscured and more easily cleaned since the main contaminants are point sources and galactic emission, which can be masked and removed, respectively, based on their distinct frequency dependencies. This means that there is a clear path to improve PNG~measurements with E-mode observations for $\ell \lesssim 8000$ or beyond as long as the lensing impact can be mitigated through delensing.

\paragraph{Implicit Likelihood Inference}~\\
Recent advances in implicit-likelihood inference~(ILI; also referred to as likelihood-free or simulation-based inference)~\cite{Alsing:2018eau, Alsing:2019xrx, Jeffrey:2020itg}, see~\cite{Cranmer:2019eaq} for a recent review, open up new and potentially more powerful ways of extracting information on~PNG from cosmological data sets, and addressing the challenges posed by secondary anisotropies, other foregrounds and astrophysical or observational systematics. These ILI~methods compute constraints by comparing statistics computed on data with statistics computed on simulated mock observations.

The potential of these methods derives from multiple points. First, there is considerable flexibility in the choice of statistics, i.e.\ analyses can use~(combinations of) $n$-point functions, both in configuration space and Fourier space, or other measures. More importantly, any type of filtering, cuts or masking can be applied to the data as long as it can also be applied to the simulations since the methods make only very weak assumptions about the form of the likelihood. In addition, the effects of non-Gaussianity on secondary anisotropies and astrophysical or observational systematics can be taken into account non-perturbatively at the level of numerical simulations, Finally, these methods can~(at least in principle) directly perform inference without choosing a particular combination of statistics, but directly using the observations.

The choice of statistical summaries is driven by a combination of their information content and the fidelity of the simulated observations. There are several challenges to simulation accuracy, in particular realizing physically accurate simulations of non-Gaussian initial conditions and the ability to simulate realistic observables. While in some cases~(such as local~PNG), physical realizations of the~CMB with non-Gaussian initial conditions can be generated~\cite{Elsner:2009md}, this is not possible in other cases where only low-order $n$-point functions are known perturbatively~(with $2 \le n\le 4$). Methods exist to generate initial conditions with a given combination of two-point and three-point correlation functions~\cite{Smith:2006ud}, but the physicality of higher-order $n$-point functions of those generated fields is not guaranteed. Having said that, simpler approaches exist that give rise to efficient estimators, rather than full, multi-dimensional posterior distributions. The needed optimal compression of statistics~\cite{Alsing:2017var}, or regression weights for posterior moments or marginals~\cite{Jeffrey:2020itg} can be trained on nonlinear simulations and can therefore go beyond the perturbative regime.

\subsubsection{Large-Scale Structure}

A promising observational probe of primordial non-Gaussianity is the distribution of structures on large scales and at late times. Matter overdensities grew from the initial conditions under the effect of gravity into the structures that we observe in the sky today with LSS~surveys. It is therefore natural to search for information on the initial conditions by studying how matter is distributed in the universe. The main challenges to overcome in this endeavor are gravitational nonlinearities, biasing, redshift space distortions and observational systematic effects. Gravitational nonlinearities arise due to the fact that gravitational growth is a nonlinear process which we need to model beyond linear order in perturbation theory to recover the initial conditions from the late-time density field. In addition, we only measure the clustering of luminous objects which trace the distribution of dark matter in a nonlinear way, and we observe these biased tracers in redshift space, which introduces yet another nonlinear mapping between the observables and the underlying dark matter distribution. Finally, the robust extraction of this signal relies on careful modeling/subtraction of large-scale systematics, such as variations in the properties of the target sample of spectroscopic surveys~\cite{Rezaie:2021voi}. Having said that, LSS~observations are able to explore the three-dimensional density field as opposed to its two-dimensional projection on the last-scattering surface probed by the~CMB. This is a major advantage of LSS~probes and is projected to result in tighter PNG~constraints given the anticipated large increase in the number of available modes with a high signal-to-noise ratio.

The large-scale structure of the universe can be explored through a variety of different probes which is another advantage. At low redshift, the statistical distribution of galaxies can be mapped through spectroscopic and photometric surveys, and observations of weak gravitational lensing directly study the distribution of dark matter. These are established LSS~probes. Line intensity mapping~(LIM) is an emerging technique~\cite{Kovetz:2017agg, Liu:2019awk}, which uses fluctuations in the intensity of spectral emission lines at different frequencies, and provides an additional way to map the~LSS over a wide range of scales and over many redshift epochs~(see the dedicated Snowmass~2021 White Papers on \SI{21}{cm} and mm-wavelength~LIM~\cite{Snowmass2021:21cm, Snowmass2021:LIM}). Particular targets are the \SI{21}{cm}~hyperfine transition of neutral hydrogen, rotational lines of carbon monoxide and the fine-structure line of ionized carbon. In contrast to galaxy surveys, which measure individually resolved galaxies, LIM~relies on detecting cumulative light from an ensemble of sources or the intergalactic medium which allows to efficiently map the~LSS over very large comoving volume. As a biased tracer of~LSS, all PNG~signatures imprinted in galaxy statistics are also contained in line intensity clustering statistics. LIM~therefore complements upcoming wide-field galaxy surveys at redshifts $z \lesssim 2$ and provides a detailed spectroscopic probe of~LSS at higher redshifts.

\paragraph{PNG from the Power Spectrum of Tracers}~\\
The local bispectrum is somewhat special in our effort to constrain~PNG since it probes the coupling between large and small scales. Importantly, local~PNG also induces a unique scale dependence of the bias~$b(\mathbf{k}, \fnl)$ of LSS~tracers, which sets the relation between the overdensity of biased tracers~$\delta_b(\mathbf{k})$ and the matter overdensity~$\delta(\mathbf{k})$ on large scales via $\delta_b(\mathbf{k}) = b(\mathbf{k}, \fnl)\hskip1pt \delta(\mathbf{k})$~\cite{Dalal:2007cu, Matarrese:2008nc, Slosar:2008hx}~(see~\cite{Desjacques:2016bnm, Biagetti:2019bnp} for recent reviews). The physical effect of local~PNG is a modulation of short-wavelength overdensity modes by long-wavelength modes of the Bardeen potential and its spatial derivatives. This results in a scale-dependent enhancement~(or suppression, depending on the sign of~$\fnl$) of the power spectrum on large scales. Such a scale dependence is not produced by other processes~(with the exception of projection effects on our past light cone which can however be calculated~\cite{Alonso:2015uua, Castorina:2021xzs}) and can be modeled mostly using linear physics since it affects large scales.

The strongest LSS~constraints on local~PNG to date, $-31 < \fnl^\mathrm{local} < 22$ at~95\%~C.L., have been inferred from a population of quasars between redshifts of~0.8 and~2.2 by the extended Baryon Oscillation Spectroscopic Survey~(eBOSS) collaboration by exploiting this scale-dependent-bias effect~\cite{Castorina:2019wmr, Mueller:2021tqa}. A similar scale-dependent bias can be observed in the galaxy-shape power spectrum which is measured in weak-lensing observations as an intrinsic shape alignment. While the imprint of massive particles with spin on the galaxy power spectrum is highly suppressed~\cite{MoradinezhadDizgah:2017szk, Cabass:2018roz}, these intrinsic alignments provide a unique opportunity to constrain these particles~\cite{Chisari:2013dda, Schmidt:2015xka, Chisari:2016xki, Kogai:2018nse, Akitsu:2020jvx, Kogai:2020vzz}.

Further improvements in the constraining power on local~PNG from the scale-dependent bias will be available with future multi-tracer observations which will allow us to mitigate sample variance~\cite{Seljak:2008xr}. The large variance of the matter field on large scales limits constraints on~$b(\mathbf{k}, \fnl)$ using a single tracer field. This sample variance can however be removed by taking the ratio of multiple biased tracers of the same underlying matter field~(each with their own deterministic bias) or of a biased tracer and the underlying matter field~(see also the Snowmass~2021 White paper~\cite{Snowmass2021:StaticProbes} on LSS~cross-correlations).\footnote{It is even possible to mitigate sample variance using a single tracer by applying certain reconstruction techniques~\cite{Darwish:2020prn}.} While this technique has not been applied to observational data yet, it in principle allows measurements of the scale-dependent bias to be only limited by shot noise. Forecasts using different tracer samples of the same galaxy survey or combinations of galaxy surveys with CMB-lensing or kSZ~measurements show the potential to improve constraints by almost an order of magnitude~(cf.\ Fig.~\ref{fig:png_forecasts}), with $\sigma\!\left(\fnl^\mathrm{local}\right) \sim 1$ being achievable by~SPHEREx or cross-correlating Vera Rubin Observatory photometric data with \mbox{CMB-S4}~lensing, for instance~\cite{Dore:2014cca, Schmittfull:2017ffw, Munchmeyer:2018eey}.

\paragraph{PNG from the Bispectrum of Tracers}~\\
For primordial non-Gaussianity beyond the local type, the leading LSS~observable is the three-point correlation function of tracers since the effect of most other PNG~types, for example the shape induced by the presence of extra massive particles as part of the cosmological collider, on the power spectrum of tracers is challenging to probe~\cite{Gleyzes:2016tdh}, but can potentially be constrained with the LSS~bispectrum~\cite{MoradinezhadDizgah:2018ssw, MoradinezhadDizgah:2018pfo, Kogai:2020vzz}. Even for the local shape, the bispectrum of biased tracers should provide significantly tighter constraints than the power spectrum~\cite{MoradinezhadDizgah:2020whw}, which can be understood as follows: (i)~the bispectrum of biased tracers captures both the scale-dependent bias and directly probes the matter bispectrum, (ii)~it intrinsically provides more information by capturing the imprints of~PNG on more modes in contrast to the power spectrum for which the imprint of local~PNG appears on the largest scales that are dominated by sample variance and~(iii) some of the parameter degeneracies that limit the power spectrum constraints are broken in the bispectrum. This therefore also implies that jointly analyzing the power spectrum and bispectrum should be the standard approach. As anticipated above, the main challenge for bispectrum analyses is that gravitational nonlinearities also generated a large non-vanishing bispectrum, which therefore acts as a source of confusion for the primordial signature. Modeling these nonlinearities implies introducing more nuisance parameters as we try to push the model to more nonlinear scales.

Recent progress~\cite{Cabass:2022wjy, DAmico:2022gki, Cabass:2022ymb}~(see also e.g.~\cite{Baldauf:2010vn, Baldauf:2014qfa, Angulo:2015eqa, Assassi:2015jqa} in these modeling efforts has allowed us to put the first constraints on both~$\fnl^\mathrm{local}$, and on~$\fnl^\mathrm{equil}$ and~$\fnl^\mathrm{ortho}$ from the measurement of the bispectrum of BOSS~DR12~galaxies, constraining these PNG~parameters to~$O(30)$ and~$O(100)$, respectively. While these limits are not competitive with current CMB~bounds, the calculations and tests performed on BOSS~data pave the way for future surveys with much larger constraining power.

\paragraph{Map-Level Inference of PNG from Large-Scale Structure Data}~\\
Recent progress has been made in considering alternatives to using low-order correlation functions~(such as the power spectrum and bispectrum) as summary statistics to constrain primordial non-Gaussianity. While it would intuitively seem natural to ``observe a primordial bispectrum through a late-time bispectrum'', secondary non-Gaussianities are dominant for these statistics and degenerate with the primordial ones~(especially for LSS~observables), as previously discussed. The physical processes producing these non-Gaussianities are however very different: while primordial non-Gaussianities are inherently non-local in space, late non-Gaussianities are local in space (but non-local in time). For this reason, it appears promising to challenge the problem directly at the map level in configuration space, where locality is explicit~\cite{Baumann:2021ykm}.

Methods based on the analysis of survey data at the map level may therefore have an important role to play in the search for~PNG in~LSS. Employing~ILI might be a promising tool, for instance, since we can use non-perturbative, numerical simulations~(in redshift space) to model non-primordial effects due to gravitational growth and hydrodynamic processing of the initial perturbations. Moreover, ILI~can in principle ``discover'' the optimal combination of features of the evolved~LSS to disentangle processing from primordial effects.

It is intriguing to note that the information content of~LSS on~PNG has so far only been quantified perturbatively due to the nonlinearity of the problem. Recent work has begun to use~ILI to determine the cosmological information content of the nonlinear dark matter distribution~\cite{Schmittfull:2017uhh, Seljak:2017rmr, Taylor:2019mgj, Dai:2020ekz, Modi:2020dyb, Makinen:2021nly, Villaescusa-Navarro:2021pkb, Villaescusa-Navarro:2021cni, Hassan:2021ymv, Cole:2021gwr, Dai:2022dso} and even of state-of-the-art hydrodynamical simulations~\cite{Villaescusa-Navarro:2021pkb, Villaescusa-Navarro:2021cni}. Given the rapid advances in this field, it is anticipated that these techniques will also be applied to the problem of extracting inflationary physics from LSS~data~(cf.\ e.g.~\cite{Coulton:2022qbc, Jung:2022rtn, Coulton:2022rir}). Additional approaches that would exploit the same idea are also being developed in the direction of formulating field-level likelihoods for biased tracers~\cite{Jasche:2012kq, Wang:2013ep, Ata:2014ssa, Schmittfull:2018yuk, Elsner:2019rql, Cabass:2019lqx, Cabass:2020nwf, Schmidt:2020viy, Cabass:2020jqo, Schmittfull:2020trd, Andrews:2022nvv} and the application of computational topology methods to find features at various coarse graining scales~\cite{Biagetti:2020skr, Biagetti:2022qjl}.

\paragraph{Prospects and Challenges}~\\
The prospects for constraining~$\fnl^\mathrm{local}$ down to order unity are bright in the near future, cf.~Fig.~\ref{fig:png_forecasts}. Upcoming spectroscopic surveys probing $z\lesssim2$, such as~DESI and~Euclid, are expected to provide constraints comparable to those of Planck from power-spectrum-only measurements and potentially approaching the target sensitivity of $\fnl^\mathrm{local} = O(1)$ using the bispectrum~\cite{Karagiannis:2018jdt}. Similar potential may also be offered by other LSS~tracers, such as fast radio bursts~\cite{Reischke:2020cgd}. At the same time, SPHEREx~is anticipated to push the limits from both the power spectrum and bispectrum due to its wide frequency coverage and its ability to perform multi-tracer analyses by splitting their galaxy sample into several subsamples with different galaxy biases~\cite{Dore:2014cca}.

There are several challenges to overcome in order to reach this level of constraining power for~$\fnl^\mathrm{local}$, and to substantially improve the bounds on~$\fnl^\mathrm{equil}$ and~$\fnl^\mathrm{ortho}$. First, it will be important to obtain a fully consistent perturbative model for the two-loop power spectrum, one-loop bispectrum and tree-level trispectrum for galaxies in redshift space and in the presence of~PNG. It would also be greatly beneficial for PNG~constraints if we were able to impose stronger priors on the values of the nuisance parameters of the model, e.g.\ through information from high-fidelity simulations~\cite{Barreira:2021ueb}. This applies to both the EFT~and the galaxy bias parameters. The latter are more of a concern in case of a detection of~PNG since their amplitude is almost perfectly degenerate with the primordial signal~\cite{Barreira:2020kvh, MoradinezhadDizgah:2020whw, Barreira:2021ueb, Barreira:2022sey}. Future surveys will observe hundreds of triangle configurations with good signal-to-noise ratios which implies a very high-dimensional covariance matrix with large off-diagonal elements that can significantly affect the constraining power of the bispectrum~(see e.g.~the PUMA~forecasts in~\cite{Floss:2022wkq}). Recent substantial developments in covariance matrix estimation and data compression techniques are however paving the way towards a consistent likelihood analysis for future surveys~\cite{Wadekar:2020hax, Philcox:2020zyp, Biagetti:2021tua}. Finally, exquisite control of the large-scale systematics for the power spectrum and bispectrum constraints has to be achieved. While several methods are available for the former~(e.g.~\cite{Kalus:2018qsy, Rezaie:2021voi}), it is still an open problem how to best handle observational and instrumental effects in the bispectrum. In conclusion, while important advances have recently been made in all areas, as briefly summarized above, it remains to be seen whether this is adequate for upcoming surveys given their significantly larger signal-to-noise ratio.

Looking further into the future, wide-field LIM~surveys at \SI{21}{cm}~and/or mm~wavelengths can further improve the expected PNG~constraints from spectroscopic galaxy surveys~\cite{MoradinezhadDizgah:2018zrs, MoradinezhadDizgah:2018lac, Bandura:2019uvb, Karagiannis:2019jjx, Sailer:2021yzm}. Proposed experiments such as~PUMA or~\mbox{mm-LIM} will cover the largest volume with the lowest noise~(at least in principle) and could represent the ultimate probes of~PNG in the post-reionization era. Realizing the promise of~LIM however relies on multiple factors and the feasibility of this observational technique has yet to be demonstrated~(see the dedicated Snowmass~2021 White Papers~\cite{Snowmass2021:3dLSS, Snowmass2021:21cm, Snowmass2021:LIM} for more details).\medskip

We conclude with Figure~\ref{fig:png_forecasts}, which summarizes forecasts for a range of upcoming and proposed surveys, using a vast array of probes. In principle, SPHEREx~\cite{Dore:2014cca}, the galaxy survey~MegaMapper~\cite{Schlegel:2019eqc}, the \SI{21}{cm}~experiment~PUMA~\cite{Bandura:2019uvb, Castorina:2020zhz} and \si{mm}-wave~LIM~\cite{Snowmass2021:LIM} are projected to provide very interesting constraints on primordial non-Gaussianity. It is however clear that significant progress in theory, experiment and data analysis will be necessary to fully achieve the potential to extract PNG~information. Nonetheless, the future looks bright, with many advances in all directions towards decoding this treasure of cosmological information.
\clearpage
% !TEX root = whitepaper.tex

%%%%%%%%%%%%%%%%
\section{Primordial Features}
\label{sec:features}
%%%%%%%%%%%%%%%%

In addition to primordial gravitational waves and primordial non-Gaussianity, primordial features are a separate signal of physics beyond the standard models of cosmology and particle physics. These inflationary imprints are a manifestation of primordial dynamics that exhibit a significant departure from scale invariance and arise in broad classes of models, including of both inflation and its alternatives. Finding such inflationary signatures in cosmological observables would be a groundbreaking discovery that would open an entirely new window into the primordial universe~(see e.g.~\cite{Chen:2010xka, Chluba:2015bqa, Slosar:2019gvt} for previous reviews).\medskip

Vanilla models of inflation predict almost Gaussian fluctuations with a nearly scale-invariant power spectrum as we have discussed in the previous sections. However, many models of the very early universe beyond the simplest incarnations of single-field slow-roll inflation generically predict departures from scale invariance. Since these deviations from the minimal power-law power spectrum of initial fluctuations are strongly scale-dependent, primordial features are typically oscillatory and/or localized in momentum space. They are ubiquitous in theoretical attempts to connect inflationary modeling to fundamental physics, can be introduced by a wide variety of phenomena and carry valuable information about the nature of the primordial universe~(cf.\ Fig.~\ref{fig:scales}). In addition, similar signals may be imprinted in observables during the cosmic evolution after the hot big bang and reveal unique information about the particles and forces at play in the universe.\footnote{Oscillatory features may not only be imprinted in the cosmological observables in the primordial universe, but also as a result of the dynamics of the primordial plasma after the hot big bang. For instance, an interaction between all or a fraction of the dark matter and dark radiation (neutrinos or relativistic particles beyond the Standard Model) would result in so-called dark acoustic oscillations~\cite{Mangano:2006mp, Ackerman:2008kmp, Feng:2009mn, Serra:2009uu, McDermott:2010pa, CyrRacine:2012fz, Cyr-Racine:2013fsa, Petraki:2014uza, Buckley:2014hja, Tulin:2017ara}. Features may also be the result of non-standard components affecting the expansion history at early times~\cite{Poulin:2018cxd, Smith:2019ihp}. A detection of these and similar signals may therefore provide a unique probe of the existence of new particles and their interactions present in the universe. In the following, we will however focus on the inflationary origin of potential features in cosmological observations and refer to the dedicated Snowmass~2021 White Paper on early-universe model building~\cite{Snowmass2021:EarlyModels} for additional information.}\medskip

Observationally, primordial features could be imprinted in the spectrum of the cosmic microwave background, its anisotropies, all tracers of the large-scale structure of the universe and the stochastic gravitational wave background. So far, these observations however point to an almost scale-invariant primordial spectrum with a slight red tilt with no evidence for primordial features. Nevertheless, the future of primordial feature searches in all observables is bright. In the near term, CMB~and especially galaxy surveys will provide significant advances in sensitivity. In the more distant future, line intensity mapping and gravitational wave observatories are projected to lead to tremendous improvements in constraining power over a large part of parameter space. This means that future cosmological observations offer the potential for a dramatic discovery about the nature of cosmic acceleration in the very early universe.
% !TEX root = whitepaper.tex

% Primordial Features

\subsection{Theoretical Background}
\label{sec:features_theory}

The simplest models of inflation predict initial fluctuations of the hot big bang cosmology to follow (almost)~Gaussian statistics with a nearly scale-invariant power spectrum, $\Delta_{\mathcal{R},0}^2(k) \equiv \As\,(k/k_*)^{\ns-1}$, cf.~\eqref{eq:primordial_power_spectrum}, consistent with cosmological observations. Achieving this power spectrum within more fundamental models, e.g.~those taking into account interactions between the inflaton and other degrees of freedom, requires the introduction of symmetries. These symmetries are however known to be broken in a theory of quantum gravity, with inflation being sensitive to the related effects~(see e.g.~\cite{Baumann:2009ds, Baumann:2014nda}). Remnants of this tension can still be present in models, which avoid the most severe quantum gravity effects. As a consequence, sub-leading violations of scale invariance are imprinted on the primordial power spectrum in the form of features:
\begin{equation}
	\Delta_\mathcal{R}^2(k) = \Delta_{\mathcal{R},0}^2(k)\, [1 + \delta\Delta_\mathcal{R}^2(k)]\, .	\label{eq:feature_spectrum}
\end{equation}
The emergence of~$\delta\Delta_\mathcal{R}^2(k)$ can be studied within effective models of inflation taking into account many degrees of freedom and/or non-canonical interactions that represent effects from quantum gravity. The detection of these features would provide a unique insight into the physics of the very early universe. It could also provide evidence for particular models of inflation or one of its alternatives, or identify the existence of new particles and forces in the early universe.\medskip

From the point of view of ultraviolet completions in inflationary model building, the presence of primordial features is rather natural. The landscape of the inflationary potential is expected to be a complicated function of many fields and a very high-dimensional space. The inflaton trajectory is only one of the many low-energy low-dimensional subspaces embedded in this landscape. It therefore seems natural to expect the presence of various features that could violate the assumption of single-field slow-roll inflation. These features may break the smoothness of the inflaton potential or the internal space of the inflaton, or may excite heavy fields along the main inflaton trajectory.

A strongly scale-dependent correction~$\delta\Delta_\mathcal{R}^2(k)$ arises in broad classes of inflationary models since they may occur if any background quantity~$B(t)$ involved in the linear evolution of the curvature perturbation~$\mathcal{R}(t,\mathbf{k})$ experiences departures from a slow-roll evolution of the form $|\dot B / B H| \ll 1$~(even if~$B$ remains small). Here, $B(t)$~can refer to one or a combination of the following classes of background quantities commonly encountered in the building of inflationary models: (a)~background quantities parametrizing the evolution of the scale factor, such as the slow-roll parameters, (b)~background functions parametrizing departures from canonical inflation, such as the sound speed of~$\mathcal{R}(t,\mathbf{k})$, or (c)~background couplings describing the interaction between the inflaton and other degrees of freedom. While it is natural to expect that these three classes of time-dependent backgrounds should occur simultaneously, we note that there are well-motivated models in which features can emerge without slow-roll being interrupted~\cite{Achucarro:2010da, Palma:2020ejf}.

Primordial features imprinted in the power spectrum generically have an oscillatory form which encapsulates the fact that a small component of the density perturbations significantly departed from scale invariance. They are typically found to lie within two main classes related to their mechanism of generation:
\begin{itemize}
	\item \textbf{Sharp features} are produced by the momentary departure of a background parameter~$B$ from the attractor solution, $|\dot B / B H| \ll 1$~\cite{Starobinsky:1992ts}. These take the form of a~(typically transient) oscillatory component in the power spectrum which can be described by linear oscillations, $\delta \Delta_\mathcal{R}^2(k) = f(k)\hskip1pt \cos(2 k / k_0 + \phi)$, where $f(k)$~is a model-dependent envelope function, $k_0$~is the value of the comoving wavelength that crossed the horizon at the time of momentary departure and $\phi$~is a constant phase. Examples of models in which this class of features appear include single-field inflation with a non-smooth inflaton potential~\cite{Starobinsky:1992ts, Adams:2001vc, Ashoorioon:2006wc, Chen:2006xjb, Chen:2008wn, Bean:2008na, Achucarro:2010da, Miranda:2012rm, Bartolo:2013exa, Hazra:2014goa, Ashoorioon:2014yua}, multi-field models with sudden turns of the inflationary trajectory~\cite{Ashoorioon:2008qr, Achucarro:2010da, Cespedes:2012hu, Palma:2020ejf} and effective field theory models with sudden variations of the sound speed~\cite{Achucarro:2010da, Cespedes:2012hu, Park:2012rh, Achucarro:2012fd, Achucarro:2013cva}. In some cases, such as models with a resonant production of particles~\cite{Kofman:1997yn, Chung:1999ve, Kofman:2004yc, Romano:2008rr, Green:2009ds, Barnaby:2009mc, Barnaby:2009dd, Flauger:2016idt, Kim:2021ida}, the local interaction generates a bump in the power spectrum and the envelop of the sinusoidal running decays very quickly towards increased scales. In practice, the feature can therefore be represented by just a bump, e.g.~$\delta \Delta_\mathcal{R}^2(k) \sim k^3 e^{-\pi k^2/(2 k_*^2)}$, in some of these scenarios. In addition to a single sharp feature, a periodic~\cite{Green:2009ds} or random~\cite{Green:2014xqa, Amin:2015ftc, Garcia:2019icv, Garcia:2020mwi} distribution of these imprints may occur which can however usually be treated as a sum of oscillations or localized structures in Fourier space.

	\item \textbf{Resonant features} are produced by the periodic oscillation of a background quantity around the attractor solution with a super-Hubble frequency since it resonates with the subhorizon quantum modes~$\mathcal{R}$~\cite{Chen:2008wn}. The signal is characterized by an oscillatory feature with a constant frequency extending over a wide range in $\log k$-space: $\delta \Delta_\mathcal{R}^2(k) \propto \cos(\Omega \log(2 k) + \phi)$, with constants~$\Omega$ and~$\phi$. Examples of this class of features include inflationary models with oscillatory components in the potential, such as in axion monodromy~\cite{Silverstein:2008sg, McAllister:2008hb, Flauger:2009ab, Flauger:2010ja} or in other backgrounds~\cite{Chen:2010bka, Behbahani:2011it}, which may include possible runnings of the frequency~$\Omega$~\cite{Flauger:2014ana}. Resonant features can also arise if inflationary models start from certain non-Bunch-Davies vacua~\cite{Easther:2002xe, Martin:2003kp} or be used to boost the tensor-to-scalar ratio~\cite{Cai:2021yvq}.
\end{itemize}
The physics responsible for these scenarios is often deeply tied to the fundamental origin of the respective model. We can exemplify this with the resonant scenario of axion monodromy~\cite{Silverstein:2008sg, McAllister:2008hb, Flauger:2009ab, Flauger:2010ja}. While the shift symmetry of axion fields make them attractive candidates to be the inflaton, this discrete symmetry is generically broken in string theory or in the presence of multiple interacting axions. In consequence, the period of the underlying axion potential is smaller than the field range which results in resonant features in the scalar spectra with a model-dependent amplitude and envelope function of~$\delta\Delta_\mathcal{R}^2(k)$.

\begin{figure}[t]
	\centering
	\includegraphics{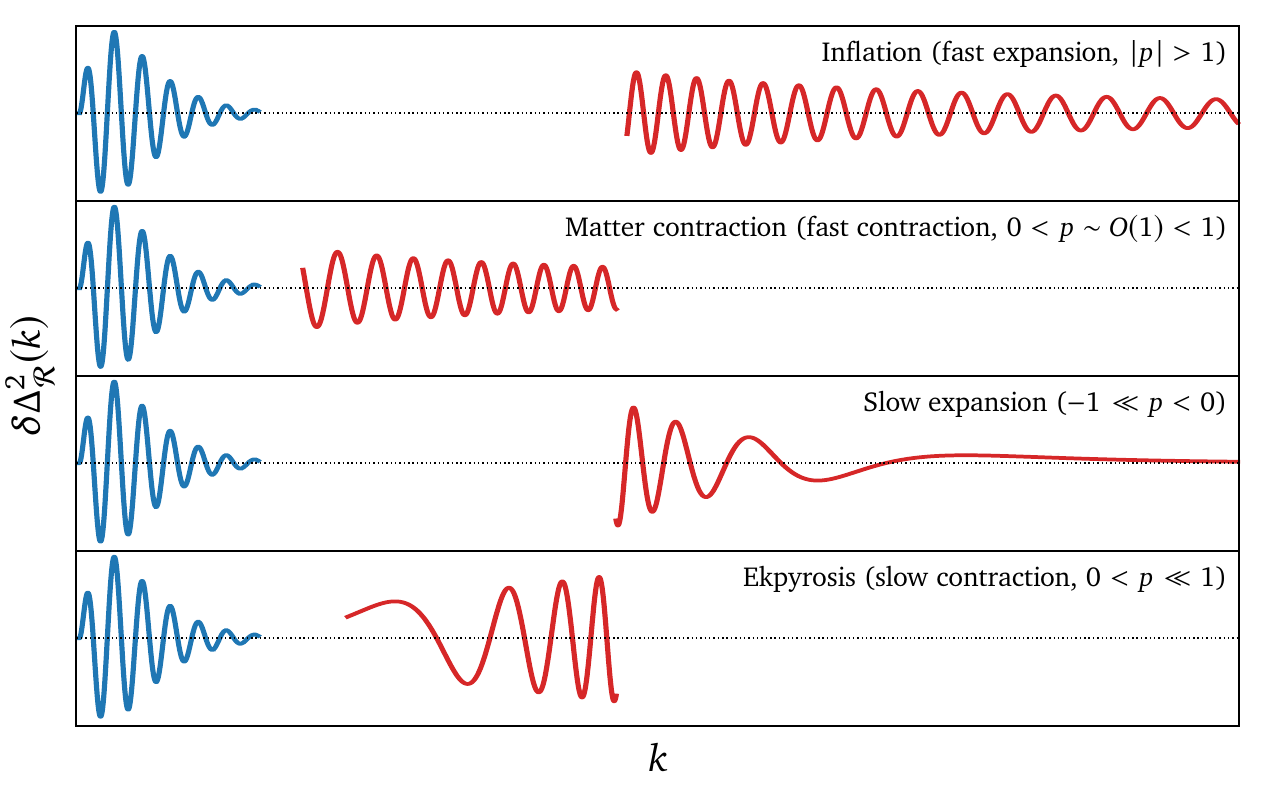}
	\caption{A qualitative sketch of signals of classical primordial standard clocks in various primordial universe scenarios that we phenomenologically parameterize by $a(t)\propto t^p$~(adapted from~\cite{Chen:2014cwa}). This includes inflation in the top panel characterized by a fast expansion with $|p| > 1$. The blue curves represent sharp feature signals, which are qualitatively similar for all scenarios. The red curves depict clock signals that are induced by a classically oscillating heavy field, with their phases directly encoding the information of~$a(t)$.}
	\label{fig:CPSCsketch}
\end{figure}

A class of features that combines both types of signals is encountered within the framework of classical primordial standard clocks~\cite{Chen:2011zf, Chen:2011tu, Chen:2012ja, Battefeld:2013xka, Gao:2013ota, Noumi:2013cfa, Saito:2012pd, Saito:2013aqa, Chen:2014joa, Chen:2014cwa, Huang:2016quc, Domenech:2018bnf, Braglia:2021ckn, Braglia:2021sun, Braglia:2021rej}. In these scenarios, classical oscillations of some massive fields~(with masses~$m \gg H$) are excited by sharp features that are encountered by the inflaton along its trajectory, such as sharp turns of trajectories in field space or tachyonic falling over potentials. The sinusoidal-running signal, generated by the sharp feature, smoothly connects with the resonant-running signal imprinted by the oscillation of the massive field~(the so-called ``clock'' signal). The phase of the oscillatory clock signal directly records the time dependence of the background scale factor~$a(t)$. This remarkable property remains valid beyond the inflationary scenario and applies to various alternative scenarios to inflation that we phenomenologically parameterize in Fig.~\ref{fig:CPSCsketch}.
This gives rise to a unique opportunity to experimentally test the very definition of the inflationary paradigm, complementary to the approach of primordial gravitational waves that we discussed in Sec.~\ref{sec:pgw}.

The dynamics underlying the appearance of features in the power spectrum may also lead to features in higher-point correlation functions and, therefore, in non-Gaussianities~(cf.\ Sec.~\ref{sec:png})~\cite{Chen:2006xjb, Chen:2008wn, Flauger:2010ja, Chen:2010bka, Achucarro:2012fd, Achucarro:2013cva, Gong:2014spa, Palma:2014hra, Mooij:2015cxa, Appleby:2015bpw, Torrado:2016sls}. As an example, the bispectrum will receive a strongly scale-dependent correction of the form $S(k_1, k_2, k_3) = S_0(k_1, k_2, k_3)\, [1 + \delta S(k_1, k_2, k_3)]$, which is analogous to~\eqref{eq:feature_spectrum}. Here, the function~$S_0$ represents the standard single-field prediction for the bispectrum discussed in~\textsection\ref{sec:png_theory}. Since both~$\delta \Delta_\mathcal{R}^2$ and~$\delta S$ have the same origin, a certain degree of correlation is expected, which in general takes the following form:
\begin{equation}
	\delta S(k_1,k_2,k_3) = f_0\, \delta\Delta_\mathcal{R}^2(k) + f_1\hskip1pt \frac{\d}{\d k} \delta\Delta_\mathcal{R}^2(k) + f_2\hskip1pt \frac{\d^2}{\d k^2} \delta\Delta_\mathcal{R}^2(k)\, ,	\label{eq:features_bispectrum}
\end{equation}
where $k \equiv (k_1 + k_2 + k_3)/2$ and $f_i = f_i (k_1,k_2,k_3)$,~$i=0,1,2$, are model-dependent functions with a smooth dependence on the three momenta~$k_j$ (see~\cite{Achucarro:2012fd, Palma:2014hra}). Observational constraints of these functions~$f_i$ would provide additional powerful tools to distinguish different models of inflation~\cite{Achucarro:2014msa}. In addition, there are also scenarios of heavy particle production or resonant models which induce correlated higher-point functions of potential observational significance~\cite{Flauger:2016idt, Leblond:2010yq}.\medskip

So far, we have focused on primordial features in the context of a small amplitude relative to the leading-order scale-invariant spectrum,~$\delta\Delta_\mathcal{R}^2(k) \ll 1$, which is motivated by the observational constraints on scales $k \ll \SI{1}{\per\Mpc}$ from CMB~and LSS~data~(cf.~\textsection\ref{sec:features_observations}). On smaller scales, $k \gg \SI{1}{\per\Mpc}$, the amplitude and shape of the primordial power spectrum remains however largely unconstrained. This consequently allows for the possibility that the curvature fluctuations are significantly larger. In fact, this enhancement can even dominate over the featureless spectrum by orders of magnitude in some cases~\cite{Leach:2000ea,Kohri:2007qn,Alabidi:2009bk,Linde:2012bt,Kawasaki:2012wr,Kohri:2012yw,Bugaev:2013fya,Clesse:2015wea,Garcia-Bellido:2017mdw,Germani:2017bcs,Motohashi:2017kbs,Hertzberg:2017dkh,Pi:2017gih,Cai:2018tuh,Cai:2019bmk}. While the scale-dependent oscillatory pattern still depends on its generation mechanism and can be classified in the same way as discussed above, in particular as sharp and resonant features, the phenomenology of features with a large amplitude is richer.

To illustrate this, we consider sharp large-amplitude features on small scales. In this case, the enhancement of the leading-order power spectrum~$\Delta_{\mathcal{R},0}^2(k)$ leads to an enhancement of the oscillatory part~$\delta \Delta_\mathcal{R}^2(k)$ which results in large $O(1)$~oscillations~\cite{Palma:2020ejf, Fumagalli:2020adf, Fumagalli:2020nvq, Braglia:2020taf}. This can be understood as a consequence of the sharp feature dynamically inducing an effective excited state for the curvature perturbation~$\mathcal{R}$~(and potentially other entropic fluctuations in a multi-field inflation setting) with moderate to copious particle production~\cite{Fumagalli:2020nvq}. Since such a significant enhancement of fluctuations leaves potentially observable signatures in the form of induced gravitational waves and primordial black holes, this scenario allows to observationally scrutinize inflation also on small scales~(cf.~\textsection\ref{sec:features_gwb}). More generally, this exemplifies that small-scale large-amplitude features can give rise to new physical phenomena compared to large-scale features with an amplitude that is constrained to be small.

\bigskip
To summarize, departures from the minimal power-law power spectrum of primordial fluctuations occur ubiquitously in theoretical attempts to connect the inflationary modeling to fundamental physics. In addition, it is possible to extract broader lessons from the discussed scenarios for low-energy effective field theory and data analysis. Since there are however no useful theoretical priors on the scale or amplitude of primordial features, which is related to the lack of our understanding of fundamental physics, cosmological searches for these inflationary signatures should cover as much of parameter and model space as possible. Conversely, the past and future extensive observational hunts, which we will discuss in the following, can inform the theoretical modeling. The combination of theory and observations may therefore offer an exciting opportunity to not only reveal a portion of rather detailed evolutionary history of inflation, but also provide direct model-independent evidence for the inflationary paradigm.
% !TEX root = whitepaper.tex

% Primordial Features

\subsection{Observational Imprints}
\label{sec:features_observations}

All cosmological observables that are sensitive to fluctuations in the universe will contain signals from features in the primordial spectra, if present. It is useful to employ all these observables since they probe complementary scales and have different advantages. On large scales, the leading constraints come from observations of the cosmic microwave background anisotropies and the large-scale structure of the universe. Spectral distortions of the CMB~black body spectrum and the stochastic gravitational background provide an entirely complementary window on the primordial power spectrum and features on small scales.

\subsubsection{Cosmic Microwave Background}

Analyses of CMB~data have been the cornerstone of primordial feature searches. The primary CMB~anisotropies have been extensively employed to constrain these inflationary imprints on large scales, while spectral distortions of the black body spectrum can put bounds on these departures from scale invariance on small scales.

\paragraph{CMB~Anisotropies}~\\
As with the searches for primordial gravitational waves, primordial non-Gaussianity and other inflationary signatures, the primary CMB~temperature anisotropies and polarization signal have been at the forefront of the observational sensitivity to features in the primordial spectra. They are imprinted in the CMB~spectra after convolution with the transfer functions as detailed in~\eqref{eq:clscalar} for the power spectra~$C_\ell^{XY}$. The particular sensitivity of the CMB~anisotropies to these oscillatory signatures is due to the following advantages: (i)~they probe the largest accessible scales, (ii)~their physics is entirely linear and, therefore, under complete theoretical control, and (iii)~they are extremely well measured. On the other hand, projection and transfer-function effects decrease the primordial signal in the CMB~data: the linear transformation between the primordial spectra and the observed spherical power and higher-order spectra intrinsically averages oscillatory imprints which in particular impedes searches for high-frequency features. In addition, since Planck has already measured the temperature power spectrum to the cosmic variance limit up to $\ell\sim1600$~\cite{Planck:2015bpv}, future CMB~experiments will only bring relatively incremental improvements of factors of a few at most with polarization measurements approaching the cosmic variance limit~(see e.g.~\cite{Finelli:2016cyd, Hazra:2017joc, Sohn:2019rlq, Beutler:2019ojk}).\footnote{We however note that the Planck~likelihood for the CMB~TT, TE and EE~power spectra employed sub-optimal pseudo-$C_\ell$ estimators at high multipoles~$\ell$. It might not be widely appreciated that this sub-optimality can lead to excess variance in the pseudo-$C_\ell$ estimates at the 10-20\%~level. An optimal re-analysis could therefore improve the power spectrum constraints on primordial features and other cosmological parameters by a similar amount.}\medskip

Extensive template searches for primordial features have been performed in the CMB power spectra, in particular as measured by the WMAP~and Planck~satellites. In these analyses, the particular functional forms of sharp, resonant and similar features is typically incorporated in the model for the power spectrum to be constrained by the data. This may also include particular aspects of certain inflationary models as discussed in~\textsection\ref{sec:features_theory}. However, in the most general case, features represent any component that modulates a smooth ``background'' given by the near power-law power spectrum produced by slow-roll, $\Delta_{\mathcal{R},0}^2(k)$. Two template models are linear oscillations,
\begin{equation}
	\Delta_\mathcal{R}^2(k) = \Delta_{\mathcal{R},0}^2(k) \left[ 1 + A_\mathrm{lin} \cos(\omega_\mathrm{lin}\, k + \phi_\mathrm{lin}) \right] ,	\label{eq:linear_features}
\end{equation}
which modulate the minimal slow-roll power-law spectrum by a sinusoidal fluctuation with a certain relative amplitude~$A_\mathrm{lin}$, frequency~$\omega_\mathrm{lin}$ and phase~$\phi_\mathrm{lin}$, and logarithmic oscillations,
\begin{equation}
	\Delta_\mathcal{R}^2(k) = \Delta_{\mathcal{R},0}^2(k) \left[ 1 + A_\mathrm{log} \cos\left(\omega_\mathrm{log} \log(k/k_*) + \phi_\mathrm{log} \right) \right] ,	\label{eq:logarithmic_features}
\end{equation}
with the same three parameters. These linear and logarithmic feature templates encapsulate the general form of sharp and resonant features discussed in~\textsection\ref{sec:features_theory}, but with a constant amplitude~$A_X$. Apart from dedicated model predictions, linear oscillations can also be a useful basis in which to look for features since they can capture large parts of model space because these oscillations form an orthogonal basis of functions over a given range of wavenumbers~$k$, i.e.\ similar to a time-series analysis problem. We also note that the restricted range of scales over which the observations have an appreciable signal-to-noise ratio may be interpreted as an implicit envelope function over the constant-amplitude oscillations of~\eqref{eq:linear_features} and~\eqref{eq:logarithmic_features}. On the other hand, this approach has its limitations and dedicated analyses should be performed when these shapes significantly deviate and/or tentative signals are found in the~future since the feature templates are not random fields, but instead have well-defined shapes (or phase relations in decomposition).

To date, no significant detection has been made in CMB~data, but constraints on the feature amplitudes have been reported at the percent level relative to the primordial scalar amplitude~$\As$~\cite{Pahud:2008ae, Hazra:2010ve, Aich:2011qv, Meerburg:2011gd, Adshead:2011jq, Peiris:2013opa, Meerburg:2013cla, Meerburg:2013dla, Easther:2013kla, Achucarro:2013cva, Miranda:2013wxa, Fergusson:2014tza, Planck:2015sxf, Hazra:2016fkm, Ballardini:2018noo, Planck:2018jri, Planck:2019kim, Beutler:2019ojk, Bayer:2020pva, Braglia:2021ckn, Hamann:2021eyw}. Having said that, some potentially interesting candidates for such departures have been reported at marginal statistical significance, including a dip in the temperature power spectrum around $\ell \sim 20 - 40$ and a potential oscillatory feature around multipoles of $\ell \sim 700 - 800$~\cite{WMAP:2003syu, Planck:2015sxf, Planck:2018jri}~(see also~\cite{Snowmass2021:CosmologyIntertwined}). Analyses of CMB~temperature and polarization data have considered these and other feature-like anomalies in cosmological datasets, and revealed new potential candidates for oscillatory imprints~\cite{Mortonson:2009qv, Achucarro:2013cva, Hu:2014hra, Miranda:2014fwa, Torrado:2016sls, Domenech:2019cyh, Domenech:2020qay, Braglia:2021sun, Braglia:2021rej, Hazra:2022rdl, Antony:2022ert}. As discussed in~\textsection\ref{sec:features_theory}, many inflationary models also predict correlated features in the power spectrum and bispectrum (and additional higher-point spectra), but combined analyses of these spectra, which include the look-elsewhere effect, have also not found any significant deviations from a featureless spectrum~\cite{Fergusson:2014hya, Fergusson:2014tza, Meerburg:2015owa, Planck:2018jri}. In general, dedicated feature searches are either based on the templates~\eqref{eq:linear_features} and~\eqref{eq:logarithmic_features} or other more model-dependent templates, but might not capture all theoretically predicted aspects, such as the momentum dependence of the bispectrum phase of~\eqref{eq:features_bispectrum}. At the same time, new approaches to the data, such as the use of an estimator that resums $n$-point functions in the context of features generated by heavy particle production~\cite{Munchmeyer:2019wlh}, show the future potential of ongoing developments of efficient numerical and data analysis techniques~(see e.g.~\cite{Lewis:1999bs, Lewis:2002ah, Skilling:2006gxv, Feroz:2007kg, Feroz:2008xx, Blas:2011rf, Hazra:2012yn, Howlett:2012mh, Audren:2012wb, Feroz:2013hea, Handley:2015fda, Handley:2015vkr, Brinckmann:2018cvx, GAMBITCosmologyWorkgroup:2020htv}).

\begin{figure}[t]
	\centering
	\includegraphics{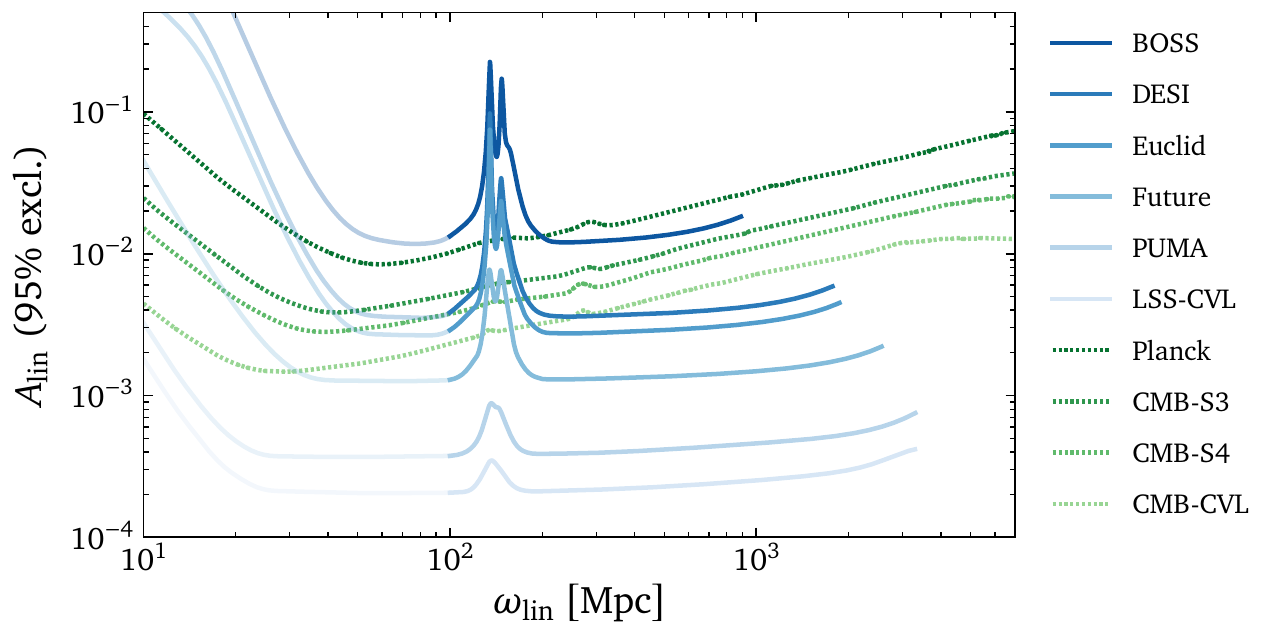}
	\caption{Forecasted sensitivity for the `feature spectrometer' of linear features. The potential reach of various CMB~(dashed) and LSS~(solid) experiments to constrain the feature amplitude~$A_\mathrm{lin}$ at a confidence level of~95\%~(under the assumption that the true amplitude is zero) is presented as a function of their frequency~$\omega_\mathrm{lin}$~(adapted from~\cite{Beutler:2019ojk, Ansari:2018ury}). The positive-semi-definite nature of~$A_\mathrm{lin}$ is taken into account in the displayed estimates. The underlying experimental specifications for planned surveys are similar to those of~BOSS~\cite{BOSS:2016hvq}, DESI~\cite{DESI:2016fyo}, Euclid~\cite{EUCLID:2011zbd}, Planck~\cite{Allison:2015qca}, a CMB-S3-like experiment~(e.g.~\cite{SimonsObservatory:2018koc}) and CMB-S4~\cite{Abazajian:2019eic}, respectively~(see also~\cite{Baumann:2017gkg}). To illustrate the potential future reach, we additionally show the expected sensitivity of a future LSS~survey with \num{e8}~objects up to redshift $z_\mathrm{max} = 3$ over half the sky using a maximum wavenumber $k_\mathrm{max} = \SI{0.5}{\h\per\Mpc}$, which yields a slightly less sensitive sensitivity than the proposed galaxy survey MegaMapper~\cite{Schlegel:2019eqc}~(cf.~\cite{Sailer:2021yzm}), and the proposed \SI{21}{cm}~experiment PUMA~\cite{Slosar:2019gvt}. In addition, we include cosmic-variance-limited~(CVL) observations of~LSS up to $z_\mathrm{max}=6$, over half the sky with $k_\mathrm{max}=\SI{0.75}{\h\per\Mpc}$, and of the~CMB up to $\ell_\mathrm{max}^T = 3000$ and $\ell_\mathrm{max}^P = 5000$ over 75\%~of the sky. The LSS~forecasts with $\omega_\mathrm{lin} \lesssim \SI{100}{\Mpc}$ should be treated cautiously since these low frequencies are more sensitive to the details of signal modeling. We also note that a reconstruction efficiency of~50\% was assumed which should be surpassed in the future thanks to further theoretical developments. Overall, LSS~surveys have the potential to improve over the~CMB by more than an order of magnitude, while the~CMB will always dominate the reach in feature frequency. We refer to~\cite{Beutler:2019ojk} for further details.}
	\label{fig:feature_forecast}
\end{figure}

To forecast the sensitivity of cosmological surveys to primordial features, it is useful to estimate the bounds on a linear feature model and decompose other feature models into a sum of linear oscillations as discussed above. While the precise bounds for specific models might differ from these estimates, this `feature spectrometer' allows to easily compare the sensitivity of different probes and experiments. We display such forecasts in Fig.~\ref{fig:feature_forecast},
which show that future measurements of the CMB~anisotropies will be able to gradually improve over the current constraints from Planck and dominate the sensitivity for the smallest and especially the largest feature frequencies~$\omega_\mathrm{lin}$.\medskip

In addition to these dedicated template searches, several approaches to directly infer the primordial power spectrum, and therefore also any potential features, through non-parametric reconstruction have been developed since~\cite{Bridle:2003sa}. To this end, penalized likelihood reconstruction~\cite{Gauthier:2012aq}, Bayesian reconstruction~\cite{Vazquez:2012ux, Aslanyan:2014mqa, Canas-Herrera:2020mme}, cubic spline reconstruction~\cite{Guo:2011re}, Richardson-Lucy reconstruction~\cite{Shafieloo:2003gf, Hazra:2014jwa, Chandra:2021ydm, Chandra:2021qay}, generalized slow-roll methods~\cite{Kadota:2005hv, Dvorkin:2009ne, Hu:2011vr}, principle component analysis~\cite{Leach:2005av, Dvorkin:2010dn, Dvorkin:2011ui} and further methods have been applied to CMB~data. As in the template searches, these techniques also point to a featureless power spectrum within current error bars and over the range of scales accessible in the~CMB~\cite{Dvorkin:2010dn, Dvorkin:2011ui, Planck:2013jfk, Achucarro:2014msa, Planck:2015sxf, Aghamousa:2017uqe, Obied:2017tpd, Planck:2018jri, Obied:2018qdr}.

\paragraph{CMB~Spectral Distortions}~\\
Spectral distortions of the CMB~black body spectrum provide an entirely complementary window on the primordial power spectrum and its potential departures from a power law on small scales since they are sensitive to the primordial amplitude at scales of $k \simeq \SIrange{1}{e4}{\per\Mpc}$~(cf.\ e.g.~\cite{Barnaby:2009dd, Chluba:2012we, Chluba:2015bqa, Chluba:2019kpb}). These scales cannot be accessed in the anisotropy signal of the~CMB and would be very challenging to reliably extract from LSS~data. Departures from a featureless power spectrum~$\Delta_{\mathcal{R},0}^2(k)$ will result into a net energy release or deficit that is potentially detectable as a net distortion signal. In particular bumps or troughs in the primordial power spectrum can therefore be constrained by spectral distortion measurements. On the other hand, oscillatory features usually result in a small average effect since power enhancements and deficits cancel out. This implies that spectral distortions might only be observable for low-frequency oscillations that span a significant range of wavenumbers~\cite{Chluba:2015bqa}. In the future, interesting constraints on primordial features might be derived from measurements by an experiment such as PIXIE~\cite{Kogut:2011xw} or PRISM~\cite{Andre:2013nfa}. Additional theory and analysis development could therefore shed more light on the potential constraining power of CMB~spectral distortions on small-scale features.

\subsubsection{Large-Scale Structure}

Primordial features are also imprinted in the large-scale structure of the universe. The information in the primordial power spectrum~$\Delta_\mathcal{R}^2(k)$ is transferred to the (linear)~matter power spectrum according to $P(k,z) \propto k\, T(k)^2 D(z)^2\, \Delta_\mathcal{R}^2(k)$, which is the usual linear evolution from the initial conditions with linear growth rate~$D(z)$ and transfer function~$T(k)$. The inflationary imprints are therefore more directly imprinted in LSS~observables than in the CMB~anisotropies, cf.~\eqref{eq:clscalar}, but are additionally processed by the nonlinear gravitational evolution in the late universe. This has to be taken into account in any LSS~search. Another advantage of LSS~probes is the fact that the number of available signal-dominated LSS~modes grows approximately as~$k_\mathrm{max}^3 V_\mathrm{survey}$, with maximum wavenumber~$k_\mathrm{max}$ and survey volume~$V_\mathrm{survey}$, compared to~$\ell_\mathrm{max}^2$ for the~CMB, with the maximum multipole~$\ell_\mathrm{max}$. These points explain why analyses of current LSS~data have started to overtake the sensitivity of CMB~searches over a certain range of feature frequencies and will dominate the constraining power for a decisive part of parameter space in the future~\cite{Beutler:2019ojk}. In the following, we will separately discuss optical galaxy surveys, which are the current frontier in the search for primordial oscillations, and line intensity mapping, which will dominate the sensitivity in the more distant future~\cite{Ansari:2018ury}. As in the case of CMB~anisotropies, it is useful to constrain both dedicated models and the model-agnostic templates of~\eqref{eq:linear_features} and~\eqref{eq:logarithmic_features}.

\paragraph{Galaxy Surveys}~\\
The cosmological use of the large number of modes that are in principle accessible in galaxy surveys is usually limited by gravitational nonlinearities, baryonic physics on small scales and observational shot noise. While forecasts had indicated improvements in sensitivity for future LSS~surveys in combination with CMB~experiments, these limitations restricted the observational reach~\cite{Huang:2012mr, Chantavat:2010vt, Hu:2014hra, Benetti:2016tvm, Chen:2016vvw, Ballardini:2016hpi, Fard:2017oex, Palma:2017wxu, LHuillier:2017lgm, Ballardini:2017qwq, Zeng:2018ufm, Debono:2020emh}. However, recent advances in the theoretical understanding of these effects on the feature imprints in the LSS~spectra now allow to employ not only linear scales, but also those in the (weakly)~nonlinear regime: large-scale gravitational bulk flows can be resummed in perturbation theory and treated analytically, leading to an exponential damping of the primordial signal~\cite{Vasudevan:2019ewf, Beutler:2019ojk}~(cf.~\cite{Vlah:2015zda, Ballardini:2019tuc, Chen:2020ckc} for simulation-based confirmations) and small-scale nonlinearities should not impact these inflationary oscillations as long as their frequency is large enough~\cite{Beutler:2019ojk}. In addition, it suffices to model the oscillatory part of the power spectrum, i.e.\ we do not need to model its full shape, which is an easier problem and can be achieved to smaller scales than the full nonlinear treatment of biased tracers~\cite{Beutler:2019ojk}. Consequently, LSS~constraints that are independent of and competitive with those from the CMB~anisotropies can be inferred from present data for the currently accessible feature frequencies~\cite{Beutler:2019ojk}. In addition, significantly better bounds should be achievable in the upcoming decade~\cite{Beutler:2019ojk, Schlegel:2019eqc, Ballardini:2019tuc, Sailer:2021yzm} with the next generation of LSS~surveys~\cite{Snowmass2021:3dLSS}, as illustrated in~Fig.~\ref{fig:feature_forecast}.

These forecasts show that LSS~surveys have a smaller dynamical range in the feature frequency~$\omega_\mathrm{lin}$ than CMB~experiments. This is in particular due to (i)~the largest available scales in real space being intrinsically smaller since a comoving scale per radian is considerably larger at the surface of last scattering and (ii)~a smaller range of scales available from the fundamental mode to the onset of nonlinear evolution. On the other hand, LSS~observations have several advantages. While the large number of usable signal-dominated modes drives the overall sensitivity of LSS~feature searches, the shape of the LSS~transfer function~$T(k)$ is smoother than that of the~CMB. This results in a larger intrinsic signal in LSS~observables compared to the~CMB which implies that primordial features are in principle easier to find in the matter spectra than in the CMB~spectra. In addition, spectroscopic galaxy surveys can probe very large volumes and have a full three-dimensional sampling of the underlying density fluctuations, which means that the maximum oscillation frequency is limited entirely by the volume of the survey since a larger volume implies a smaller fundamental frequency and, in turn, a higher maximal~$\omega_\mathrm{lin}$. In comparison, photometric surveys have in general the same maximal~$\omega_\mathrm{lin}$, but their sensitivity is restricted by the smearing of the primordial signal by the large radial kernels for weak lensing and galaxies with photometric errors, i.e.\ a similar effective averaging effect as for the anisotropies of the two-dimensional CMB~sky. Since these surveys are able to observe significantly more objects than spectroscopic surveys by several orders of magnitude, they may however remain competitive on the largest scales due to the raw number of objects.

The current best limits inferred from galaxy clustering data of~BOSS~DR12 alone are comparable to, but slightly stronger than those derived from current Planck~CMB~data for the accessible range of feature frequencies~\cite{Beutler:2019ojk}~(see~\cite{Ballardini:2022wzu} for the analogous BOSS~two-point correlation function analysis). This was made possible by the discussed theoretical advances which allowed to use all signal-dominated modes and, therefore, the full statistical power of~BOSS. In addition, since the predictions for these primordial signals in the statistical distributions of the CMB~anisotropies and LSS~probes are strictly correlated, these searches can also be combined, which has in particular been explored in~\cite{Hu:2014hra, Benetti:2016tvm, Zeng:2018ufm, Beutler:2019ojk}.

An ongoing field of theoretical development pertains to density field reconstruction~(see e.g.~\cite{Wang:2017jeq, Schmittfull:2017uhh, Hada:2018fde, Schmidt:2018bkr, Birkin:2018nag, Sarpa:2018ucb, Modi:2019hnu, Zhu:2019gzu}) which has the potential to effectively further decrease the nonlinear damping scale induced by large-scale bulk flows. In turn, this will increase the primordial signal in the LSS~power spectrum. For instance, an improvement in the forecasted sensitivity to~$A_\mathrm{lin}$ of factors in excess of 3, 2 and~1.5 for~BOSS, DESI and~LSS-CVL could be achieved with an increase of the reconstruction efficiency from about~50\% to~100\%~\cite{Beutler:2019ojk}. This is a considerable improvement compared to the results shown in Fig.~\ref{fig:feature_forecast} and substantially larger than what we expect for the BAO~frequency, i.e.\ the search for primordial features might give new motivation to develop more efficient reconstruction techniques. While the methods developed in the context of the BAO~frequency are theoretically expected to reconstruct any nonlinearly damped peak in the matter correlation function, this was recently shown explicitly for a set of low-frequency feature models~\cite{Li:2021jvz}.

\paragraph{Line Intensity Mapping}~\\
Future \SI{21}{cm}~and other line intensity mapping~(LIM) surveys, such as the Stage-\textsc{ii} experiment PUMA~\cite{Ansari:2018ury, Bandura:2019uvb, Castorina:2020zhz}, millimeter-wave intensity mapping experiments probing rest-frame infrared lines from dusty star-forming galaxies~\cite{Snowmass2021:LIM} or surveys targeting the era prior to reionization~(c.f.~\cite{Snowmass2021:3dLSS, Snowmass2021:21cm}), hold the promise to improve the constraints from galaxy surveys by another few orders of magnitude~\cite{Chen:2016zuu, Xu:2016kwz, Ansari:2018ury, Beutler:2019ojk, Sailer:2021yzm}. The primordial signal is imprinted in these observables in the same way as in other LSS~tracers, but these experiments are able to observe to higher redshifts. This means that the raw statistical power is significantly larger since there is three times more comoving volume available in the redshift range $z=2-6$ compared to $z<2$, for instance. This also implies an important increase in the maximum feature frequency that can be probed. In addition, the universe is more linear at earlier times and can therefore be described using (resummed)~perturbation theory to smaller scales which allows an increase by a factor of at least two in the maximum wavenumber in data analyses.

On the other hand, the full three-dimensional LIM~information is obscured by the foreground wedge and interloper lines for \SI{21}{cm}~and \si{mm}-wave~observations, respectively. More generally, it remains a challenge to separate the cosmic signal from the galactic and extragalactic foregrounds which are larger by many orders of magnitude. There are however ongoing efforts to mitigate the foregrounds and extract the cosmic signal~(see e.g.~\cite{Snowmass2021:21cm}). For instance, cross-correlations between different probes, such as intensity maps with various spectral lines or intensity maps with galaxy/CMB~surveys, can significantly help in foreground cleaning since different observables have different~(presumably uncorrelated) foregrounds and systematics.

The \SI{21}{cm}~signal is in principle observable out to redshift $z\sim 200$ which would provide the ultimate goal for primordial feature searches. At redshifts beyond $z \sim 6$, even more comoving volume is available and the nonlinear scale is beyond the Jeans scale for redshifts $z \gtrsim 30$. For this reason, features could be observed on scales far beyond the reach of CMB~and galaxy surveys~\cite{Chen:2016zuu}. However, close-packed interferometers with large baselines~($>\SI{10}{km}$) are required to observe the \SI{21}{cm}~signal out to these redshifts with sufficient resolution and sensitivity. Because the ionosphere becomes opaque at the frequencies corresponding to this redshifted \SI{21}{cm} signal, such experiments would have to be operated in space or on the back side of the moon~\cite{Silk:2020bsr}. While futuristic, several pilot studies are currently under consideration at both~ESA~\cite{Koopmans:2019wbn} and~NASA~\cite{Burns:2020gfh}~(see also~\cite{Snowmass2021:21cm}).

\subsubsection{Gravitational Wave Background}
\label{sec:features_gwb}

The stochastic gravitational wave background provides an observational channel that is particularly relevant to searches for primordial features at scales much smaller than those accessible in CMB~and galaxy surveys. Primordial features are imprinted in the~SGWB through the sourcing of tensor fluctuations from scalar fluctuations at nonlinear order~\cite{Acquaviva:2002ud, Mollerach:2003nq, Ananda:2006af, Baumann:2007zm}. On CMB~scales, the constraints on the amplitude of scalar fluctuations imply that the irreducible contribution to the GW~spectrum from vacuum fluctuations dominates over this scalar-induced SGWB~signal for scenarios with a tensor-to-scalar ratio~$r$ which is detectable in B-mode searches of present or upcoming CMB~experiments. However, the amplitude of scalar fluctuations is much less constrained on small scales, $k \gg \SI{1}{\per\Mpc}$, and permits a potentially observable scale-dependent feature contribution to the~SGWB.\medskip

Tensor modes associated with a primordial feature are produced twice during the history of the universe: (i)~at the time that the feature is produced during inflation and (ii)~when the scalar fluctuations affected by the feature re-enter the horizon in the post-inflationary era.\footnote{We neglect the mixed component which receives contributions from both periods since~$\Omega_\mathrm{GW}^\mathrm{inf}(k)$ and~$\Omega_\mathrm{GW}^\mathrm{post}(k)$ dominate the signal and are therefore most relevant for observations~\cite{Fumagalli:2021mpc}.} In the context of GW~observations, the relevant quantity is the gravitational wave energy density fraction per $\log(k)$~interval, which we denote by~$\Omega_\mathrm{GW}^\mathrm{inf}(k)$ and~$\Omega_\mathrm{GW}^\mathrm{post}(k)$ for the inflationary and post-inflationary contributions, respectively. These background quantities exhibit an oscillatory modulation similar to the primordial signal in~$\Delta_\mathcal{R}^2(k)$.

\begin{figure}[t]
	\centering
	\includegraphics{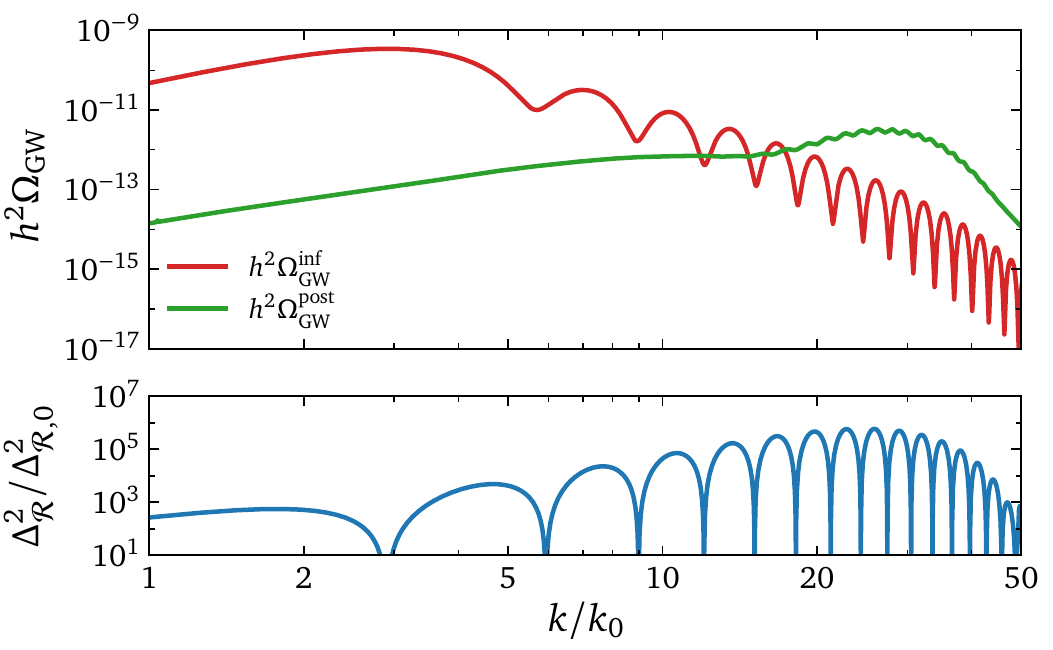}
	\caption{\emph{Top:}~Contributions to the stochastic gravitational wave background sourced during inflation~($\Omega_\mathrm{GW}^\mathrm{inf}$) and induced in the post-inflationary era during a phase of radiation domination~($\Omega_\mathrm{GW}^\mathrm{post}$) due to a sharp feature which induces an excited state with copious particle production. Both spectra exhibit oscillatory modulations albeit with different amplitudes. \emph{Bottom:}~Scalar power spectrum associated with the same sharp feature which exhibits $O(1)$~oscillations and follows the form of the linear feature template~\eqref{eq:linear_features} with a scale-dependent envelope.}
	\label{fig:GWfeature}
\end{figure}

The inflationary contribution~$\Omega_\mathrm{GW}^\mathrm{inf}(k)$ due to a sharp feature was computed explicitly in ~\cite{Fumagalli:2021mpc}~(see also e.g.~\cite{Peng:2021zon, Cai:2021wzd} for explicit realizations and~\cite{An:2020fff} for similar GW~profiles arising from instantaneous sources during inflation). For the phenomenologically most-relevant case of the feature sufficiently amplifying~$\Omega_\mathrm{GW}^\mathrm{inf}(k)$ to be potentially observable, the GW~spectrum was found to exhibit $O(1)$~modulations in the UV~tail with frequency~$\omega_\mathrm{lin} = 2/k_0$~(see the red curve in Fig.~\ref{fig:GWfeature}).
The origin of this oscillation is the same as for the modulations in~$\Delta_\mathcal{R}^2(k)$: The oscillation with frequency $\omega_\mathrm{lin} = 2/k_0$ is a direct consequence of the sharp feature inducing an effective excited state which further implies an $O(1)$~oscillation amplitude if it is associated with moderate to copious particle production.

The post-inflationary contribution~$\Omega_\mathrm{GW}^\mathrm{post}(k)$ due to sharp, resonant and standard clock features has been analyzed in~\cite{Fumagalli:2020nvq, Braglia:2020taf, Fumagalli:2021cel, Dalianis:2021iig, Witkowski:2021raz}. The oscillation in~$\Omega_\mathrm{GW}^\mathrm{post}(k)$ can be understood as a superposition of several resonance peaks from the resonant amplification of the peaks associated with the modulation in~$\Delta_\mathcal{R}^2(k)$~\cite{Fumagalli:2020nvq}. As these resonance peaks have a finite width in wavenumber, their superposition averages the oscillation so that $\Omega_\mathrm{GW}^\mathrm{post}(k)$~only exhibits $O(10\% - 20\%)$ modulations, even if the associated scalar power spectrum~$\Delta_\mathcal{R}^2(k)$~(and~$\Omega_\mathrm{GW}^\mathrm{inf}$) contains $O(1)$~oscillations~(see the green curve in Fig.~\ref{fig:GWfeature}). Interestingly, the contribution~$\Omega_\mathrm{GW}^\mathrm{post}(k)$ also encodes information about the post-inflationary era when the~GWs are induced, such as the frequency of the oscillation in the GW~spectrum being $\omega_\mathrm{lin}^\mathrm{GW}= c_\mathrm{s}^{-1} \omega_\mathrm{lin}$, with the propagation speed~$c_\mathrm{s}$ of scalar fluctuations~\cite{Witkowski:2021raz}. The amplitude of the oscillation and the shape of the envelope of~$\Omega_\mathrm{GW}^\mathrm{post}(k)$ are additionally sensitive to the equation of state when the~GWs are induced, e.g.~$w = c_\mathrm{s}^2 = 1/3$ during radiation domination. For a resonant feature, the post-inflationary GW~spectrum exhibits a superposition of two oscillatory terms, one with frequency~$\omega_\mathrm{log}$ and one with frequency $2 \omega_\mathrm{log}$~\cite{Braglia:2020taf, Fumagalli:2021cel}, with the respective amplitudes depending on $\omega_\mathrm{log}$, the shape of the peak in~$\Delta_\mathcal{R}^2(k)$ and the equation of state~$w$ during the GW~sourcing. The templates that have been derived in~\cite{Fumagalli:2020nvq, Braglia:2020taf, Fumagalli:2021cel, Witkowski:2021raz} for the post-inflationary contribution to the~SGWB capture these aspects.\medskip

Observationally, the amplitude of features has to be much larger to be detectable in the stochastic gravitational wave background compared to the discussed CMB~anisotropy and LSS~measurements. On the other hand, the SGWB is sensitive to scales $k \gg \SI{1}{\per\Mpc}$ and could potentially cover up to 23~e-folds of unchartered territory since GW~observations may probe scales from $k \sim \SI{e6}{\per\Mpc}$ with pulsar timing arrays~\cite{RoperPol:2022iel} to $k \sim \SI{e16}{\per\Mpc}$ with next-generation ground-based interferometers~(e.g.~Cosmic Explorer~\cite{Reitze:2019iox} and the Einstein Telescope~\cite{Punturo:2010zz}, cf.~\textsection\ref{sec:pgw_gwb}). The sensitivity of the space-based interferometer~LISA~\cite{LISA:2017pwj} will peak in the \si{mHz}~regime which corresponds to inflationary features that occurred some 30~e-folds after CMB~modes exited the horizon or $k \sim \SI{e12}{\per\Mpc}$. An initial investigation of the prospects to detect these imprints with~LISA suggests that a possible detection of $O(10\%)$~oscillations from the post-inflationary contribution requires an overall amplitude of $h^2 \Omega_\mathrm{GW}^\mathrm{post} > \numrange{e-12}{e-11}$~\cite{Fumagalli:2021dtd}. This would imply $\Delta_\mathcal{R}^2 \gtrsim \num{e-3}$ since $h^2 \Omega_\mathrm{GW}^\mathrm{post} \sim \num{e-5} \Delta_\mathcal{R}^4$ during radiation domination. To measure the oscillations in the inflationary contribution with~LISA, the peak amplitude of $h^2 \Omega_\mathrm{GW}^\mathrm{inf}$ would need to be even higher because the oscillations only occur in the ultraviolet tail. At the same time, the next-generation space-based GW~observatories, such as DECIGO~\cite{Kawamura:2011zz}, will however further increase the signal-to-noise ratio and bridge the frequency gap between~LISA and the ground-based observatories. This will allow us to probe the details of these feature signals and their inflationary origin.
\clearpage
% !TEX root = whitepaper.tex

\section{Conclusions}
\label{sec:conclusion}

The standard model of cosmology rests on three pillars: the visible sector, the dark sector and the initial conditions. 
We observe the universe through observations of the visible sector, gathering information about the distribution of light and matter in the universe. In turn, we can infer the presence of an additional, mysterious dark sector which dominates the energy budget of the universe today, and consists of dark matter and dark energy. Dark matter provides the scaffolding around which galaxies clump and largely dictates the large-scale structure of the universe, while dark energy affects the geometry of the universe and drives its current accelerated expansion. Finally, observations indicate that we need very special initial conditions that imprinted tiny fluctuations at the big bang. These three pillars therefore provide us with the best picture of the history, composition and structure of the universe. In this paper, we focus on the initial conditions of the universe and how observations of the visible sector enable their inference to ultimately probe the main paradigm that explains their origin, cosmic inflation.\medskip

Since all models of inflation predict the existence of primordial gravitational waves and primordial non-Gaussianity, and many include deviations from the almost scale-invariant power spectrum of primordial fluctuations, these are the three signatures we center our attention on. We note that in a large class of models, important thresholds are within reach in the next decade given the ever increasing sensitivity of cosmological surveys. The relevant observational probes are (i)~the cosmic microwave background and tracers of the large-scale structure, which are sensitive to the largest scales of the primordial spectrum, and (ii)~spectral distortions of the CMB~black body spectrum and direct observations of the stochastic gravitational wave background, which will provide valuable information on smaller scales.

Historically, the detailed measurement of the anisotropies and polarization of the~CMB has been leading our observational insights. Next-generation experiments are projected to tremendously increase our sensitivity to many of these imprints. In particular, an experiment such as CMB-S4, which covers both large scales and large sky area, will be instrumental in reaching important thresholds on all three fronts, and uniquely for primordial gravitational waves. For primordial non-Gaussianity and features, the sensitivity of the~CMB is currently complemented by galaxy surveys, but a range of different observations of tracers of the large-scale structure of the universe will lead the observational sensitivity for a number of important signatures in the future. On smaller scales, CMB~spectral distortions and direct gravitational wave observations have started to add valuable information for our understanding of inflation on a broad range of scales and are projected to significantly improve in the future.\medskip

Given the anticipated observational advances, in the following, we give a bird's eye view on the implications of a detection of any one of these signatures of new physics~(or absence thereof), any of which would tremendously impact our understanding of the early universe and high-energy physics:
\begin{itemize}
	\item \textbf{Primordial gravitational waves}: In simple inflationary models, the amplitude of~PGWs reveals the energy scale at which inflation occurred. Given the expected reach of planned CMB~experiments, a detection of~PGWs would imply an energy scale near the scale of grand unification. The quantum origin of these fluctuations of the metric itself implies that a detection would provide evidence for quantum gravity. Specifically, current and planned CMB~experiments targeting primordial B-mode polarization will cross important thresholds for the tensor-to-scalar ratio~$r$: $r \simeq 0.01$ and $r \simeq 0.001$. A detection of $r \gtrsim 0.01$ would provide evidence for the existence of an approximate shift symmetry in quantum gravity; a detection of $r \gtrsim 0.001$ would provide evidence for the simplest models of inflation which naturally predict the observed values of the spectral tilt~$\ns$ and have a characteristic scale that exceeds the Planck scale. In turn, a non-detection of~PGWs at the sensitivity of upcoming experiments will vastly restrict the space of viable inflationary models and provide important insights into what nature accommodates at these extreme scales. The key challenge to obtaining robust upper limits on or a decisive detection of~$r$ are astrophysical foregrounds and weak gravitational lensing of the~CMB, which require continued developments in modeling, simulations and analysis.

	\item \textbf{Primordial non-Gaussianity}: A detection or constraint of~PNG will teach us about the interactions of primordial fluctuations. Interpreting a detection depends on the overall amplitude~$\fnl$ and the PNG~shape function. For local non-Gaussianity, models with \mbox{$\fnl^\mathrm{local} > 1$} typically point to the existence of extra light species active during or after inflation. For equilateral and orthogonal non-Gaussianity, models with $\fnl > 1$ tend to favor scenarios with a strong breaking of boost symmetries of the inflationary background~(or small sound speed of the scalar perturbations). Moreover, around the squeezed limit, there is the exciting possibility of imprints left by new particles, with masses all the way to the inflationary Hubble scale. This is much heavier than any direct detection experiment currently available. Learning which new particles and symmetries play a role during inflation will open new avenues in building inflationary models and connections to the Standard Model of particle physics. Conversely, stronger bounds on these shapes of non-Gaussianity will constrain large classes of inflationary models and point to favored directions in ``theory space''. The main observational challenges ahead lie in separating the primordial signal from late-time effects in CMB~and LSS~probes, and mitigating sources of noise and nuisance. Theoretically modeling~(or simulating) the late-time observables for a given PNG~signal is also an important challenge \mbox{for the immediate future.}

	\item \textbf{Primordial features:} Deviations from the minimal power-law power spectrum of primordial fluctuations indicate that new energy scales play an important role in the inflationary dynamics. This occurs pervasively when connecting models of inflation to theories of fundamental physics. Relatedly, there are currently no useful theoretical priors on the scale or amplitude of primordial features, which necessitates broad cosmological searches covering as much of parameter and model space as possible. A potential detection of these primordial imprints could have profound implications for our understanding of fundamental physics, while upper limits can still inform model building efforts and narrow the vast theoretical possibilities provided by physics at the highest energies. Analyses already exploit that the separation of the primordial signal from late-time effects in CMB~and LSS~observations is easier for oscillatory features than for primordial non-Gaussianities. While the observational prospects for feature searches are therefore bright, further theoretical advances will however be required, in particular in efficient data analysis techniques to even more efficiently extract information across various cosmological probes on even smaller scales and with higher significance.
\end{itemize}
Achieving these targets with further theoretical advances in combination with the next generation of CMB~and LSS~surveys, which will yield unprecedented maps of the universe, will therefore provide us with an exquisite window into the primordial universe.\medskip

The future for inflationary cosmology is very bright, with clear directions for further observational and theoretical explorations. Upcoming surveys of various observational probes with increasing instrumental sensitivity, together with new developments in theory, simulations, modeling and analysis, will enable us to probe this unique and spectacular epoch of our cosmic history. Discoveries teaching us about the highest accessible energy densities of the universe can be within reach in the near future. In our quest to understand the beginning of the universe, nature has also given us the opportunity to learn about physics at the most microscopic scales by making observations at the largest distances.

\vskip20pt
\paragraph{Acknowledgments}~\\
A.\,A.~acknowledges support by the Netherlands Organization for Scientific Research~(NWO), by the Basque Government~(\mbox{IT-979-16}) and by the Spanish Ministry~MINECO~(FPA2015-64041-C2-1P). M.\,Bi.~acknowledges support from the Netherlands Organization for Scientific Research~(NWO), which is funded by the Dutch Ministry of Education, Culture and Science~(OCW), under VENI~Grant~\mbox{016.Veni.192.210}. M.\,Br.~is supported by the Spanish Atracci\'on de Talento Contract No.~\mbox{2019-T1/TIC-13177} granted by Comunidad de Madrid, the \mbox{I+D}~Grant~\mbox{PID2020-118159GA-C42} of the Spanish Ministry of Science and Innovation and the \mbox{i-LINK~2021}~Grant~\mbox{LINKA20416} of~CSIC. R.\,C.~is supported in part by US~Department of Energy Award No.~\mbox{DE-SC0010386}. R.\,F.~was supported in part by the US~Department of Energy under Grant~\mbox{DE-SC0009919}, the Simons Foundation under Grant~SFARI~560536, and by~NASA under Grants~80NSSC18K1487 and~80NSSC18K0561. J.\,F.~is supported by a Contrato de Atracción de Talento~(Modalidad~1) de la Comunidad de Madrid~(Spain), number~\mbox{2017-T1/TIC-5520} and the IFT~Centro de Excelencia Severo Ochoa Grant~\mbox{SEV-2}. H.\,L.~is supported by the Kavli Institute for Cosmological Physics through an endowment from the Kavli Foundation and its founder Fred Kavli. P.\,D.\,M.~acknowledges support from the Netherlands Organization for Scientific Research~(NWO) VIDI~Grant~\mbox{639.042.730}. A.\,M.\,D.~acknowledges support from The~Tomalla Foundation for Gravity and from the SNSF~Project ``The Non-Gaussian Universe and Cosmological Symmetries'', Project Number~\mbox{200020-178787}. G.\,A.\,P.~is supported by a Fondecyt Regular Project Number~1210876~(ANID). G.\,L.\,P.~is supported by a grant of the Netherlands Organisation for Scientific Research~(NWO/OCW) and through the Delta-ITP consortium of~NWO/OCW. S.\,R.-P.~is supported by the European Research Council under the European Union's Horizon~2020 Research and Innovation Programme~(Grant Agreement No.~758792, Project~GEODESI). B.\,W.~was supported by the US~Department of Energy under Grants~\mbox{DE-SC0009919} and~\mbox{DE-SC0019035}, and the Simons Foundation under Grant~SFARI~560536. W.\,L.\,K.\,W.~is supported in part by the US~Department of Energy, Laboratory Directed Research and Development Program and as part of the Panofsky Fellowship Program at SLAC~National Accelerator Laboratory under Contract~\mbox{DE-AC02-76SF00515}.

\clearpage
\phantomsection
\addcontentsline{toc}{section}{References}
\bibliographystyle{utphysupd}
\bibliography{references}

\end{document}